\documentclass[useAMS,usenatbib,usegraphicx]{pasa}
\usepackage{booktabs}
\usepackage{courier}
\usepackage{multirow}
\usepackage[authoryear]{natbib}
\usepackage{lipsum}
\usepackage{caption}
\bibpunct{(}{)}{;}{a}{}{,}
\newcommand {\aple} {\ {\raise-.5ex\hbox{$\buildrel<\over\sim$}}\ }
\newcommand {\solmass}{\textrm{M}_{\odot}}

\usepackage{aas_macros}
\usepackage{hyperref} 
\hypersetup{colorlinks,citecolor=blue,linkcolor=blue,urlcolor=blue}

\title[Connecting the ISM to TeV PWNe and PWNe candidates.]{Connecting the ISM to TeV PWNe and PWNe candidates.}
 \author[Voisin et al]{F. J. Voisin$^{1}$\thanks{E-mail: fabien.voisin@student.adelaide.edu.au (RCB)}; G.\,P.\,Rowell$^{1}$; M.\,G.\,Burton$^{2\,3}$; Y.\,Fukui$^{4}$; H.\,Sano$^{4}$; F.\,Aharonian$^{6\,7}$; N.\,Maxted$^{5}$; C.\,Braiding$^{2}$; R.\,Blackwell$^{1}$; J.\,Lau$^{1}$; \\
\small $^1$ School of Physical Sciences, University of Adelaide, SA 5005, South Australia, Australia\\
$^2$ School of Physics, University of New South Wales, NSW\,2052, New South Wales, Australia\\
$^3$ Armagh Observatory and Planetarium, College Hill, Armagh BT\,61\,9DG, Northern Ireland \\
$^4$ Department of Physics, Nagoya University, Furo-cho, Chikusa-ku, Nagoya, 464-8601, Japan \\
$^5$ School of Science, University of New South Wales, Australian Defence Force Academy, Canberra, ACT 2600, Australia \\
$^6$ Dublin Institute for Advanced Studies, 31 Fitzwilliam Place,  Dublin 2, Ireland\\
$^7$ Max-Planck-Institut f\"ur Kernphysik, PO Box 103980, 69029 Heidelberg, Germany\\}

\begin{document} 
\begin{abstract}
We investigate the interstellar medium (ISM) towards seven TeV gamma-ray sources thought to be  pulsar wind nebulae (PWNe) using Mopra molecular line 
observations at 7mm [CS(1--0), SiO(1--0,v=0)], Nanten CO(1--0) data and the SGPS/GASS H\textsc{i} survey.
We have discovered several dense molecular clouds co-located to these TeV gamma-ray sources, which allows us to search for cosmic-rays (CRs)
coming from progenitor SNRs or, potentially, from PWNe.
We notably found SiO(1--0,v=0) emission towards HESS\,J1809$-$193, highlighting possible interaction between the adjacent supernova remnant SNR G011.0$-$0.0 and the molecular cloud at $d\sim 3.7$\,kpc.
Using morphological features, and comparative studies of our column densities with those obtained from X-ray measurements, we claim a distance  
$d\sim8.6-9.7\,$kpc for SNR\,G292.2$-$00.5, $d\sim3.5-5.6$\,kpc for PSR\,J1418$-$6058 and $d\sim1.5$\,kpc for the new SNR candidate found towards HESS\,J1303$-$631. 
From our mass and density estimates of selected molecular clouds, we discuss signatures of hadronic/leptonic components from PWNe and their progenitor SNRs.
Interestingly, the molecular gas, which overlaps HESS\,J1026$-$582 at $d\sim$5\,kpc, may support a hadronic origin. 
We find however that this scenario requires an undetected cosmic-ray accelerator to be located at $d<10$\, pc from the molecular cloud.
For HESS\,J1809$-$193, the cosmic-rays which have escaped SNR G011.0$-$0.0 could contribute to the TeV gamma-ray emission.
Finally, from the hypothesis that at most 20\% the pulsar spin down power could be converted into CRs, we find that, among the studied PWNe, only those from PSR\,J1809$-$1917 could potentially contribute to the TeV emission.\\


\end{abstract}
\maketitle

{\center }

\bigskip

\section{Introduction}
Pulsar wind nebulae (PWNe) represent the majority of the identified  TeV gamma-ray sources in the Galactic plane  \citep{HGPS2018,HESSPWN2018}. 
The TeV emission is generally believed to be  of leptonic origin, where high energy electrons are accelerated after crossing  the pulsar termination shock, and scatter soft photons to produce 
inverse-Compton (IC) radiation at TeV gamma-ray energies.
The interstellar medium (ISM) greatly affects the morphology of the PWN observed from radio up to gamma-ray energies.
For instance, the interaction between the progenitor SNR and a nearby molecular cloud (MC) leads to an offset position of the pulsar with respect to the TeV peak along the pulsar-MC axis \citep{Blon2001}.

In this paper, we investigate the ISM towards seven TeV PWNe and PWNe candidates \citep{FermiPWN,HGPS2018}.
We make use  of the Nanten CO(1--0) survey \citep{Fukui}  to illustrate the wide-field morphology of the diffuse molecular gas towards the PWNe and 7mm Mopra spectral line observations to probe the dense molecular, and possibly shocked, gas along with star forming regions.
An accurate description of the ISM  can help explain the morphology of the PWN (e.g Vela X, \citealt{VelaCO,Velaslane}).
Linking each TeV source to its local ISM can also provide additional constraints on its distance.
We can  identify target material for hadronic components (i.e. cosmic-rays) escaping the TeV source, for example from a progenitor supernova remnant (e.g.\,\citealt{voisin2016a}).
Lastly, by combining  ISM mapping with the improved sensitivity and angular resolution of the next generation Cerenkov Telescope Array (CTA, \citealt{CTAproject,CTAscience}), we will be able to study the diffusion
process of high energy particles towards and inside the ISM clouds.

In section\,\ref{sec:ISManalysis}, we briefly outline the technical properties of the Mopra and Nanten telescopes as well as the analysis used for our 7mm Mopra data reduction.
We provide the results towards the different individual sources in section\,\ref{sec:result} and discuss the nature of the TeV source in section\,\ref{sec:discussion}.

\section{ISM Data and Analysis Procedure}
\label{sec:ISManalysis}
\subsection{7mm Mopra data and Analysis}
We conducted $20^{\prime}\times20^{\prime}$ and $10^{\prime}\times10^{\prime}$ observations of the ISM towards several HESS TeV sources as part of our `MopraGam' survey\footnote{http://www.physics.adelaide.edu.au/astrophysics/MopraGam/}.
The 7mm observations towards these studied sources were carried out between 2012 and 2015.
Table \ref{coverage} indicates the position and size of the different observations towards the seven regions which were mapped.

We used the Mopra spectrometer MOPS in `zoom' mode, which records 16 sub-bands, each consisting of 4096 channels over a 137.5 MHz bandwidth.
The `on the fly mapping' (OTF) Nyquist sampled  these regions with a 1$^{\prime}$ beam size and a velocity resolution of $\sim 0.2$\,km/s.
We can thus simultaneously observe tracers that can help understand the structure and morphology of molecular clouds.
Among the tracers observed in this paper, both the carbon monosulfide and cyanoacetylene transitions CS(1$-$0) and HC$_3$N(5$-$4,F=4$-$3) can be found in dense molecular gas and star forming regions \citep{Irvine1987}. 
Silicon monoxide SiO emission can also be detected inside shocked dense molecular clouds \citep{Schilke,Gusdorf} and is a good signpost to claim physical association between a SNR and a nearby molecular cloud (e.g see \citealt{Nicholas7mm}). 
Finally, the 44\,GHz methanol class I maser CH$_3$OH(I) generally indicates nearby active star forming regions \citep{Voronkov}.  \newline

\noindent{\emph{\underline{Data reduction}}}\\
We used {\tt Livedata}\footnote{http://www.atnf.csiro.au/computing/software/Livedata/}  to produce the spectra of each observation, calibrated by an OFF position, and subtracted the baseline using a linear fit.
Then, we used {\tt Gridzilla}\footnote{http://www.atnf.csiro.au/computing/software/Gridzilla/} to obtain a 3D cube showing the variation of the antenna temperature $T_{\text{A}}^{*}$ as a 
function of position and line of sight velocity $v_\text{lsr}$.
Each region was mapped with a pixel spacing of $15^{\prime\prime}$.
After a careful look at the data, we recursively performed linear and sinusoidal fits  on each pixel in order to remove baseline ripples.
Notably, each sinusoidal fit  was performed with different initial wavelengths to account for ripples of various sizes.
Velocity ranges with significant emission were masked during these fits.
The cleaned data cubes were then smoothed via a Gaussian with a Full-Width-Half-Maximum (FWHM) of  $1.25^{\prime}$ so as to mitigate spatial fluctuations.
Finally, the cubes were Hanning smoothed using 5 channels in order to remove spikes, reduce the $T_\text{rms}$, and highlight broader CS(1--0) emission.
An overview of the reduction can be found in Appendix\,\ref{sec:integratedmap}.\\

\noindent{\emph{\underline{Producing CS(1--0) integrated intensity maps}}}\\
We used the 7mm CS(1--0) transition  to probe the denser gas. 
The standard method to produce CS(1--0) integrated intensity maps is to sum the temperature $T_\text{A}^{*}$ of each channel within a given velocity range.
However, in the case where detections span a large velocity range ($\Delta v_\text{lsr}\sim10\text{ to }30$\,km/s), any narrow CS(1--0) detections are likely to fall below the noise level, as a large portion of the noise may be included.
In order to deal with this issue, we adopted the method described in Appendix\,\ref{sec:integratedmap}.
This method, although more complex, provides a lower $T_\text{rms}$ and cleaner integrated intensity maps, which could be used for multi-wavelength studies towards TeV sources.\newline

 \begin{table}
  \centering
  \caption{Mopra 7mm coverage of the TeV sources studied in this paper. The central position in (RA, Dec) and size is given for all observations undertaken towards 
  the TeV sources.}
  \begin{tabular}{cp{2.5cm}c}
  \toprule
  \toprule
  TeV sources & central position & Area \\
  & ($\alpha^{\circ}$, $\delta^{\circ}$)& $\Delta l\times\Delta b$\\
  & (J2000.0)\\
  \midrule
  \midrule
  HESS\,J1018--589B& (154.7$^{\circ}$,-58.9$^{\circ}$)& $20^{\prime}\times20^{\prime}$ \\
  \midrule
  \multirow{4}{*}{HESS\,J1026--582} & (156.8$^{\circ}$,-57.8$^{\circ}$)& $20^{\prime}\times20^{\prime}$\\ 
   & (156.4$^{\circ}$,-58.1$^{\circ}$)& $20^{\prime}\times20^{\prime}$ \\
   & (156.1$^{\circ}$,-58.4$^{\circ}$)& $20^{\prime}\times20^{\prime}$ \\
   & (155.9$^{\circ}$,-58.0$^{\circ}$)& $10^{\prime}\times10^{\prime}$\\
   \midrule
   HESS\,J1119--614& (169.7$^{\circ}$,-61.4$^{\circ}$)& $20^{\prime}\times20^{\prime}$ \\
   \midrule
   HESS\,J1303--631& (195.4$^{\circ}$,-63.1$^{\circ}$)& $20^{\prime}\times20^{\prime}$ \\
   \midrule
   HESS\,J1418--609& (215.1$^{\circ}$,-60.8$^{\circ}$)& $20^{\prime}\times20^{\prime}$ \\
   \midrule
   HESS\,J1420--607& (214.4$^{\circ}$,-60.9$^{\circ}$)& $20^{\prime}\times20^{\prime}$ \\
   \midrule
   \multirow{4}{*}{HESS\,J1809--193}& (272.4$^{\circ}$,-19.2$^{\circ}$)& $20^{\prime}\times20^{\prime}$ \\
   & (272.2$^{\circ}$,-19.5$^{\circ}$)& $20^{\prime}\times20^{\prime}$ \\
   & (272.5$^{\circ}$,-19.7$^{\circ}$)& $20^{\prime}\times20^{\prime}$ \\
   & (272.7$^{\circ}$,-19.4$^{\circ}$)& $20^{\prime}\times20^{\prime}$ \\
   \bottomrule
  \end{tabular}
  \label{coverage}
\end{table}

\noindent{\underline{\emph{Physical properties of CS(1--0) regions}}}\\
From the CS(1--0) integrated intensity maps, we have selected clumps whose angular diameter were equal or greater than our Mopra CS beam size $\theta_\text{FWHM}\sim1.6^{\prime}$.
At the component velocity range, we fitted the CS(1--0) emission with Gaussian distributions.

 We then used the Galactic rotation curve model from \citet{Blitz} to obtain near/far kinematic distance estimates based on the peak velocity of each detection.
In the case where the isotopologue C$^{34}$S(1--0) is detected, we use Eq.\,\ref{tauCS} to derive the averaged optical depth $\tau_{\text{CS(1--0)}}$, using the isotopologue ratio $\alpha=^{32}$S/$^{34}S\sim24$, 
based on terrestrial measurements \citep{Fink1981}. Otherwise, an optically thin scenario $\tau_{\text{CS(1--0)}}=0$ was adopted.
We obtained the column density of the upper state $N_{\text{CS}_1}$ using Eq.\,\ref{CSeq2}.
Assuming the gas to be in local thermal equilibrium, we thus obtained the total CS column density $N_{\text{CS}}$ using Eq.\,\ref{CSeq3}, assuming a kinetic temperature $T_{\text{kin}}$=10\,K typical of cold dense molecular clouds. 
The CS to H$_2$ abundance ratio $X_{\text{CS}}$ inside dense molecular clumps varies between $10^{-9}$ and $10^{-8}$ as suggested by \citet{Irvine1987}. In this work we chose $X_{\text{CS}}=4\times10^{-9}$ as per  
\citet{Zitchenko1994} who studied molecular clouds associated with star-forming regions. As a result, we expect our H$_2$ column density estimates to systematically vary by a factor of 2.
Finally, we use Eqs.\,\ref{H2mass} and \ref{H2dens} to determine the total mass $M_{\text{H}_2}\left(\text{CS}\right)$ and H$_2$ density $n_{\text{H}_2}\left(\text{CS}\right)$.
The Gaussian fits for the CS(1--0) components towards the regions of the individual TeV sources can be found in Tables \ref{J1809table} to \ref{tableJ1018} and their derived physical properties in Tables \ref{MASSJ1809} to \ref{MASSJ1018}.

\subsection{CO(1--0) Data and Analysis}
While the Mopra CO(1--0) survey is well under way (see \citealt{Mopra2015,Braiding2018}), we used for this work the 4 metre Nanten CO(1--0) survey \citep{Fukui}, as it encompassed all our sources.
 The Nanten telescope CO(1--0) survey covered the Galactic plane with a $4^{\prime}$ sampling grid, a velocity resolution $\Delta v=0.625$\,km/s, and an averaged 
noise temperature per channel of $\sim0.4$\,K. 
 
The CO(1--0) emission probes  the more diffuse molecular gas surrounding PWNe.
As PWNe expand inside a low density medium, we seek velocity ranges where the CO(1--0) emission spatially anti-corresponds with the TeV emission as that would support the PWN scenario (see Section\,\ref{sec:highenergyelectrons}).
Molecular gas may also provide sufficient target material to produce hadronic TeV emission. Therefore, probing extended CO(1--0) emission overlapping the TeV emission is also 
a powerful means to test the hadronic scenario in the vicinity of a cosmic-ray (CR) accelerator.

We arbitrarily selected CO regions based on the prominence of the CO(1--0) integrated emission compared to the rest of the maps and their proximity with the TeV gamma-ray source, as these 
can be helpful to test the hadronic/leptonic scenario (see section\,\ref{sec:discussion}).
In the case where the pulsar is offset from the observed TeV gamma-ray source (i.e HESS\,J1026$-$583), we highlight molecular regions which could produce such asymmetry along the pulsar-molecular cloud axis.
 We also derive physical parameters towards some extended regions defined from our CS detections (e.g see HESS\,J1809$-$193 and HESS\,J1418$-$609). Assuming 
all the gas traced by CO is embedded within this CS region, we are able to derive an upper-limit on the averaged density $n_{\text{H}_2}$.

We fitted the CO components at the velocity range of interest, with Gaussian distributions.
We then used the X-factor $X_{\text{CO}}=2\times10^{20}$\,cm$^{-2}$/(K km/s) to convert the CO integrated intensity into H$_2$ column density.
\citet{BolattoXco} have argued that this value is correct to $\sim30$\% across the Galactic plane.
We then used Eq.\,\ref{H2mass} to obtain the total molecular mass $M_{\text{H}_2}\left(\text{CO}\right)$ of the cloud, accounting for a 20\%  contribution from Helium. 
We finally assumed a prolate geometry to obtain the H$_2$ volume occupied by the molecular gas, and the averaged particle density $n_{\text{H}_2}\left(\text{CO}\right)$ using Eq.\,\ref{H2dens} (see Appendix\,\ref{sec:Mass}). 
The Gaussian fits for the CO(1--0) components towards the regions of the individual TeV source can be found in Tables \ref{J1809table} to \ref{tableJ1018} and their derived physical properties in Tables \ref{MASSJ1809} to \ref{MASSJ1018}.

\begin{figure*}
\centering
 \begin{minipage}{\textwidth}
  \includegraphics[width=\textwidth]{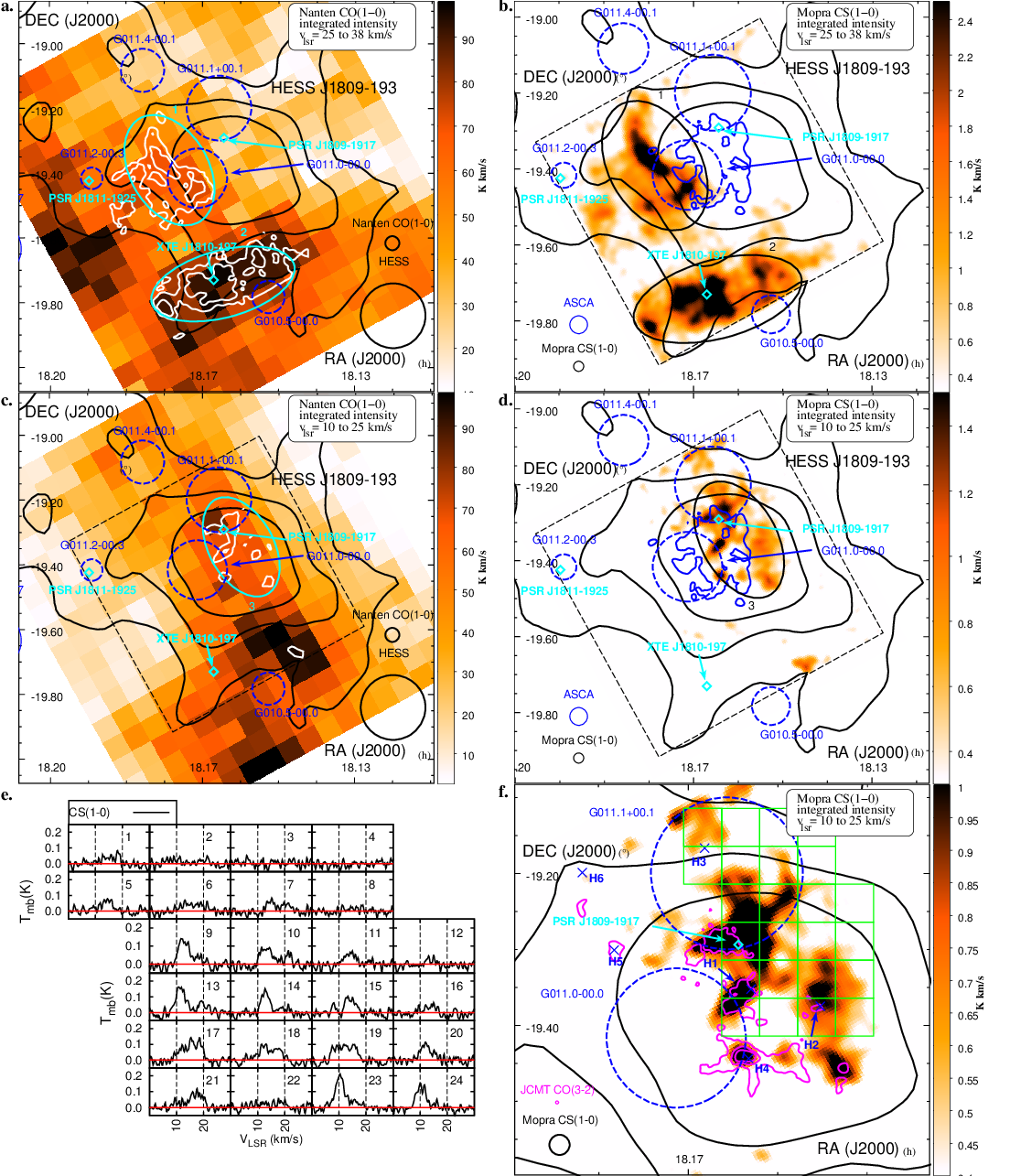}
  \caption{Nanten CO(1--0) and Mopra CS(1--0) integrated intensity maps across two velocity bands $v_\text{lsr}=25\text{ to }38$\,km/s (panels a. b.) and $v_\text{lsr}=10\text{ to }25$\,km/s (panels c. d.)
   towards HESS\,J1809$-$193 overlaid by the TeV gamma-ray counts in black contours \citep{HESSJ1809}.
  The dashed black box represents the area covered during our 7mm survey. 
  The ellipses selected for CO and CS analyses (see Section\,\ref{sec:sectionJ1809}) are shown in cyan and black respectively. 
  The SNRs are shown as dashed blue circles while the detected pulsars are shown as cyan diamonds.
   The ASCA hard X-ray (2 to 10\,keV) contours are displayed on the right panels in blue while the CS white contours overlays are shown on the left panels.
  A zoomed image of the CS(1--0) integrated intensity emission at $v_{\text{lsr}}=10\text{ to } 25$\,km/s is shown in panel f. overlaid by the JCMT CO(2--1) integrated intensity contours in magenta. The position of H\textsc{ii} regions `H1$\rightarrow$H5' are shown in blue crosses.
  The averaged CS(1--0) emission over the green grid of boxes in panel f. is displayed in panel e. (see colour version online).}
  \label{J1809COCS}
 \end{minipage}
\end{figure*} 

\subsection{HI analysis with SGPS/GASS surveys}
From the Southern Galactic Plane Survey (SGPS) and GASS H\textsc{i} surveys \citep{SGPS2005,GASS}, we also obtained the atomic H\textsc{i} column density $N_{\text{H}\textsc{i}}$  assuming the (optically thin) conversion factor $X_{\text{H}\textsc{i}}=1.8\times10^{18}$\,cm$^{-2}$/(K km/s) \citep{Dickey1990}.
H\textsc{i} self absorption may however occur towards colder H\textsc{i} regions and, as a result, the optically thin assumption may underestimate $N_{\text{H}\textsc{i}}$ by a factor of 2 \citep{FukuiHIgas2015}.
Comparing the total column density $N_{\text{H}}=N_{\text{H}\textsc{i}}+2N_{\text{H}_2}$ to the absorbed X-ray column density from X-ray counterparts can  provide further constraints on the distance to the TeV source.
The images showcasing the evolution of the total column density as a function of $v_\text{lsr}$ and distance towards several TeV sources can be found in Appendix\,\ref{sec:distance}.
Atomic gas should also be accounted for while testing the hadronic scenario towards TeV sources.
Consequently, as per the CO(1--0) emission, we derived the atomic masses $M_{\text{H}_I}$ towards the selected CO regions.
These masses can be found in Tables \ref{MASSJ1809} to \ref{MASSJ1018}. 

\section{The ISM found towards the TeV sources}
\label{sec:result}
\begin{figure*}
\centering
 \begin{minipage}{\textwidth}
 \includegraphics[width=0.87\textwidth]{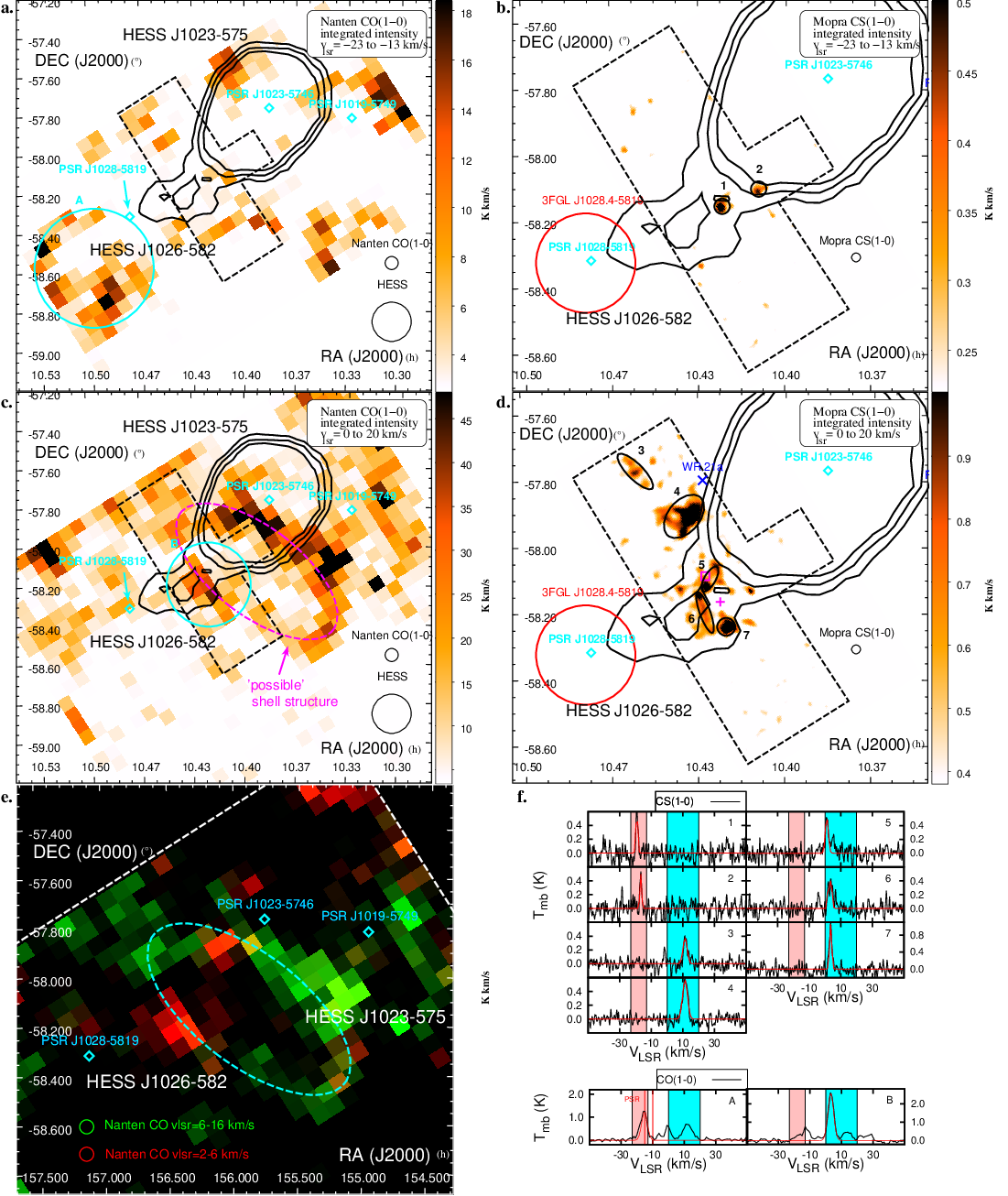}
  \caption{Nanten CO(1--0) and Mopra CS(1--0) emission between $v_{\text{lsr}}=-23\text{ to }13$\,km/s and  $v_{\text{lsr}}=0\text{ to }20$\,km/s towards \mbox{HESS\,J1026--582} and \mbox{HESS\,J1023--575} whose TeV gamma-ray counts are shown in black contours. The position of the pulsars PSR\,J1028$-$5819, PSR\,J1023$-$5746 and PSR\,J1019$-$5749 are indicated as cyan diamonds.
   The GeV emission 3FGL\,J1028$-$5819 is shown as a red circle.
  The cyan ellipses indicate the selected regions (labelled A and B) from our CO analysis while The black circles (labelled 1 to 7 in panels b. and d.) show the position of selected CS(1--0) regions.
  The location of WR\,21a is shown as a blue cross in panel d. while  the purple square and cross indicate the position of the H\textsc{ii} region GAL\,284.65$-$00.48 and the reflection nebula GN\,10.23.6 respectively.
  Panel e. is  a two-colours image showing the Nanten CO(1--0) integrated intensity at $v_{\text{lsr}}=2 \text{ to }6$\,km/s (red) and  $v_{\text{lsr}}=6 \text{ to }16$\,km/s (green) overlaid by the HESS
   TeV contours in white towards HESS\,J1026$-$582. The cyan dashed ellipse represent the possible molecular ring structure (see Section 3.2).
   Panel f. illustrates the averaged CO(1--0) and CS(1--0) emission  from the selected regions. The red lines indicate the fit used to model the emission (see Table\,\ref{tableJ1026} for fit parameters). 
 The two red vertical lines show the dispersion measure distance of the pulsar PSR\,J1018$-$5819. The blue and  pink regions indicate the velocity range shown in panels a. to d.
  }
  \label{J1026COCS}
 \end{minipage}
\end{figure*}

We list here various detections from our 7mm observations.
For each source (see Figure\,\ref{J1809COCS} to \ref{J1018CO}) we labelled the regions with CS(1--0) detections in numerical order  (e.g. `1') while the regions with SiO(1--0, v=0), HC$_3$N(5--4, F=4--3), 
CH$_3$OH(I) detections are labelled `S', `HC' and `CH' respectively.
In this section however, we focus on combining  our CS(1--0) detections with the Nanten CO(1--0) and primarily focus on the morphology of the various molecular regions surrounding our TeV PWNe and PWNe candidates, 
which is useful to understand their nature and morphology.
\subsection{HESS\,J1809-193}
\label{sec:sectionJ1809}
\begin{figure}
 \centering
 \includegraphics[width=0.5\textwidth]{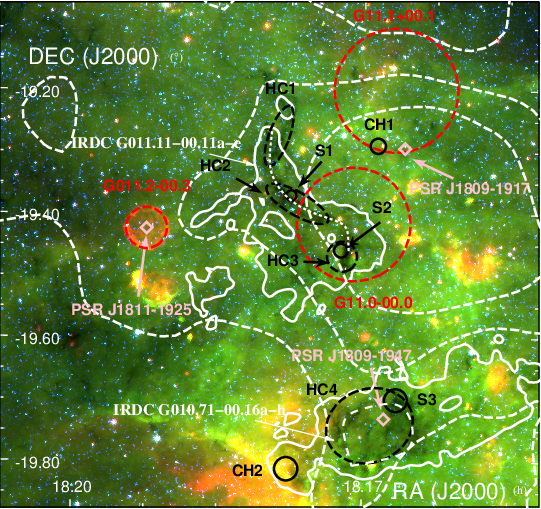}
  \caption{Three colours image showing the MIPSGAL 24\,$\mu$m and GLIMPSE 8\,$\mu$m and 4.6\,$\mu$m in red, green and blue respectively towards HESS\,J1809$-$193 overlaid by the HESS TeV gamma-ray counts in dashed white contours and CS(1--0)
   integrated intensity between $v_{\text{lsr}}=25\text{ to }38$\,km/s contours (0.6 K) in solid white. The SNRs are shown in red dashed circles while the pulsars position are indicated in pink diamonds.
   The black dashed ellipses labelled `HC' indicates the positions of HC$_3$N(5--4,F=4--3) detections, while the black solid circles labelled `CH' and `S' respectively indicate CH$_3$OH and SiO(1--0,v=0) detections. The spectra of these regions can be found in Figure\,\ref{J1809spectra}.
   The white dotted lines represent the extent of the infra-red dark clouds IRDC\,G010.71$-$00.16a$-$h and IRDC\,G011.11$-$00.11a$-$e. (see colour version online)}
     \label{J1809zoom}
\end{figure}

  \mbox{HESS\,J1809--193} is a bright and extended TeV source   whose position is coincident with several potential cosmic-ray (CR) accelerators \citep{HESSJ1809}.
  HAWC has also recently detected the TeV source 2HWC\,J1809$-$190, associated with HESS\,J1809$-$193 \citep{HAWC2017}. \citet{Araya2018} recently 
  found extended GeV emission associated with HESS\,J1809$-$193.
  
  ASCA \citep{J1809Bamba} and Suzaku observations \citep{J1809Suzaku} revealed non-thermal X-rays likely associated with the pulsar PSR\,J1809$-$1917 (shown as a cyan diamond in Figure\,\ref{J1809COCS} and pink in Figure\,\ref{J1809zoom}), 
  with spin down power $\dot{E}_{\text{SD}}=1.8\times10^{36}$\,erg\,s$^{-1}$, a characteristic age $\tau_c=51$\,kyr and a dispersion measure distance \mbox{$d\sim3.7$\,kpc} \citep{Cordes}.
  The presence of two SNR shells G011.0$-$0.0 and G011.1$+$0.1 (shown as blue circles in Figure\,\ref{J1809COCS}) both observed at 330\,MHz and 1465\,MHz \citep{J1809SNRBrogan2004,J1809radio2016} adds
  more complexity to the picture.
  Additionally, the $\sim 2\,$kyr old millisecond pulsar PSR\,J1811$-$1925 with spin down energy  $\dot{E}_{\text{SD}}=6.4\times10^{36}$\,erg\,s$^{-1}$ \citep{Torii1999} and its progenitor SNR\,G011.2$-$0.3, located at $d\sim4.4\text{ to }7$\,kpc,
 are also positioned adjacent to HESS\,J1809$-$193 (see Figures\,\ref{J1809COCS} and \ref{J1809zoom}) and thus might also contribute to  the TeV emission.
 Finally, the anomalous X-ray magnetar XTE\,J1810$-$197 with period $P=5.54$\,s and period derivative $\dot{P}=1.8\times10^{-12}$\,s\,s$^{-1}$ is located $\sim 0.35^{\circ}$ south of HESS\,J1809$-$193.

\begin{figure*}
 \begin{minipage}{\textwidth}
  \centering
  \includegraphics[width=\textwidth]{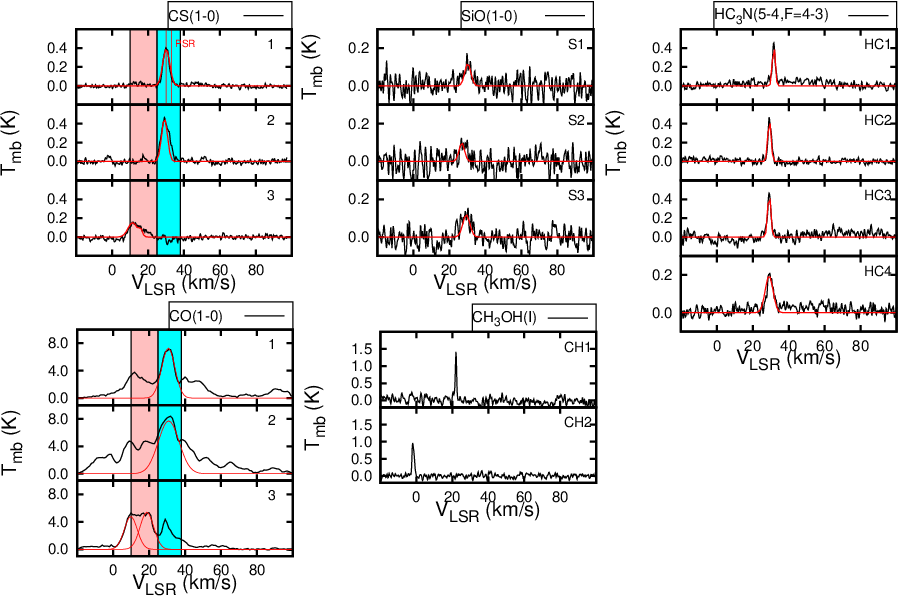}
  \caption{Averaged CS(1--0), CO(1--0), SiO(1--0,v=0), CH$_3$OH(I) and  HC$_3$N(5--4,F=4--3) spectra towards the emission from the selected regions in Figs.\,\ref{J1809COCS} and \ref{J1809zoom} towards HESS\,J1809$-$193.
   The solid red lines represent the Gaussian fit of the emission whose parameters are shown in Table \ref{J1809table}. The two red vertical lines indicate the pulsar PSR\,J1809$-$1917 dispersion measure distance 
   converted to kinematic velocity.
   The pink and cyan regions represent the velocity range for the CS(1--0) and CO(1--0)  integrated intensity maps displayed in Figure\,\ref{J1809COCS}.}
   \label{J1809spectra}
 \end{minipage}
\end{figure*}

 Based on the kinematic distance of PSR\,J1809$-$1917 ($d\sim3.7$\,kpc), we here focus on components at $v_\text{lsr}=10\text{ to }38\,$km/s (see blue and pink regions in Figure\,\ref{J1809spectra}), 
 although our Nanten CO(1--0) data have revealed several other components in the line of sight (see CO(1--0) emission in Figure\,\ref{J1809spectra}).

  At this velocity range, we remark that the molecular gas generally overlaps the TeV emission shown in black contours.
  Notably, at $v_{\text{lsr}}=25\text{ to }38$\,km/s ($d\sim3.7$\,kpc), prominent CO emission is found south and east of the TeV emission, while the prominent CO emission spatially overlaps the TeV emission at $v_{\text{lsr}}=10\text{ to }25$\,km/s ($d\sim2.7$\,kpc). 
  Our Mopra CS(1--0) integrated intensity maps (see Figure\,\ref{J1809COCS} panels b., d., f.) however provides a clear view of the dense gas.

  At v$_{\text{lsr}}=25\text{ to }38$\,km/s (Figure.\,\ref{J1809COCS}, panels a., b.), the molecular clouds in the regions labelled `1' and `2' and located east and south of SNR\,G011.0$+$00.0 respectively appear very extended.
  From our CO and CS analyses, the masses derived in region `1' attain $M_{\text{H}_2}\left(\text{CO}\right)=8.1\times10^{4}\solmass$ and $M_{\text{H}_2}\left(\text{CS}\right)=3.2\times10^{4}\solmass$ 
  while we obtain $M_{\text{H}_2}\left(\text{CO}\right)=2.3\times10^{5}\solmass$ and $M_{\text{H}_2}\left(\text{CS}\right)=3.2\times10^{4}\solmass$ in region `2'.
   A significant fraction of the molecular gas in region `1' and `2'  is therefore concentrated in clumps.
   The CS(1--0) emission in region `1' also appears to anti-correspond with the ASCA X-ray diffuse emission shown in blue, supposedly produced by high energy electrons from the pulsar PSR\,J1809$-$1317 (see \citealt{J1809Suzaku}).
  
  Interestingly, we have found embedded dense filaments in region `1' from C$^{34}$S(1--0) and HC$_3$N(5--4, F=4--3) detections (see Figure\,\ref{extraJ1809figure} in Appendix\,\ref{J1809appendix}).
  In fact, the positions of the HC$_3$N detections labelled `HC1 to HC3' in Figure\,\ref{J1809zoom} coincide with the Spitzer infra-red (IR) dark cloud IRDC\,G011.11$-$00.11a$-$e (\citealt{Parsons2009}, see dotted white lines in Figure\,\ref{J1809zoom}), confirming the molecular gas is foreground to the IR emission.

 We have also identified two weak but broad (FWHM $\sim4$\,km/s) SiO(1--0) features in region `1' (see Figure\,\ref{J1809zoom} and Figure\,\ref{J1809spectra} for spectra), labelled `S1' and `S2'. The absence of overlapping IR continuum emission indicates the lack of active star-forming 
  regions and could indicate a possible interaction between the 
  molecular cloud and a non star-forming shock \citep{Schilke,Gusdorf}, which could here come from the adjacent SNR\,G011.0$-$0.0.
  
  Based on the potential interaction between the SNR\,G011.0$-$0.0 and the dense molecular cloud at $v_{\text{lsr}}\sim30\,$km/s, we thus suggest an alternate SNR distance $d\sim3.7$\,kpc compared to the distance $d\sim3.0$\,kpc claimed by \citet{J1809radio2016}.
  If SNR\,G011.0$-$0.0 indeed interacted with the molecular cloud at $d\sim$3.7\,kpc, it would then be located at the pulsar PSR\,J1809$-$1917 distance and quite 
  likely be its progenitor SNR.
  Using eq.\,3.33a from \citet{CioffiSNR}, we derive the ambient density required to match the small projected radius $r_\text{SNR}\sim5$\,pc with the pulsar characteristic age $\tau_c=51$\,kyr, 
  is $n_\text{amb}\sim370$\,cm$^{-3}$.
  The averaged density $n_{\text{H}_2}\left(\text{CO}\right)=440\,$cm$^{-3}$ found towards region `1' appears somewhat consistent with this estimated averaged density.

  In region `2', we have also detected extended HC$_3$N(5--4,F=4--3) emission and C$^{34}$S(1--0) emission overlapping the infra-red dark cloud IRDC G010.71$-$00.16a$-$h \citep{Parsons2009}, 
  highlighting another dense region.
  The morphology of the IR dark cloud and our HC$_3$N detection appear elliptic and embed the anomalous X-ray magnetar XTE\,J1810$-$197 (see \citealt{AXPJ1809} and references therein), 
  suggesting their potential physical association.
  Additionally, a prominent SiO(1--0) detection (see region `S3' in Figure\,\ref{J1809zoom}) was also found inside this dark cloud.
  This molecular cloud may thus be disrupted by another shock, perhaps caused by the progenitor SNR of XTE\,J1810$-$197.
    
  At $v_{\text{lsr}}=10-25$\,km/s (Figure\,\ref{J1809COCS} panel d.), we observe several CS(1--0) components inside the region here labelled `3'.
  We notably find that the CS(1--0) prominent emission corresponds with the JCMT CO(3--2) peaks found by \citet{J1809radio2016}  
 and the H\textsc{ii} regions (purple contours and blue crosses in Figure\,\ref{J1809COCS} panel f. respectively).
 
   We observe that the  CS(1--0) emission averaged over the grid regions (Figure\,\ref{J1809COCS} panels e-f.) exhibits considerable variation of the peak velocity ranging between v$_{\text{lsr}}=10-22$\,km/s.
  For example, the two peaks at $v_{\text{lsr}}\sim12$ and $18$\,km/s inside `boxes 9 and 10'  merge to a single peaked emission with $v_{\text{lsr}}\sim15$\,km/s in `box 15'.
  This dense molecular region appears to host several H\textsc{ii} regions (see `H1 to H3' in Figure\,\ref{J1809COCS}.f).
  Consequently, the two spectral components may actually probe the same molecular gas, disrupted by the driving motion forces from various H$\textsc{ii}$ regions.
  We also note that the molecular cloud  anti-corresponds with the two SNRs. The disrupted gas could consequently be caused by one of these SNRs.
  As a result, we cannot rule out the SNR\,G011.0$-$0.0 distance $d\sim 3.0$\,kpc suggested by \citet{J1809radio2016}.

\subsection{HESS\,J1026--583}

The TeV source HESS\,J1026$-$583 was discovered from energy dependent morphology studies  towards HESS\,J1023$-$591 \citep{J1026HESS}. The latter source is thought to be powered by the
colliding winds from  Wolf-Rayet stars within the massive stellar cluster Westerlund\,2 at $d=5.4^{+1.1}_{-1.4}$\,kpc \citep{J1026CCC}.
\citet{FermiPWN} catalogued HESS\,J1026$-$583 as a PWN candidate based on the detection of a nearby radio quiet gamma-ray  pulsar PSR\,J1028$-$5810 \citep{Ray1} responsible for the GeV emission towards 3FGL\,J1028.4$-$5819 (shown as a red circle in Figure\,\ref{J1026COCS}).
 PSR\,J1028$-$5819 has a spin down power $\dot{E}_{\text{SD}}=8.3\times10^{35}$\,erg\,s$^{-1}$, characteristic age $\tau=89$\,kyr, and a dispersion measure distance $d=2.3\pm0.7$\,kpc.
However, HESS\,J1026$-$583 shows a hard VHE spectral index $\Gamma_\gamma=1.94$. It also does not exhibit  any X-rays that are spatially coincident with the TeV source. 
Additionally,  diffuse GeV gamma-ray emission has been detected with Fermi-LAT towards Westerlund\,2 \citep{YangWesterlund2_2017}.
The authors have  argued for a hadronic origin based on its 200\,pc extension of the 1 to 250\,GeV  emission.
Consequently, a clear identification remains to be seen.
Because of its proximity to HESS\,J1023$-$575, several ISM features have already extensively been studied \citep{J1026Dame,FukuiJ1023,J1026CCC,FurukawaJ1023,Hawkes2014}.

Figure\,\ref{J1026COCS} shows the Nanten CO(1--0) integrated intensity at $v_{\text{lsr}}=-23\text{ to }-13$\,km/s (panel a.) and $0\text{ to }20$\,km/s (panel c.), inferring a distance $d\sim2.3$\,kpc and
$d\sim4.9$\,kpc respectively, positioning the detections along the Carina arm (see Figure\,\ref{galactic_model} in Appendix\,\ref{sec:galacticmodel}).
At $v_{\text{lsr}}=-23\text{ to }-13$\, km/s, we note that the CO emission does not overlap the TeV emission, nor the pulsar's position.
In fact, the molecular region with prominent CO emission, that we labelled `A',  is located east of the pulsar position, and could support the crushed PWN scenario (see \citealt{Blon2001}).
Assuming a kinematic distance $d\sim2.3$\,kpc, we obtain a total mass $M_{\text{H}_2}\left(\text{CO}\right)=9.7\times10^{3}\solmass$.
Lastly, from our CS survey, we have also identified two unresolved CS(1--0) features labelled `1' and `2' in Figure\,\ref{J1026COCS} panel b.
\begin{figure*}
\centering
 \begin{minipage}{\textwidth}
  \includegraphics[width=\textwidth]{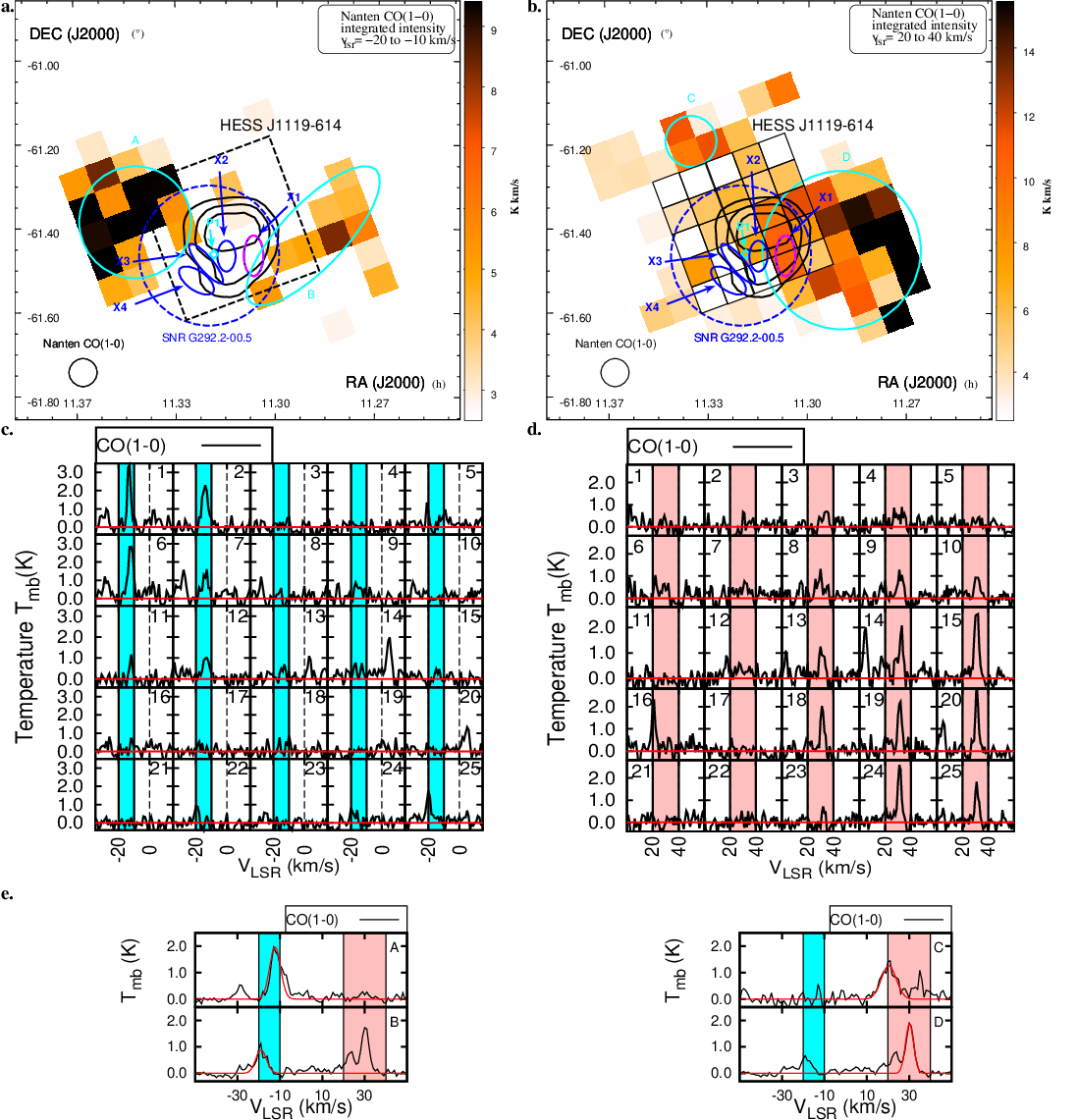}
   \caption{Nanten CO(1--0) emission between $v_{\text{lsr}}=-20\text{ to }-10$\,km/s (panel a.) and  $v_{\text{lsr}}=20\text{ to }40$\,km/s (panel b.) towards HESS\,J1119$-$614 whose TeV gamma-ray emission is shown as solid black contours.
    The progenitor SNR\,G292.2$-$0.5 is delimited by the blue  dashed circle. 
    The solid blue and dashed blue-pink ellipses labelled `X1' to `X4' (see text) highlight bright X-ray regions studied by \citet{KumarJ1119} with XMM Newton and Chandra.
   The pulsar \mbox{PSR\,J1119--6127} (P1)'s position is indicated as a cyan diamond.
  The cyan ellipses show the selected regions (labelled A$\text{ to }$D) for our CO analysis.
  Panels c. and d. show the variation of the averaged CO(1--0) spectra over the black grid of boxes shown in panel (b). The cyan and pink regions indicate the velocity ranges mapped in panels a. and b. 
  Panel e. shows the averaged CO(1--0) emission from the  selected regions  towards HESS\,J1119$-$164. The red lines indicate the fit used to parametrise the emission. The fit parameters are displayed in Table\,\ref{J1303table}.
  The pink and cyan regions show the velocity range used for the above integrated intensity maps.
  }
  \label{J1119COCS}
 \end{minipage}
\end{figure*}

At $v_{\text{lsr}}=0\text{ to }20$\,km/s ($d\sim4.9$\,kpc, see Figure\,\ref{J1026COCS} panel c.), we observe prominent CO emission at $v_{\text{lsr}}\sim4$\,km/s, labelled `B', spatially coincident with the HESS\,J1026$-$582 TeV peak.
The molecular cloud in region `B', with total mass $M_{\text{H}_2}\left(\text{CO}\right)=4.9\times10^{4}\solmass$ appears next to a possible shell like structure (see dashed purple and cyan ellipses in Figure\,\ref{J1026COCS} panels c. and e. respectively) which overlaps the HESS\,J1023$-$575 TeV emission. 
The molecular region is however located at the tangent of the Sagittarius arm (see Figure\,\ref{galactic_model}), thus it is possible that the various molecular clouds found in Figure\,\ref{J1026COCS} panel c.) may be unrelated.
From Figure\,\ref{J1026COCS} panel d., it nonetheless appears that the CO(1--0) emission at $v_{\text{lsr}}=2\text{ to }6$\,km/s fills in some missing segment of the shell-like structure and might form a ring (indicated as a dashed-cyan ellipse) whose centre is located south-west of HESS\,J1026$-$582
 observed at $v_{\text{lsr}}=6\text{ to }16$\,km/s. 
 The molecular structure at $v_\text{lsr}=6\text{ to }16$\,km/s could therefore be physically connected to the molecular cloud in region `B'.

Among the CS(1--0) features found at $v_{\text{lsr}}=0\text{ to }20$\,km/s, the gas clumps labelled `5' to `7', embedded in the molecular gas in region `B' form a partial shell structure 
overlapping the TeV peak emission with a combined mass $M_{\text{H}_2}\left(\text{CO}\right)\sim5.0\times10^{3}\solmass$, very similar to the CO mass obtained in region `B'.
No massive stars have been catalogued towards the centre of this structure. However, we note the presence of the H\textsc{ii} region GAL\,284.65$-$00.48 towards region `5' and the reflective 
nebula GV\,10.23.6 at the centre of the shell (see purple box and cross in Figure\,\ref{J1026COCS} panel d. respectively).

\subsection{HESS\,J1119--614}
\citet{Arache2009} and \citet{HGPS2018} reported the detection of the TeV source HESS\,J1119$-$614 (see solid black contours  in Figure\,\ref{J1119COCS}).
It is thought to be associated with either the PWN, powered by the radio pulsar PSR\,J1119$-$6127 (see cyan diamond in Figure\,\ref{J1119COCS}), with 
spin down period $P=400$\,ms \citep{CamiloJ1119} and characteristic age $\tau_c=1.9$\,kyr \citep{WelJ1119}; or its progenitor SNR\,G292.2$-$0.5 
(see dashed blue circle in Figure\,\ref{J1119COCS}).
The $3^{\prime\prime}\times3^{\prime\prime}$ PWN has been resolved in X-ray with Chandra \citep{J1119SafiGonzalez2005,J1119Safi2008}.
The shell of progenitor SNR has been observed both in the radio band with ATCA \citep{Crawford2001}, and in X-ray with ASCA between 0.4-10\,keV \citep{J1119Pivovaroff2001}. 
Follow-up studies with Chandra and XMM-Newton \citep{KumarJ1119,J1119XMM} revealed additional information about the nature of the X-ray emission.
Indeed, from the four regions studied by \citep{KumarJ1119} (here labelled `X1' to `X4' in Figure\,\ref{J1119COCS}), the three regions `X1' to `X3' shown as
blue ellipses, have a non-thermal component.
However, the X-ray emission in region `X4' shown as a blue-pink ellipse, is prominent and is thought to be of thermal origin.

\citet{Caswell2004} estimated a distance $d\sim 8.4\,$kpc for the pulsar and its progenitor SNR, based on the H\textsc{i} and magnetic field studies.
However, reconciling the characteristic age $\tau_c=1.9\,$kyr of PSR\,J1119$-$6127 and the $\sim 25$\,pc diameter of the progenitor SNR requires 
a low density medium \citep{J1119XMM} and a massive progenitor star.
\citet{J1119SafiGonzalez2005} argued for a SNR distance at $d=3.6\text{ to }6.3$\,kpc by modelling the X-ray spectrum.
Our gas study thus aims to provide additional constraints on the distance and how it could affect the nature of the TeV source.

No extended CS emission have been found towards this region.
From our Nanten CO data, several components have been detected along the line of sight at $v_\text{lsr}\sim-30\,$km/s, $v_\text{lsr}\sim-10\,$km/s 
(near/far distance $d\sim2.6/4.5$\,kpc, near Carina-arm, see Figure\,\ref{galactic_model}), $v_\text{lsr}\sim20\,$km/s  (distance $d\sim8.6$\,kpc, far Carina-arm),
$v_\text{lsr}\sim30\,$km/s  (kinematic distance $d\sim9.7$\,kpc, far Carina-arm).

At $v_\text{lsr}=-20\text{ to }-10\,$km/s, we observe two molecular clouds (labelled `A' and `B' in Figure\,\ref{J1119COCS} panel a.) positioned 
north-east and west to the SNR respectively, and  with respective masses $M_{\text{H}_2}\left(\text{CO}\right)=2.3\times10^{4}\solmass$ 
  and $M_{\text{H}_2}\left(\text{CO}\right)=7.1\times10^{4}\solmass$.
  We also note that the molecular ISM anti-corresponds with all X-ray regions.
  
 At $v_\text{lsr}\sim20\text{ to }40\,$km/s we observe that the bulk of the molecular gas is found in two regions labelled `C' and `D', with  masses $M_{\text{H}_2}\left(\text{CO}\right)=1.3\times10^{4}\solmass$ 
  and $M_{\text{H}_2}\left(\text{CO}\right)=2.3\times10^{5}\solmass$ respectively.
  The morphology of the gas overlaps the TeV gamma-ray detection.
  It also corresponds with the thermal X-ray in region `X1' both at $v_{\text{lsr}}\sim20\,$km/s and $v_{\text{lsr}}\sim30\,$km/s (see Figure\,\ref{J1119COCS} panel d.), 
  highlighting potential SNR-MC interaction.
  The molecular gas also anti-corresponds with the other X-ray regions `X2' to `X4', thought to have non-thermal components likely produced by 
  the high energy electrons from both the SNR and the PWN.
  Therefore, we argue that the morphology of the CO(1--0) both at $v_{\text{lsr}}\sim20\,$km/s and $v_{\text{lsr}}\sim30\,$km/s appears consistent with the X-ray results,
  inferring a source kinematic distance $d=8.6\text{ to }9.7$\,kpc. 
  We finally remark that these two components may be physically connected and highlight motion caused by the progenitor star.
  
  We also  compare the column densities, shown in Figure\,\ref{J1119distance}, towards the four X-ray regions.
  Towards the regions `X1' and `X2', we match the X-ray modelled column density $N_\text{H}$ (grey regions in Figure\,\ref{J1119distance} top panels) 
  at $v_{\text{lsr}}=22 \text{ to } 30$\,km/s, also suggesting a distance $d=8.6-9.6$\,kpc, positioning the source in the far-Carina arm.
  Towards the X-ray regions `X3' and `X4' however, our column densities  do not reach the X-ray one, 
  or yield to unreasonable large distance $d\sim15.5$\,kpc.
  Clumps unresolved by the Nanten CO(1--0) may account for the missing components.
  Our column density studies thus provide a lower-limit kinematic distance $d>8.6$\,kpc, consistent with our morphological studies and with \citet{Caswell2004}. 
  To explain both the large SNR radius and the enhanced thermal X-ray emission in region `X4' , we argue that the SNR once propagated in a low density medium \citep{J1119XMM}, until it reached the denser gas in region `D'.

  \begin{figure*}
\centering
 \begin{minipage}{\textwidth}
  \includegraphics[width=\textwidth]{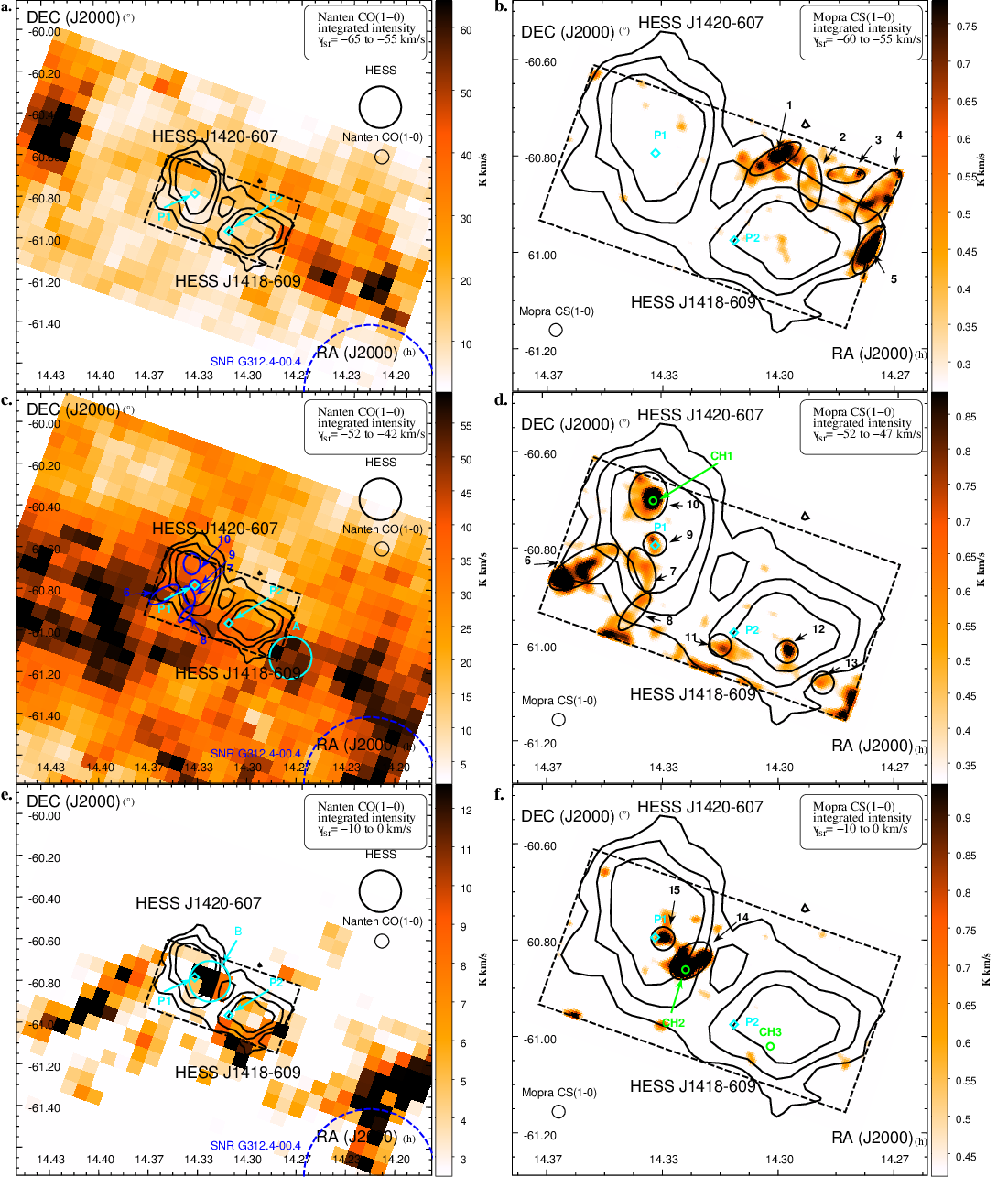}
  \caption{Nanten CO(1--0) and Mopra CS(1--0) emission between $v_{\text{lsr}}=-65\text{ to }-55$\,km/s, $-52\text{ to }-42$\,km/s and $-10\text{ to }0$\,km/s  towards \mbox{HESS\,J1420--607} and \mbox{HESS\,J1418--609} shown in black contours. 
   The black dashed box indicates our Mopra 7mm coverage. The position of the pulsars PSR\,J1420--6048 and PSR\,J1418--6058 (labelled P1 and P2 here) are shown  as cyan diamonds while the nearest SNR\,G312.4--0.04 is indicated as a blue dashed circle.
  Our CO regions `A' and `B' are shown in cyan in panels c. and e.
  The blue ellipses on panel c. and the black ellipses on the right panels indicate the position of the CS regions (`1' to `15').
  The position of the CH$_3$OH(I) detections labelled `CH1' to `CH3' are shown in green circles.
  }
  \label{J1420COCS}
 \end{minipage}
\end{figure*}

 \subsection{Kookaburra and Rabbit}
 The two  TeV sources HESS\,J1418$-$609 and HESS\,J1420$-$607 have been classified as PWNe based on their spatial coincidence with the X-ray \citep{RobertsJ1418} and GeV gamma-ray counterparts \citep{FermiPWN}.
  \citet{NgRobertsJ1418Chandra}  indicated that two diffuse non-thermal X-ray sources were associated with the pulsar PSR\,J1420$-$6048 (labelled P1 in Figure\,\ref{J1420COCS}), a 68.2 ms period pulsar with spin down power
 $\dot{E}_{\text{SD}}=1.0\times10^{37}$\,erg\,s$^{-1}$, characteristic age $\tau_c=13$\,kyr and dispersion measure distance $d\sim5.6$\,kpc ; and the 108\,ms radio-quiet gamma-ray pulsar PSR\,J1418$-$6058 (labelled P2), with a characteristic
  age $\tau_c=1.6$\,kyr.
  It has also been noted that P2's location is offset ($\sim8.4^{\prime}$) from the HESS\,J1418$-$609 TeV peak.
  \begin{figure}
\centering
  \includegraphics[width=0.5\textwidth]{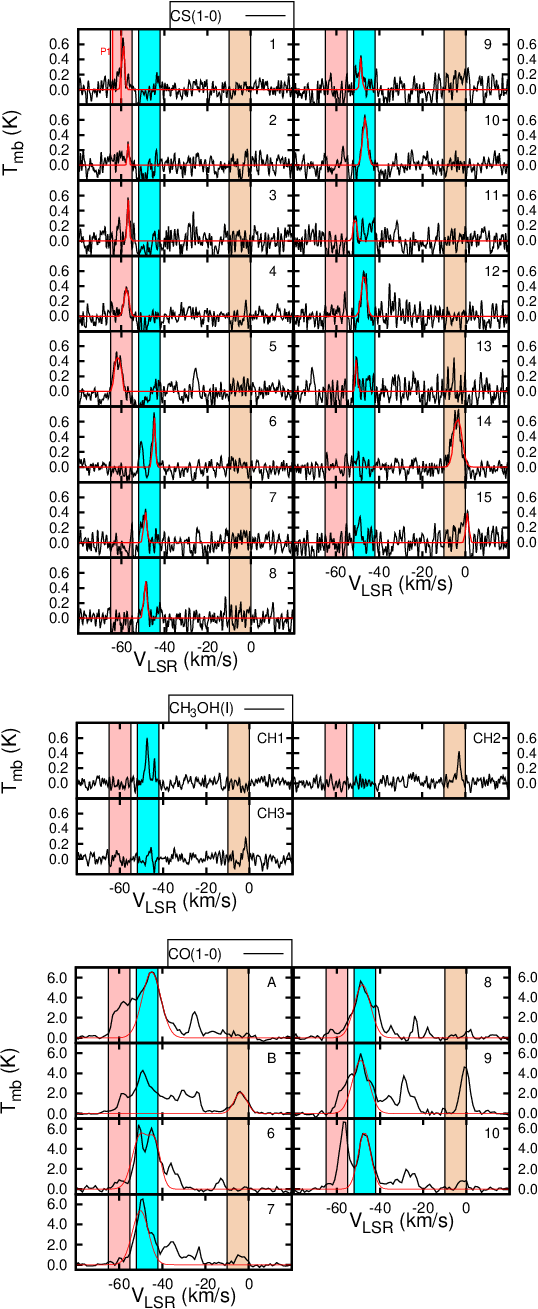}
  \caption{Averaged CS(1--0), CH$_3$OH(I) and CO(1--0) emission from the regions towards HESS\,J1418$-$609 and HESS\,J1420$-$607 shown in Figure\,\ref{J1420COCS}. The red lines indicate the fit used to model the emission at the velocity ranges where the regions are shown in Figure\,\ref{J1420COCS} 
   (see Table\,\ref{tableJ1418} for fit parameters).
   The red vertical lines show the dispersion measure distance of the pulsar PSR\,J1420$-$6048 (P1 in Figure\,\ref{J1420COCS}). The pink, cyan and brown regions show the velocity ranges for the integrated intensity maps in Figure\,\ref{J1420COCS}. }
  \label{J1420spec}
\end{figure}
  \citet{RobertsRPWNJ1418} argues that the expected cometary shape of the PWN produced by a pulsar with a high space velocity, would explain the offset position of the gamma-ray emission with respect 
  to the pulsar.
  The distance to P2 have not yet been constrained.
  \citet{NgRobertsJ1418Chandra} has argued a PWN distance $d=2\text{ to }5$\,kpc, while \citet{PSRJ1826distance} has claimed a much smaller distance $d=1.4\text{ to }1.9$\,kpc.
  Thus, by looking at the gas distribution in various velocity ranges, we aim to highlight  the distance which would support the PWN scenario.

  From our CO(1--0) observations, we have detected several molecular complexes along the line of sight. We mostly focus at the velocity ranges $v_{\text{lsr}}=-65\text{ to }-55$\,km/s, ($d\sim5.6$\,kpc), $\sim50$\,km/s (Scutum Crux arm, distance $d\sim3.5$\,kpc), and 
   $\sim-5$\,km/s (local arm, $d\sim0.1$\,kpc) respectively, as shown in Figure\,\ref{J1420COCS}.
   In Figure\,\ref{J1420spec}, we also note a CO component at $v_\text{lsr}\sim-25$\,km/s. 
   As opposed to the other velocity ranges, no prominent CS detections have been found at $v_\text{lsr}\sim-25$\,km/s in this region, and thus we suggest that the molecular cloud is located 
   on the far distance.

   At $v_{\text{lsr}}=-65\text{ to }-55$\,km/s (see Figure\,\ref{J1420COCS}a.), we note the bulk of the  CO(1--0) emission is located towards the west and north side of HESS\,J1418$-$609.
   We also remark that the molecular emission shows little overlap with any of the TeV sources at this velocity range.
   We observed extended CS(1--0) emission, labelled `1' to `5', north of HESS\,J1418$-$609.
   Assuming a distance $d=5.6$\,kpc, the combined mass of these clumps attains $M_{\text{H}_2}\left(\text{CS}\right)=1.1\times10^{4}\solmass$.
  Notably, the bulk of the CS(1--0) emission appears to wrap around  the TeV emission (as expected for leptonic IC emission).

   At $v_\text{lsr}=-52\text{ to }-42$\,km/s, we note that the CO(1--0) emission overlaps  the two TeV sources, and peaks west of HESS\,J1018$-$609 and south east of HESS\,J1420$-$607.
   However, due to the small velocity separation between the components and the components at $v_\text{lsr}=-65\text{ to }-55$\,km/s, it is somewhat difficult to accurately describe the morphology of the 
   diffuse molecular gas at these velocities.
   We particularly observe that the prominent CO emission in region `A', with mass $M_{\text{H}_2}\left(\text{CO}\right)=3.5\times10^{4}\solmass$ and averaged density $n_{\text{H}_2}\left(\text{CO}\right)=5.6\times10^{2}$\,cm$^{-3}$,  
   anti-corresponds with the TeV source HESS\,J1418$-$609.
   Our CS(1--0) results highlight  a filamentary structure (see regions `6' to `10' in Figure\,\ref{J1420COCS}d.), with a combined CS mass $M_{\text{H}_2}\left(\text{CS}\right)=5.2\times10^{3}\solmass$ (CO mass
   $M_{\text{H}_2}\left(\text{CO}\right)=3.1\times10^{4}\solmass$) which crosses HESS\,J1420$-$607.
   Interestingly, the dense gas in regions `6' to `8' also appears in a shell-like arrangement centred towards the south-east of HESS\,J1420$-$607.
   CS(1--0) clumps, labelled `11' to `13' are also found towards HESS\,J1418$-$609. The regions `11' and `12' are in fact coincident with the TeV peak emission.
 
    At $v_{\text{lsr}}=-10\text{ to }0\,$km/s (bottom panels), we found prominent CO emission  south of HESS\,J1418$-$609, east of HESS\,J1420$-$607, and particularly between the two pulsars.
 Notably, the latter molecular region (labelled `B') with mass $M_{\text{H}_2}\left(\text{CO}\right)=6.8\solmass$ nests an extended dense clumps (see region `14') with mass $M_{\text{H}_2}\left(\text{CS}\right)=2.7\solmass$ and 
 with averaged density reaching $n_{\text{H}_2}\left(\text{CS}\right)=3.8\times10^{4}$\,cm$^{-3}$. 
 
   In Figure\,\ref{J1418emissiondistance}, we compare the column density from CO and H\textsc{i} measurements  with the column density $N_{\text{H}}=2.7\pm0.2\times10^{22}$\,cm$^{-2}$ towards the pulsar PSR\,J1418$-$6058 derived from X-ray measurements \citep{J1418Suzaku}. 
 Assuming that all components, except that at $v_\text{lsr}\sim25$\,km/s, are in the near distance, we observe in Figure\,\ref{J1418emissiondistance} that the column density value from X-ray measurements is matched at v$_{\text{lsr}}\le-50$\,km/s, which 
 infers a distance range $d\sim3.5$\,kpc.
 As a result, this column density X-ray study appears to favour a PWN location at $d\sim3.5\,$kpc ($v_\text{lsr}=-52\text{ to }-42$\,km/s), while our morphological study of the 
  molecular ISM favours a PWN distance at $d\sim$5.6\,kpc ($v_\text{lsr}=-65\text{ to }-55$\,km/s), as it anti-corresponds with the TeV gamma-ray emission (see Figure\,\ref{J1420COCS}).
   
   \begin{figure*}
\centering
 \begin{minipage}{\textwidth}
  \includegraphics[width=\textwidth]{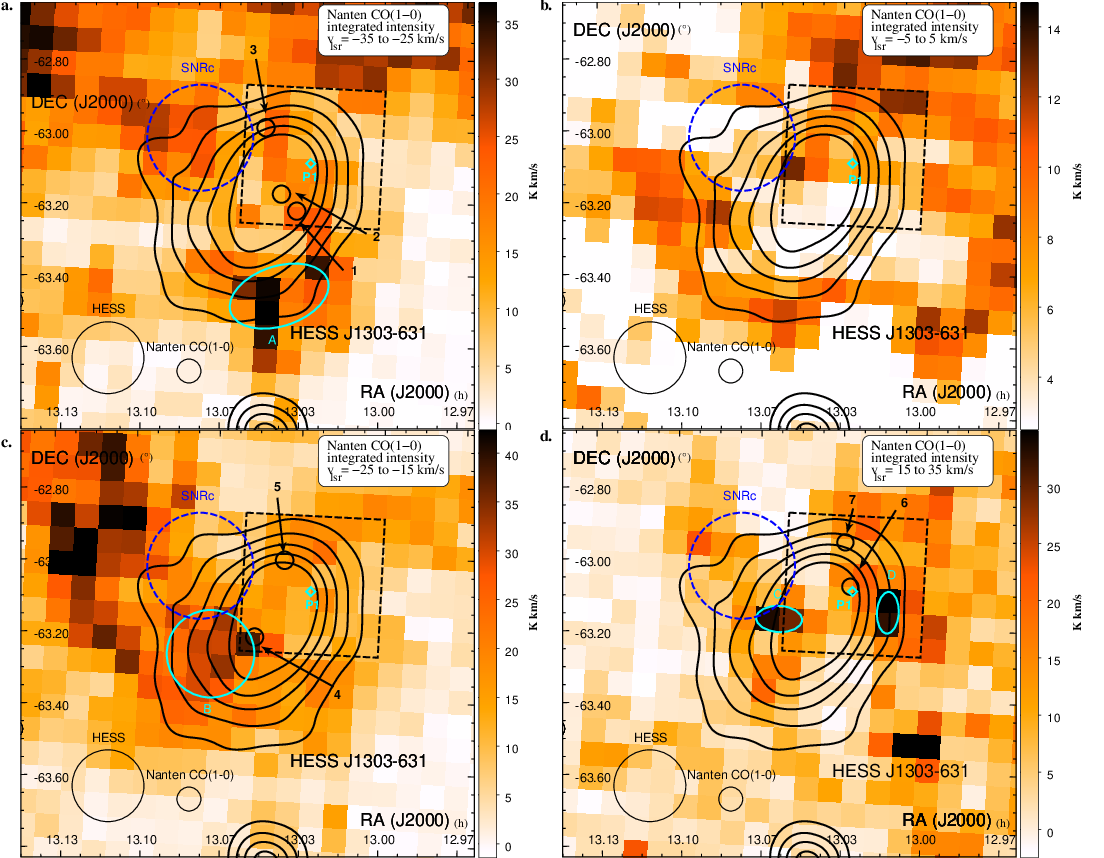}
  \caption{Nanten CO(1--0) integrated intensity map towards \mbox{HESS\,J1303--631} (shown in black contours) between \mbox{$v_{\text{lsr}}=-35\text{ to }-25$\,km/s}, $-25\text{ to }-15$\,km/s, 
  $-5\text{ to }+5$\,km/s and $15\text{ to }35$\,km/s. The 7mm map is shown as a black-dashed box, while the position and size of the SNR candidate from \citet{HESSJ1303SNR2017} is indicated by a blue dashed circle. The position of the pulsar PSR\,J1301$-$6305 (P1). 
  is shown in cyan diamond. The regions labelled `1' to `7' where CS(1--0) was detected are shown in black circles. The positions of prominent CO detections slightly overlapping the TeV emission are shown in cyan ellipses.
  }
  \label{J1303COCS}
 \end{minipage}
\end{figure*}

\subsection{HESS\,J1303-631}

HESS\,J1303$-$631 was first  classified as a `dark source' due to its  lack of any counterparts at other wavelengths \citep{HESSJ1303discovery1,FermiPWN}.
However, energy dependent morphology of the TeV source (see \citealt{J1303PWN}) unambiguously highlighted its association with the pulsar PSR\,J1301$-$6305 (P1 in Figure\,\ref{J1303COCS}) with spin down energy $\dot{E}_{\text{SD}}=2.6\times10^{36}$\,erg\,s$^{-1}$, 
a rotation period $P=184$\,ms, and a characteristic age $\tau_c=11$\,kyr. 
Follow-up observations with XMM-Newton revealed  diffuse X-ray emission towards the pulsar with a power law spectral index $\Gamma_X=2.0^{+0.6}_{-0.7}$\citep{J1303PWN}.
\citet{FermiPWN} detected a GeV counterpart with a gamma-ray spectral index $\Gamma_\gamma=1.7$.
Finally, from the 1.384\,GHz ATCA observations, \citet{HESSJ1303SNR2017} recently announced the presence of a plausible SNR radio shell  with radius $\sim12^{\prime}$  next to the pulsar PSR\,J1301$-$6305, although this association 
 appears unlikely.
\begin{figure}
 \centering
 \includegraphics[width=0.5\textwidth]{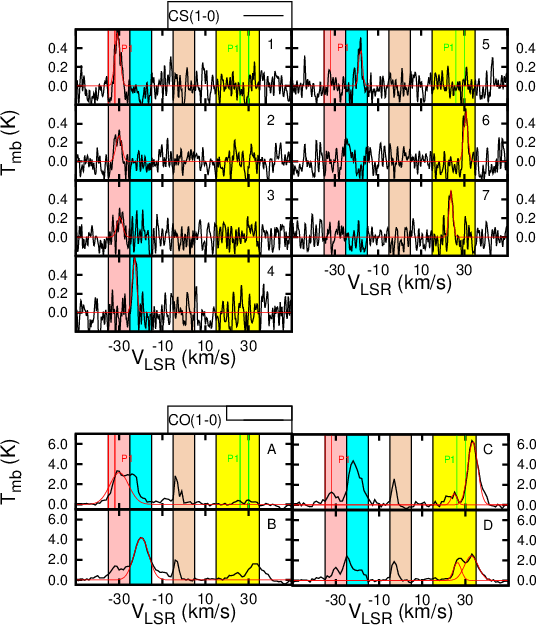}
 \caption{The averaged CS(1--0) and CO(1--0) emission from the different regions shown in Figure\,\ref{J1303COCS} towards HESS\,J1303$-$631. The red lines indicate the fit used to parametrise the emission. The fit parameters are displayed in Table\,\ref{J1303table}.
   Finally, the red and green vertical lines represent the dispersion measure distance of the pulsar P1 as predicted by \citet{Cordes} and \citet{Taylorcordes1993} respectively.
    The pink, cyan, brown and yellow regions indicate the velocity range of the integrated intensity maps shown in Figure\,\ref{J1303COCS}.}
   \label{J1303spec}
\end{figure}

 Based on the dispersion measure  of this pulsar, \citet{Cordes}  suggested a distance $d\sim6.6$\,kpc, much closer than the previous distance $d\sim12.6$\,kpc \citep{Taylorcordes1993}.
From the Nanten CO(1--0) components identified (see Figure\,\ref{J1303spec}), we focus on several molecular complexes in the line of sight at 
$v_{\text{lsr}}=-35\text{ to }-25$\,km/s (distance $d\sim6.6$\,kpc, Scutum Crux arm ), $v_{\text{lsr}}=-25\text{ to }-15$\,km/s (distance $d\sim1.5$\,kpc, near Sagittarius-Carina arm), $v_{\text{lsr}}=-5\text{ to }5$\,km/s ($d\sim0.1$\,kpc, local arm)
 and $v_{\text{lsr}}=25\text{ to }35$\,km/s (distance $d\sim12.6$\,kpc, far Sagittarius-Carina arm) shown in Figure\,\ref{J1303COCS}.
  From our 7mm CS observations, which cover the north-west part of the SNR towards the TeV source, we have identified several molecular clumps, that we have labelled `1 to 7' (see Tables \ref{J1303table} and \ref{MASSJ1303} for 
their physical parameters) but no extended CS(1--0) emission has been detected.

 At all the aforementioned velocity ranges, it appears that some CO(1--0) emission always overlaps the TeV emission.
 It should be noted that the CO emission peaks south of HESS\,J1303$-$631 inside region `A' at $v_\text{lsr}=-35\text{ to }-25$\,km/s.
 At the local arm ($v_\text{lsr}=-5\text{ to } 5\,$km/s), most of the CO(1--0) emission is distributed north east of the TeV source.
 At $v_\text{lsr}=15\text{ to }35$\,km/s, prominent CO emission is found  overlapping HESS\,J1303$-$631 (see regions `C' and `D').
 We also notice that the CO(1--0) emission does overlap the  position of the SNR candidate represented in blue circle in Figure\,\ref{J1303COCS} at $v_{\text{lsr}}=-35\text{ to }-25$\,km/s 
 and $v_{\text{lsr}}=15\text{ to }35$\,km/s.

\begin{figure*}
\centering
 \begin{minipage}{\textwidth}
  \includegraphics[width=\textwidth]{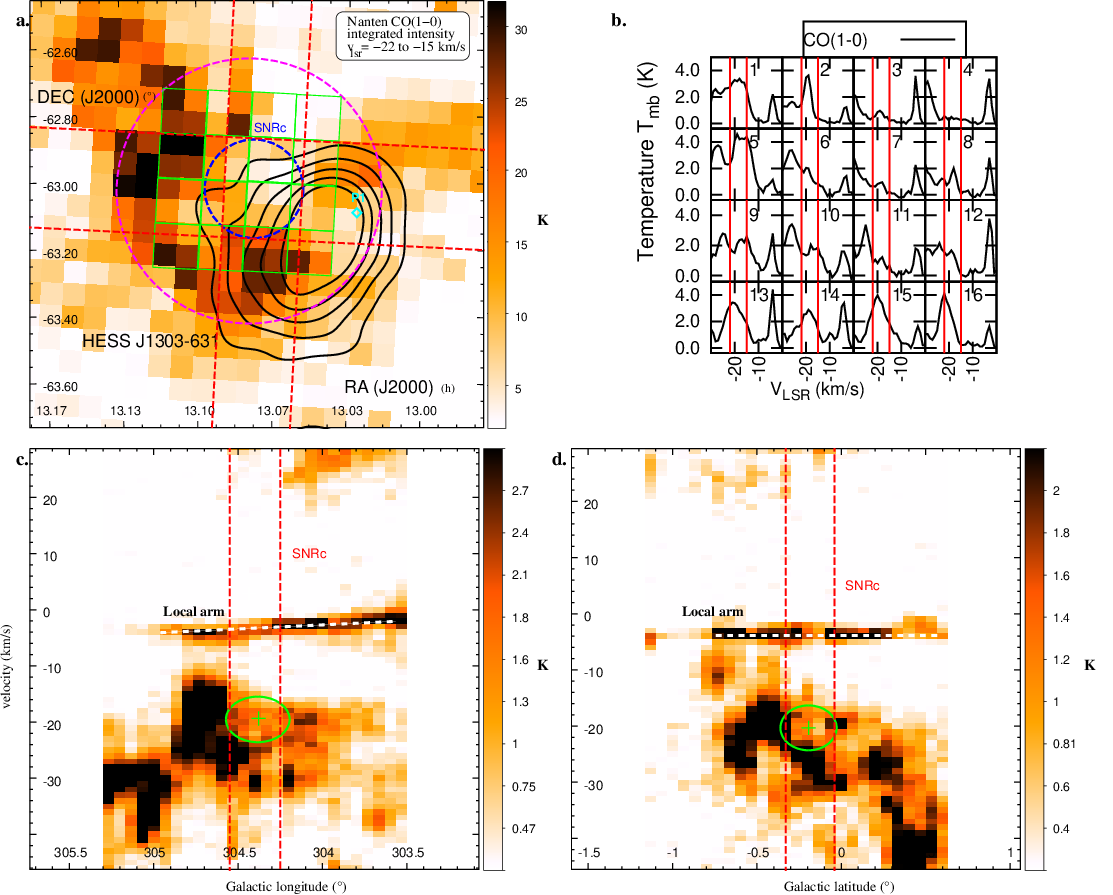}
  \caption{\textit{(panel a.)} Nanten CO(1--0) integrated intensity between $v_{\text{lsr}}=-22\text{ to }-15$\, km/s overlaid by the HESS TeV contours in black. The blue dashed circle indicates the size of the candidate SNR \citep{HESSJ1303SNR2017}.
  The cyan diamond shows the position of the pulsars PSR\,J1301$-$6305 (P1). 
  The purple dashed circle indicates the region used to compute the mass of the putative molecular shell (see discussion in section\,\ref{sec:discussion}).
  The green grid of boxes indicates the position of the displayed 
  CO(1--0) spectral lines \textit{(panel b.)}. 
  \textit{(panels c. and d.)} Galactic longitude-velocity $\left(l,v\right)$ and latitude-velocity $\left(b,v\right)$ images integrated between $l=\left[304.25^{\circ}:304.55^{\circ}\right]$ and  $b=\left[-0.34^{\circ}:-0.04^{\circ}\right]$
   respectively (shown as red dashed lines in top left panel). The green cross-hair and ellipse show the location of a putative expanding molecular shell while the red dashed lines indicate the boundaries of the candidate SNR.}
  \label{J1303PVSNR}
 \end{minipage}
\end{figure*}

 Interestingly, we have found a CO(1--0) emission dip  at $v_{\text{lsr}}=-25\text{ to }-15$\,km/s localised towards the SNR candidate.
  From the position-velocity plots shown in Figure\,\ref{J1303PVSNR}, we  observe   prominent CO emission  which in fact appears to surround the SNR position  (whose boundaries are shown in red dashed lines in Figure\,\ref{J1303PVSNR} bottom panels) between $v_{\text{lsr}}=-22\text{ to }-15$\,km/s. 
  From the CO(1--0) integrated intensity region at this velocity range, we also observe little spatial overlap between the molecular gas  and the SNR candidate.  
  Consequently, it may highlight the presence of a putative molecular shell at $d\sim1.5$\,kpc coincident with the recently observed SNR candidate. The green cross and ellipse in Fig.\,\ref{J1303PVSNR} indicate
   the position and the expansion speed ($v_\text{exp}\sim4$\,km/s) of the putative molecular shell surrounding the SNR candidate.
   From the dashed magenta circular region in Figure\,\ref{J1303PVSNR} centred towards the SNR candidate position, the maximum mass swept by the SNR candidate reaches $M_{\text{H}_2}=3.3\times10^{4}\solmass$.
 We obtain as a upper limit a required kinetic energy $E_{\text{kin}}=1/2M_{\text{H}_2}v_{\text{exp}}^{2}=5.1\times10^{47}$\,erg which represents $\sim 0.05\%$ of the total kinetic energy
 from powerful O stars stellar winds over 1 Myr time-scale (see \citealt{Weaver1977} for detailed study on interstellar bubbles).
 Consequently, the molecular shell may have been produced by the SNR progenitor star. We thus suggest this new SNR candidate is at  distance at $d\sim$1.5\,kpc, supporting its non-physical association with 
  PSR\,J1301$-$6305 claimed by \citet{HESSJ1303SNR2017}.

\subsection{HESS\,J1018--589}
\begin{figure*}
\centering
 \begin{minipage}{\textwidth}
  \includegraphics[width=\textwidth]{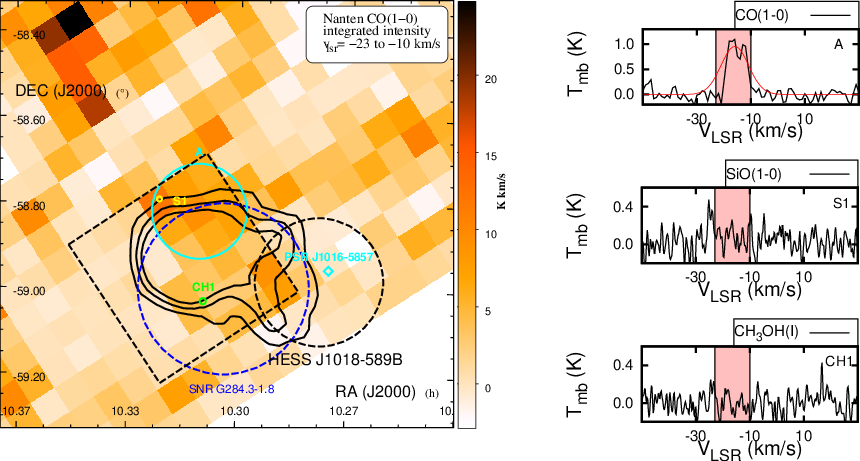}
  \caption{Nanten CO(1--0) integrated intensity map towards \mbox{HESS\,J1018--589} between $v_{\text{lsr}}=-23\text{ to }10$\,km/s overlaid by the TeV gamma-ray emission from HESS\,1018$-$589 in solid black contours. 
  The dashed black circle indicates the size and position of HESS\,J1018$-$589B.
  The SNR\,G284.3--01.8 is shown as a blue dashed circle while the pulsar PSR\,J1019--5857 is shown as  a cyan diamond. The extended CO region labelled 'A' is shown in cyan, 
  the position of the SiO(1--0,v=0) `S1' are displayed in yellow and the CH$_3$OH maser found in the region `CH1' is shown in green.
  Their respective spectral lines are displayed on the right-hand side. The pink region illustrates 
  the aforementioned velocity range.}
  \label{J1018CO}
 \end{minipage}
\end{figure*}

HESS\,J1018$-$589 was first reported by \citet{J1018HESS2} and actually consists of two distinct sources.
The gamma-ray binary 1FGL\,J1018.6$-$5856 appears to be responsible for the HESS\,J1018$-$589A TeV emission \citep{J1018HESS2} while 
HESS\,J1018$-$589B (shown in dashed black circle in Figure\,\ref{J1018CO}) is thought to be a PWN powered by the pulsar PSR\,J1016$-$587 (shown by a cyan diamond), with a rotation period $P=107$\,ms, a spin down energy $\dot{E}_{\text{SD}}=2.6\times10^{36}$\,erg,
 and a characteristic age $\tau_c=21$\,kyr \citep{PSRJ1016Camilo}.  
 It has been suggested that the pulsar, with dispersion measure distance  $d$=8\,kpc, is not associated with the nearby SNR\,G292$-$1.8 located at $d\sim2.9$\,kpc \citep{RuizMay1986}.

 From the Nanten CO(1--0) observations shown in Figure\,\ref{J1018CO}, we have  identified CO emission at \mbox{$v_{\text{lsr}}=-23\text{ to }-10$\,km/s}, inferring a near distance \mbox{$d\sim2.8$\,kpc}, matching the SNR distance.
 The CO(1--0) emission appears filamentary north of  HESS\,J1018$-$589B , and partially overlaps the northern rim of SNR\,G292$-$1.8.
 The molecular gas in region `A', with mass attaining $M_{\text{H}_2}\left(\text{CO}\right)=2.9\times10^{3}\solmass$, shows quite broad emission ($\Delta  v\sim12$\,km/s, see Table \ref{tableJ1018} and Figure\,\ref{J1018CO}).
Although no CO emission was revealed at $v_\text{lsr}\sim30$\,km/s, we note from Figure\,\ref{J1018emissiondistance} significant H\textsc{i} emission, which may suggest that the ISM surrounding HESS\,J1018$-$589B mostly 
consists of atomic gas.

    \section{Discussion of gamma-ray emission}
    \label{sec:discussion}
In this section, we use results from our ISM studies to discuss whether the CRs and/or high energy electrons interacting with the ISM  can contribute to the TeV emission or at least affect their morphology.
In this section, we first introduce the method used to check whether hadronic cosmic-rays  could contribute to the observed TeV emission.
Then, we will briefly indicate how leptonic emission can also be affected by the ISM.

\subsection{TeV emission from CRs}
\begin{figure}
  \includegraphics[height=0.5\textwidth,angle=270]{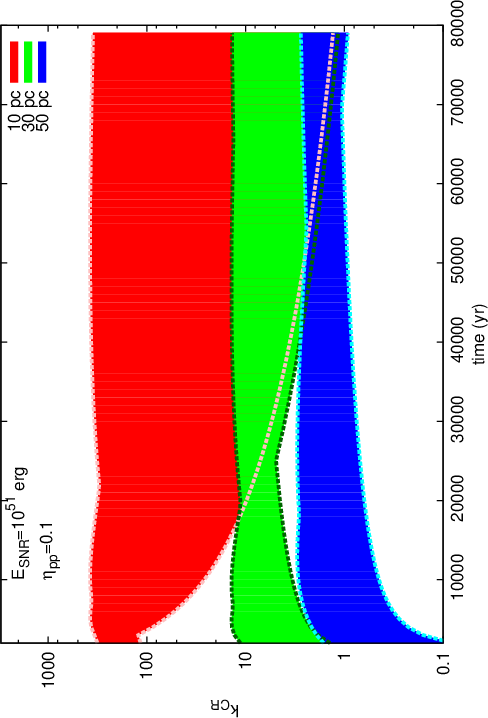}
  \caption{Evolution of the cosmic-ray enhancement factor $k_\text{CR}$ range as a function of time at a distance $d=10$\,pc (red), $d=30$\,pc (green), and $d=50$\,pc (blue)  away from an impulsive source with initial energy 
  $E_{\text{SNR}}=10^{51}$\,erg and initial CR spectral index $\alpha$=2.2.
  A energy dependent diffusion of CRs (see Section 4) has been applied with a diffusion coefficient at 10 GeV bounded between  $D_\text{10}=10^{26}\text{ to }10^{28}$\,cm$^{2}$\,s$^{-1}$. 
  }
  \label{energydensity}
  \end{figure}

Based on the mass estimates towards molecular regions overlapping the TeV sources, we use eq.\,10 from \citet{Aha1991} to derive the cosmic-ray enhancement factor 
$k_\text{CR}= w_\text{CR}/w_{\odot}$ which represents the ratio between $w_\text{CR}$ the energy density of CRs towards a molecular cloud next to a TeV source, and $w_{\odot}\sim1.0$\,eV\,cm$^{-3}$ 
the energy density found in the solar neighbourhood.
We use the combined atomic and molecular mass to obtain the total amount of target material available.
Table\,\ref{cosmicraytable} indicates the $k_\text{CR}$ values required for each molecular regions (partially) overlapping the TeV sources to account for the observed TeV fluxes.
The $k_\text{CR}$ value can then be compared to typical cosmic-ray enhancement factors predicted in the vicinity of SNRs (and potentially towards PWNe) to check the plausibility 
of hadronic contribution.\newline\newline
\noindent{\emph{\underline{Contribution from nearby SNRs?}}}\\
Nearby SNRs are the most likely viable candidates to produce  CRs energy densities up to $\sim10^{3}$\,eV\,cm$^{-3}$ (see \citealt{ReynoldsSNR} and references therein).
CRs propagate along the magnetic field lines and they scatter from their interaction with magnetic field perturbations (provided the scale of the perturbation roughly equals the CR gyroradius).
 As the magnetic field turbulence is thought to be enhanced in molecular clouds, we here assume an isotropic diffusion of CRs and electrons as a first order approximation.
 If we assume an isotropic diffusion of CRs and neglect energy losses, we can estimate the energy density distribution of CRs at a distance $R$ from 
 the SNR (here assumed as an impulsive source of CRs, see \citealt{Aha1996}):
 \begin{eqnarray}
\label{diffusionimp} n\left(E,R,t\right)&=&\frac{\eta_{\text{pp}}E_\text{SNR}}{\left(m_\text{p}c^2\right)^{2-\alpha}}\frac{E^{-\alpha}}{\pi^{3/2} R_\text{d}^{3}}\exp\left(-\left(\frac{R}{R_\text{d}}\right)^2\right)\\ 
 R_\text{d}&=&2\left(\chi D_{10}t_\text{age}\sqrt{E/10\text{ GeV}}\right)^{1/2}
 \end{eqnarray}
with $m_\text{p}$ being the proton mass, $D_{10}=10^{28}$\,cm$^{2}\,$s$^{-1}$ being the diffusion coefficient of 10 GeV CRs, $\alpha$ the spectral index of the proton distribution, $\eta_\text{pp}$ the ratio of 
the total SNR energy $E_\text{SNR}$ transferred to CRs. $R_\text{d}\left(t\right)$ represents the diffusion radius travelled by CRs after a time $t$. The diffusion suppression factor $\chi$ accounts for slower diffusion of particles which can be caused, for instance, by streaming instabilities or perturbations caused by shocks (see \citealt{Navadiffusion2016}, \citealt{Malkov2013} and references therein). 
Here we use, $\chi=0.01\text{ to } 1$ which matches  the slow and fast diffusion coefficient regimes defined by \citet{Aha1996}. 
Various studies of the cosmic-rays interacting with nearby molecular clouds W\,28 \citep{GiulianiW282010,LiChenW28Diffusion,Gabici2010W28} have, in fact, suggested a suppression factor between $\chi=0.01\text{ and }0.1$.
However, the value of $\chi$ remains poorly constrained.
From Eq.\,\ref{diffusionimp}, we can then obtain the total energy density of CRs $w_\text{CR}\left(R,t\right)$ using the following equation: 
\begin{equation}
 w_\text{CR}\left(R,t\right)=\int_{\epsilon_0}{n\left(E,R,t\right)E\text{d}E}
 \label{getenergy}
\end{equation}
with $\epsilon_0=m_\text{p}c^2$ is the proton energy at rest.
Figure\,\ref{energydensity} illustrates the range of $k_\text{CR}$  produced by SNRs with initial energy $E_\text{SNR}=10^{51}$\,erg as a function of the SNR age, using a standard proton spectral index $\alpha=2.2$ and $\eta_\text{pp}=0.1$.
From the distance between the SNRs and  the surrounding ISM regions (see Table\,\ref{cosmicraytable}), and the age of the SNRs, we can then check whether the required enhancement factors from Table\,\ref{cosmicraytable} fall within the predicted range from 
nearby SNRs.\\

  \noindent{{\emph{\underline{CR contribution from PWNe?}}}}\\
   Additionally, a few authors (see \citealt{Amato} and references therein) have also argued that high energy hadrons could 
  also be produced inside the pulsar environment, and be responsible for several features inside the PWN (e.g wisps in the Crab PWN, \citealt{gallant}). 
 Providing high energy CRs have indeed been produced inside PWNe and not suffered heavy adiabatic losses, we will thus discuss whether the  pulsars considered in this work could also generate the required cosmic-ray enhancement factors shown in Table\,\ref{cosmicraytable}. 
 In order to model the high cosmic-ray energy density potentially produced by those pulsars, we account for the evolution of the spin down power $\dot{E}_\text{SD}$ as a function of time $t$, 
 which can be described as :
 \begin{eqnarray}
  \dot{E}_\text{SD}\left(t\right)&=&\dot{E}_\text{SD}\left(t_\text{age}\right)\left(1+\left(n_\text{b}-1\right)\frac{\dot{P}\left(t-t_\text{age}\right)}{P}\right)^{-\Gamma}\\
  \Gamma&=&\frac{n_\text{b}+1}{n_\text{b}-1}
 \end{eqnarray}

with $t_\text{age}$ being the current age of the PWN,  $P\left(t\right)$ and $\dot{P}\left(t\right)$ being the current period and period derivative of a pulsar respectively at time $t$, $\dot{E}_\text{SD}\left(t\right)$ being the
pulsar spin  down power at time $t$, and $n_\text{b}$ being the pulsar braking index.
In order to obtain the density of CRs at a given radius $R$ from the pulsar, we rewrite Eq.\,\ref{diffusionimp} using the source term $S\left(t\right)=\eta_\text{pp}\dot{E}_\text{SD}\left(t\right)/\left(m_\text{p}c^2\right)^{2-\alpha}$, 
with $\eta_\text{pp}$ here being the fraction of the spin down power transferred to CRs. 
We thus numerically solve the following equation :
 \begin{equation}
  n\left(E,R,t\right)=\frac{E^{-\alpha}}{\pi^{3/2}}\int_{t_\text{age}}^{0}{-\frac{S\left(\xi-t_\text{age}\right)}{R_d^3\left(\xi\right)}\exp\left(-\frac{R^2}{R_d^2\left(\xi\right)}\right)\text{d}\xi}
  \label{diffusionpulsar}
 \end{equation}
 with $\xi=t_\text{age}-t$.
As per the SNR scenario, we obtain the CR energy density at a given distance $R$ from pulsar  using Eq.\,\ref{getenergy}.

  \begin{figure}
  \vbox{
  \includegraphics[height=0.5\textwidth,angle=270]{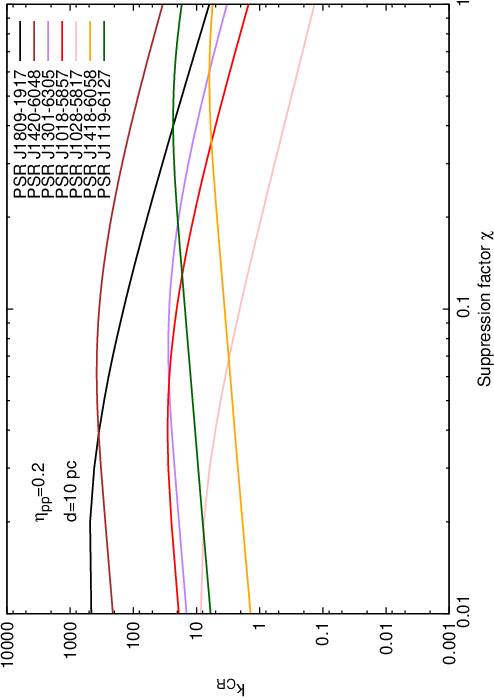}
  \includegraphics[height=0.5\textwidth,angle=270]{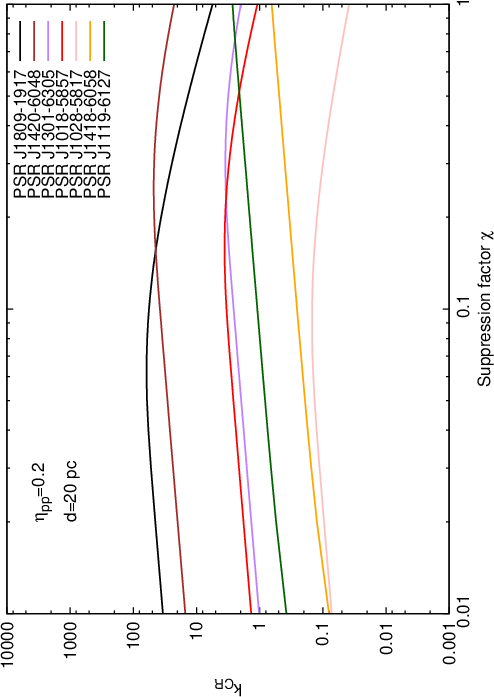}
  \caption{Predicted energy density $k_\text{CR}$ from pulsars at 10\,pc (top panel) and at 20\,pc (bottom panel) distance as a function of the 
  diffusion coefficient suppression factor $\chi$ (see colour version online).}
  \label{energydensitypulsar}
  }
  \end{figure}
  Figure\,\ref{energydensitypulsar} illustrates the CR energy density produced by the various pulsars as a function of the diffusion coefficient suppression factor $\chi$.
  Here, we assumed $n_\text{b}=3$ for all pulsars,  except for PSR\,J1119$-$6127 in which we used the measured braking index $n_\text{b}=2.68$.
  From the spectral modelling towards several PWNe, \citet{BucciantiniPWN2011} suggested that the maximum energy rate available to be transferred to CRs must be at most 0.20 
  of the total spin down power.
 We thus have used  $\eta_\text{pp}=0.20$ and  the $k_\text{CR}$ values shown in Figure\,\ref{energydensitypulsar} are consequently upper-limits.  
In the following subsections, we compare  these predictions with the various required $k_\text{CR}$ (see Table\,\ref{cosmicraytable}) derived inside the molecular regions. 
  \subsection{TeV emission from high energy electrons}
  \label{sec:highenergyelectrons}
  As opposed to CRs, high energy electrons  suffer heavy radiation losses as they diffuse inside the dense ISM because of the potentially enhanced magnetic field strength (see \citealt{Crutcher2010}).
  In the lack of intense radiation fields, synchrotron losses, with time-scale $\tau_{\text{sync}}=2.45\times10^{7}\left(B^2_{\text{mg}}E/m_\text{e}c^2\right)^{-1}$\,yr 
  ($B_\text{mg}=B/1\text{mG}$ being the magnetic field strength and $E$ being the total energy of the electron)  is likely to dominate over inverse-Compton losses inside molecular clouds.
  Although Bremsstrahlung emission, with time-scale $\tau_\text{brem}\sim4.6\times10^{7}\left(n/1\text{ cm}^{-3}\right)^{-1}$\,yr, may contribute inside molecular clouds with densities 
  $n_\text{H}>\text{ a few}\times10^{3}$\,cm$^{-3}$ in the 0.1\,to\,1\,TeV band, we expect the TeV emission above 1\,TeV to anti-correspond with the molecular ISM as the IC radiation is likely to dominate above 1\,TeV. 
  Table\,\ref{highenergytable} indicates  the synchrotron energy loss time-scale 
 $\tau_{\text{sync}}$ for electrons with energy $E=10$\,TeV, required to produce photons with energy $E_{\gamma}\sim1$\,TeV.
 To obtain $B$ values inside molecular clouds, we use the  \citet{Crutcher2010} relation :
 \begin{eqnarray}
   &B=10\,\mu\text{G} & \text{for } n_{\text{H}}<300\,\text{cm}^{-3} \\
   &B=10\left(\dfrac{n_{\text{H}}}{300\text{ cm}^{-3}}\right)^{0.6}\,\mu\text{G} &\text{for } n_{\text{H}}\ge300\,\text{cm}^{-3}
  \label{crutchereq2010}
 \end{eqnarray}

 We compare these time-scales with the age of nearby high energy sources and the diffusion time-scale $\tau_{\text{diff}}$ required for the particles with energy above 10\,TeV to fully cross a molecular region.
 We here define $\tau_{\text{diff}}=\left(\langle r\rangle+d_{\text{source}}\right)^{2}/6D\left(E,B,\chi\right)$ with $\langle r\rangle$ being the mean radius of the molecular region, and  $d_{\text{source}}$ the distance separating the centre of the 
 high energy source to the centre of the molecular region. 
 To check the leptonic scenario, we first select dense molecular regions which anti-correspond with each TeV source and/or its X-ray counterpart.
 Here, we preferably use the extended CS regions as most of the observed gas is confined.
 However, in the cases where no CS(1--0) emission was detected (i.e. HESS\,J1018$-$189B, HESS\,J1303$-$631 and HESS\,J1119$-$164), we use the CO regions.
 We again note that the densities $n_{\text{H}_2}$ from our CO analyses are upper-limits. Thus, these synchrotron time-scale values should be, in these cases,  used as upper-limits.
 Then, we derive the synchrotron time-scale and compare it to $\tau_\text{diff}$ required for the high energy electrons to cross the cloud.
 In the case where the synchrotron time-scale is much smaller than the diffusion time-scale, we expect the high energy electrons to radiate most of their energy and thus become 
 unable to produce X-rays and TeV emission inside the molecular cloud.
  An anti-correlation between the ISM and X-rays would support the physical connection between the TeV source and the ISM as well as the leptonic nature of the TeV source (see \citealt{SanoBamplification2013}).

    \subsection{Discussion of Individual Sources}
    \noindent{\emph{\underline{HESS\,J1809$-$193}}}\newline
\citet{J1809radio2016}, and more recently \citet{Araya2018}, have  discussed the hadronic scenario  in which they assume the SNR G011.0$-$0.0 to be located at a distance $d\sim$3.0\,kpc.
We focus here on how molecular clouds at $d\sim3.7$\,kpc can affect the TeV emission produced by the PWN powered by PSR\,J1809$-$1917 and SNR\,G011.0$-$0.0, in which we argue they are physically connected.

We remark in Table\,\ref{highenergytable} that the synchrotron energy loss time-scale $\tau_{\text{sync}}$ for electrons above 10 TeV is significantly smaller than the diffusion time-scale through the various ISM regions.
Consequently, most electrons producing gamma-rays above 1 TeV would not penetrate the dense molecular regions.

We also check  whether the CRs in molecular clouds may produce significant TeV emission. 
To match the observed ISM regions, the TeV flux was scaled simplistically by area ratio, that is 16\%, 16\% and 12\% towards regions `1', `2' and `3' respectively. 
We conclude that a cosmic-ray enhancement factor $k_\text{CR}=10-30$ (see  Table\,\ref{cosmicraytable}) is required to attain the TeV flux towards  region `1', `2' and `3' respectively.
As the distances between these SNRs and the molecular regions range between 10 and 20\,pc (see Table\,\ref{highenergytable}), 
 we find, based on the red and green regions in Figure\,\ref{energydensity} that, at $t_\text{age}=51$\,kyr, the predicted $k_\text{CR}$ ranges to be between $k_\text{CR}\sim1\text{ and }300$.
Consequently, the SNR\,G011.0$-$0.0  may be able to produce the required cosmic-ray enhancement and produce hadronic TeV emission.
Interestingly, we find that the PWN powered by PSR\,J1809$-$1917 (see black line in Figure\,\ref{energydensitypulsar}) could also significantly increase the predicted cosmic-ray enhancement factor towards nearby molecular clouds with $k_\text{CR}$ values peaking at $\sim270$ (for diffusion suppression $\chi=0.02$) and thus
generally exceeds the aforementioned $k_\text{CR}$ towards the molecular regions `1' and `2'.
Although these CR energy densities should be considered as  upper-limits, the PWN powered by the PSR\,J1809$-$5158 may consequently be a viable laboratory to probe possible hadronic components
 originating from the PWN.\newline

    \noindent{\emph{\underline{HESS\,J1026$-$583}}}\newline 
 It has already been stated that, based on energetics alone, the spin down energy of the pulsar PSR\,J1028$-$5819 could produce the HESS\,J1026$-$583 TeV emission \citep{J1026HESS}.

From our CO(1--0) and CS(1--0) data, we have however identified a dense molecular region spatially coincident with the TeV emission.
If the TeV emission is indeed of hadronic origin, we also require a proton spectrum $J\left(E\right)\propto E^{-1.94}$ to generate the photon spectrum towards HESS\,J1026$-$583 (see \citealt{J1026HESS}).

Taking into account this hard gamma-ray spectra, the eq.\,10 from \citet{Aha1991} must be adjusted by a factor of $1.6/0.94=1.7$. We thus obtain the cosmic-ray enhancement factor $k_\text{cr}=60$.
Using the combined mass of the regions `5' to `7' $M_{\text{H}_2}\left(\text{CS}\right)=5.0\times10^{3}\solmass$, the required cosmic-ray enhancement factor only increases to $k_\text{CR}=103$.
We find that this value can only be attained if there is a SNR located art $d<10$\,pc to the molecular region `B' (see red region in Figure\,\ref{energydensity}).
However, no SNRs nor any cosmic-ray accelerators have been detected close to this molecular cloud so far.\\

\noindent{\emph{\underline{HESS J1119$-$164}}}\newline    
The centroid of the TeV emission appears to be located north-west of PSR\,J1119$-$6127, towards regions `X1' and `X2'.
The non-thermal X-ray emission, thought to be produced by high energy electrons inside the front shock, is unlikely to have also produced the TeV emission.
The possible cause of the TeV gamma-ray emission may either come from high energy electrons from the PWN, or the cosmic-rays produced by the SNR front shock as they interact with 
the ISM in region `D`.
By modelling the GeV and TeV spectral energy distribution using Fermi-LAT and HESS data respectively,
\citet{FermiPWN} showed that electrons originating from the PWN could reproduce the observed spectrum.
The offset of the TeV emission with respect to the pulsar position could suggest that the PWN may have been crushed by the reverse shock 
of the progenitor SNR.
The Nanten CO(1--0) survey has however not revealed any significant emission  east of PSR\,J1119$-$6127 between $v_\text{lsr}=20\text{ to }40\,$km/s.
A higher resolution gas survey could resolve clumps which could be responsible for the asymmetry.

 The hadronic scenario may also be plausible as the TeV gamma-ray emission corresponds with the CO(1--0) emission west of SNR\,G292.2$-$0.5.
The thermal X-ray emission towards region `D' suggest that some of the gas in region `D' may have been heated as it interacted with the SNR.
From Table\,\ref{cosmicraytable}, the required cosmic-ray enhancement factor $k_\text{CR}$ required for the cosmic-rays to produce the TeV gamma-ray flux is $\sim 143$.
A SNR age $t=1.9\,$kyr suggests that the bulk of CRs are remained confined inside the SNR front shock.
However, the steep photon spectral index $\Gamma=2.64$  above 1 TeV \citep{HGPS2018} also indicates that CRs above 10\,TeV have already escaped the SNR. 
With the increased sensitivity of CTA, one could look at the spatial and spectral evolution of the TeV gamma-ray emission inside the ISM towards region `D'. 
A hardening of the TeV gamma-ray spectrum with increasing distance of the SNR could  favour the hadronic scenario.

Finally, as shown by the orange line in Figure\,\ref{energydensitypulsar}, we also find that the pulsar is unlikely to have produced enough CRs to produce the TeV emission towards HESS\,J1119$-$164.\newline

   \begin{table}
    \caption{Cosmic-ray enhancement factors $k_{\text{cr}}$ derived using eq.\,10 from \citet{Aha1991}, required to 
    reproduce the observed TeV emission above 1 TeV $F\left(>1 \text{ TeV}\right)=\int{N_0E_\gamma^{-\Gamma}\text{d}E_{\gamma}}$ via CR-ISM interaction. 
    }
    \small
\begin{tabular}{p{2.3cm}cccl}
 \toprule
 & Molecular\\
 & Regions & $k_{\text{cr}}$& $F\left(>1\,\text{TeV}\right)$ \\
 & & & [ph cm$^{-2}$ s$^{-1}$]  & \\
 \midrule
 \multirow{3}{*}{HESS\,J1809$-$193}& 1$^{a}$ & 32 & $^{*}6.2\times10^{-13}$\\
 & 2$^{a}$ &  12 & $^{*}6.2\times10^{-13}$ \\
 & 3$^{a}$ & 22 &$^{*}4.7\times10^{-13}$ \\
 \midrule
 HESS\,J1026$-$582 & B & 60 &$^{\dagger}1.1\times 10^{-12}$\\
 \midrule
 HESS\,J1119$-$193& D & 143 &$^{\ddagger}9.2\times10^{-13}$\\
 \midrule
 \multirow{5}{*}{HESS\,J1420$-$607}& 6 & 53 & $^{\amalg}1.3\times10^{-13}$ \\
 & 7 & 88 & $^{\amalg}6.7\times10^{-14}$ \\
 & 8 & 81 & $^{\amalg}4.2\times10^{-14}$ \\
 & 9 & 81 & $^{\amalg}3.2\times10^{-14}$ \\
 & 10 & 97 & $^{\amalg}1.1\times10^{-13}$ \\
 \midrule
 \multirow{3}{*}{HESS\,J1303$-$631} & A$^{b}$ & 257 & $^{\pm}1.7\times10^{-12}$\\
 & C$^{c}$ & 194 & $^{\pm}2.1\times10^{-13}$ \\
 & D$^{c}$ & 237 & $^{\pm}2.1\times10^{-13}$ \\
 \midrule 
 HESS\,J1018$-$589A & A & 91 &$^{\star}2.4\times10^{-13}$\\
 \bottomrule
 \multicolumn{5}{p{0.45\textwidth}}{\footnotesize{$^{*}$\citet{HESSJ1809}, $^{\dagger}$\citet{J1026HESS}, $^{\ddagger}$\citet{HGPS2018}, $^{\pm}$, $^{\amalg}$\citet{HESSJ1418}, \citet{J1303PWN},\,\,\,\, $^{\star}$\citet{J1018HESS3}}}\\   
 \multicolumn{5}{p{0.45\textwidth}}{\footnotesize{$^{\text{a}}$:We scaled down the  HESS\,J1809$-$193 photon flux by 16\%, 16\% and 12\% for regions `1', `2', `3' respectively, corresponding to the ratio between the molecular regions and the TeV emission sizes.  }}\\
\multicolumn{5}{p{0.45\textwidth}}{\footnotesize{$^{\text{b}}$:We scaled down the HESS\,J1420$-$607 photon flux by 4\%,$\sim 1$\%, $<1$\%, $<1$\% and 3\%  for the regions 6 to 10 respectively. }} \\
 \multicolumn{5}{p{0.45\textwidth}}{\footnotesize{$^{\text{c}}$: We scaled down the HESS\,J1303$-$631 photon flux by 44\%, 7\% and 7\% for region `A',`B' and `C' respectively.  }}
 \end{tabular}
 \label{cosmicraytable}
   \end{table}

   \begin{table*}
   \centering
    \begin{minipage}{\textwidth}
    \caption{Table showing the diffusion time-scale $t_\text{diff}=\left(\langle r\rangle+d_{\text{source}}\right)^2/6D\left(E,B\right)$ for a  particle with energy $E=10$\,TeV to cross the listed molecular regions with 
    mean radius $\langle r\rangle$ and positioned at $d_{\text{source}}$ from the centre of the listed sources.
    The age of the source $t_{\text{age}}$, the CR-ISM interaction time-scale $\tau_{\text{pp}}$ for protons, and the synchrotron time scale $\tau_{\text{sync}}$ for electrons with energy $E=10$\,TeV are 
    also displayed as means of comparison.
    }
     \begin{tabular}{ccccccp{0.1cm}cccc}
     \toprule 
     & &\multicolumn{4}{c}{Molecular region properties} & & \multicolumn{2}{c}{Source particle properties}\\
     \cmidrule{3-6}\cmidrule{8-9}
      &Reg.& $\langle r\rangle$ & $B^{\text{a}}$  & $\tau_{\text{pp}}^{\text{b}}$ &$\tau_{\text{sync}}^{\text{c}}$& &Source & $t_{\text{age}}$ &  $d_{\text{source}}$  & $t_{\text{diff}}\left(10\text{ TeV}\right)^{\text{d}}$  \\
      & & [pc] & [$\mu$G] & [kyr] & [kyr] & & &[kyr]& [pc]& [kyr/$\chi_{0.1}$]  \\
      \midrule
      \multirow{4}{*}{HESS\,J1809$-$193}& \multirow{2}{*}{1} & \multirow{2}{*}{9} & \multirow{2}{*}{29} & \multirow{2}{*}{69} & \multirow{2}{*}{1.4} &\multirow{2}{*}{$ \bigg\{$}& SNR\,G011.0$-$0.0 & 51 & 8 & 2 \\
      & & & & & & &PSR\,J1809$-$5158 & 51 &13 &3 \\ 
      & \multirow{2}{*}{2} & \multirow{2}{*}{10} & \multirow{2}{*}{41} & \multirow{2}{*}{35} & \multirow{2}{*}{0.6} &\multirow{2}{*}{$ \bigg\{$}& SNR\,G011.0$-$0.0 & 51 & 20 & 6 \\
      & & & & & & &PSR\,J1809$-$5158 & 51 &  28 & 9 \\ 
      \midrule
      HESS\,J1026$-$582 & A & 11 & 10 & 1200 & 12 & &PSR\,J1028$-$5819 & 89 & 5 & 1  \\
      \midrule
      \multirow{3}{*}{HESS\,J1303$-$631} & A & 12 & 18 & 26 & 3 & &PSR\,J1301$-$6305 & 11 & 45 & 14  \\
      & B & 3 & 32 & 56 & 1 & &SNRc$^{e}$ & 11 & 8 & 1  \\
      & C & 11 & 19 & 167 & 3 &  &PSR\,J1301$-$6305 & 11 & 48 & 15   \\
      & D & 24 & 10 & 555 & 11 & &PSR\,J1301$-$6305 & 11 & 26 & 4  \\
      \midrule
      \multirow{11}{*}{HESS\,J1418$-$609}& A & 6 & 33 & 41 & 1 & &PSR\,J1418$-$6058& 1.6 & 21 & 4  \\
       & 1 & 4 & 26 & 85 & 2 & & PSR\,J1418$-$6058 & 1.6 & 18 & 2 \\
       & 2 & 4 & 12 & 400 & 8 & & PSR\,J1418$-$6058 & 1.6 & 19 & 2 \\
       & 3 & 3 & 22 & 115 & 2 & & PSR\,J1418$-$6058 & 1.6 & 26 & 4 \\
       & 4 & 5 & 24 & 107 & 2 & & PSR\,J1418$-$6058 & 1.6 & 30 & 6 \\
       & 5 & 4 & 32 & 60 & 1 & & PSR\,J1418$-$6058 & 1.6 & 27 & 5 \\
       & 6 & 3 & 25 & 90 & 2 & &  PSR\,J1418$-$6058 & 1.6 & 22 & 3 \\
       & 7 & 2 & 24 & 107 & 2 & & PSR\,J1418$-$6058 & 1.6 & 15 & 1  \\
       & 8 & 2 & 35 & 49 & 1 & & PSR\,J1418$-$6058 & 1.6 & 12 & 1 \\
       & 9 & 1 & 22 & 125 & 3 & & PSR\,J1418$-$6058 & 1.6 & 16 & 1 \\
       & 10 & 3 & 30 & 41 & 1 & & PSR\,J1418$-$6058 & 1.6 & 32 & 4 \\
      \midrule
      HESS\,J1018$-$589& A  & 4 & 26 & 85 & 2 & &SNR\,G284.3$-$1.8& 10 & 8 & 1 \\
      \bottomrule
     \multicolumn{11}{p{18cm}}{\footnotesize{$^{\text{a}}$:Obtained from eq.\,21 from \citet{Crutcher2010} (see Section 4.2).}} \\
      \multicolumn{11}{p{18cm}}{\footnotesize{$^{\text{b}}$  $\tau_{\text{pp}}=6\times10^{7}/2n_{\text{H}_2}$ where $n_{\text{H}_2}$ can be found in Tables \ref{MASSJ1809} to \ref{MASSJ1018}}}\\
     \multicolumn{11}{p{18cm}}{\footnotesize{$^{\text{c}}$: $\tau_{\text{sync}}=2.45\times10^{7}\left(B_{\text{ mG}}^2E/m_\text{e}\right)^{-1}$\,yr \citep{ginzburg}. }}\\
     \multicolumn{11}{p{18cm}}{\footnotesize{$^{\text{d}}$: See eqs.\,2 and 3 from \citet{Gabicidiff2006} to obtain the diffusion time-scale. The results shown assume a diffusion coefficient suppression factor $\chi=0.1$}}\\
     \multicolumn{11}{p{18cm}}{\footnotesize{$^{\text{e}}$ SNR candidate, see text and \citet{HESSJ1303SNR2017} for further details and Figure\,\ref{J1303COCS} for location. We used standard SNR expansion 
     parameters to provide an age estimate of the SNRc if located at this distance (see text).}}
       \end{tabular}
\label{highenergytable}
    \end{minipage}
  
   \end{table*}

 \noindent{\emph{\underline{HESS\,J1418$-$609 and HESS\,J1420$-$607}}} \\  
  From Table\,\ref{cosmicraytable}, we observe $k_\text{CR}$ values ranging between 50 and 90 towards the clumps `6' to `10', whose projected distance ranges between 22\,pc and 35\,pc (see Table\,\ref{highenergytable}).
 Assuming the progenitor SNR of PSR\,J1418$-$6058 is 1.6\,kyr old, we note that escaping CRs are unlikely to produce significant TeV gamma-ray emission towards HESS\,J1420$-$607 (see blue region in Figure\,\ref{energydensity}). 
  As per HESS\,J1119$-$164, it is expected that CRs may still remain confined inside the SNR front shock.
 Similarly for the leptonic scenario (see  Table\,\ref{highenergytable}), a fast diffusion coefficient ($\chi>0.1$) is required for electrons to reach the clumps `6' to `10' and the gas in region `A'.
 \citet{J1418Suzaku} indicated that a slow diffusion of electrons inside the PWNe is required to produce the diffuse X-ray emission observed with Suzaku.
 Consequently, it is unlikely that high energy electrons escaping HESS\,J1418$-$609 can contribute to the TeV emission towards HESS\,J1420$-$607.
 
 The various clumps north of HESS\,J1418$-$609 could affect its TeV emission.
 From Table\,\ref{highenergytable}, we note that, although some clumps could impede the propagation of multi-TeV electrons ($\tau_\text{sync}\ll\tau_\text{diff}$), 
 a large diffusion coefficient would also be necessary for the high energy electrons to reach these clumps. Consequently, at this distance, the effect of the ISM on the morphology of HESS\,J1418$-$609 is also not 
 significant.\newline

\noindent{\emph{\underline{HESS J1303$-$631}}}\newline
HESS\,J1303$-$631 has clearly been identified as a PWN based on energy dependent morphology  at TeV energies \citep{J1303PWN}. 
We here check the potential contribution of the molecular regions `A' to `D' to the observed TeV gamma-ray emission.

From Table\,\ref{highenergytable}, we find that, assuming a suppression factor $\chi=0.1$, high energy electrons would only reach the molecular regions `A' (located at $d\sim6$\,kpc) and `C',`D' (both 
located at $d\sim13.6$\,kpc).
In all cases, the shorter synchrotron time-scale indicates that any high energy electrons would lose most of their energy while diffusing into these molecular regions. 
Consequently, if associated with the PWN, less TeV emission spatially overlapping with these molecular regions should be expected.

Now, we discuss whether the selected molecular regions can also affect the morphology of the TeV emission from hadronic interactions. As per the case of HESS\,J1809$-$193, we scaled down the photon flux of HESS\,J1303$-$631 to 44\%, 7\% and 7\% 
 for regions `A', `C' and `D' respectively, based on the size ratio between the molecular regions and the TeV emission. We then obtain $k_\text{CR}\sim200$ (see Table\,\ref{cosmicraytable}). 
 Based on the distances between the molecular clouds to the pulsar or SNR candidate (see Table\,\ref{highenergytable}), we find from Figures\,\ref{energydensity} and \ref{energydensitypulsar} that neither the pulsar (see purple line) nor the SNR candidate can provide sufficient CR energy density to 
 produce  hadronic TeV emission.\newline

 \noindent{\emph{\underline{HESS\,J1018$-$589}}}\newline
 We now discuss how the ISM in region `A' may affect the TeV emission towards HESS\,J1018$-$589a.
 We find that the enhancement factor $k_\text{CR}=91$ (see Table\,\ref{cosmicraytable}) is required for the cosmic-ray to produce observed TeV flux.
 A 10 kyr old SNR is capable of producing such cosmic-ray enhancement if the distance between the SNR and the cloud is $\sim 10$\,pc (see red region in Figure\,\ref{energydensity}).
 A lack of  TeV emission coincident with the region `A' may suggest that the molecular cloud may be foreground/background to the SNR and thus may not be physically connected. 
 Here we discuss only the possible contribution from CRs originating from the SNR\,G284.3$-$1.8.
 From our mass estimate in region `A' (see Table\,\ref{tableJ1018}), we find that a slow diffusion of CRs escaping the SNR might lead to a significant contribution of the hadronic TeV emission towards HESS\,J1018$-$583A.

\section{Conclusion}
In this paper, we have mapped the  molecular and atomic ISM towards pulsar wind nebulae (PWNe) and PWNe candidates, combining our new 7mm Mopra survey with the Nanten CO(1--0) data and the SGPS/GASS H\textsc{i} survey.
Except for the cases of HESS\,J1303$-$631 and HESS\,J1018$-$589B, our ISM studies provided additional information about the distance of astrophysical sources potentially related to the TeV source.
We have also identified molecular clouds which could explain the morphology of the TeV gamma-ray emission as observed by CTA \citep{CTAproject}, and 
 also highlight cosmic-rays (CRs) originating from  progenitor supernova remnants (SNRs).
 From the hypothesis that CRs could also be produced within the pulsar environment, and that at most 20\% of the spin down power could be transferred to CRs \citep{BucciantiniPWN2011}, 
we find that, among the studied PWNe, only CRs from PSR\,J1809$-$1917 could contribute to the TeV emission.

Towards HESS\,J1809$-$193, we have found several dense molecular regions at $v_{\text{lsr}}=10\text{ to }22$\,km/s and  $v_{\text{lsr}}=25\text{ to }38$\,km/s adjacent to the pulsar PSR\,J1809$-$1918 and 
the SNRs\,G011.0$-$0.0 and G011.1$+$0.1. Notably, we have detected  SiO(1--0) emission towards the dense molecular cloud south of HESS\,J1809$-$193 with no infra-red counterparts, 
suggesting a possible SNR-MC interaction at the pulsar dispersion measure distance $d\sim3.7$\,kpc.
We found that the ambient density required to reconcile the projected radius of SNR\,G011.0$-$0.0 with the characteristic age of the pulsar PSR\,J1809$-$1917 is consistent with the averaged density $n_{\text{H}_2}$ obtained from our CO analysis.
We argue that the PWN  electrons could lose most of their energy before propagating deep inside the nearby molecular clouds, and would therefore become unable to 
produce coincident TeV gamma-ray emission.
 We also note that the CRs that have escaped the pulsar's progenitor SNR may produce significant 
TeV emission towards these molecular clouds.

Towards HESS\,J1026$-$582, the molecular clouds east of the pulsar PSR\,J1028$-$5819 may explain its offset position with respect to the peak of the TeV emission at $d\sim2.7$\,kpc.
However, we have also found a partial shell structure spatially coincident with the TeV emission at $d\sim5$\,kpc and highlight a potential hadronic origin, 
powered by an unknown CR source at $d<10$\,pc from the molecular cloud.

Towards HESS\,J1119$-$164, the molecular clouds at the kinematic distances  $d\sim8.6\text{ to }9.7$\,kpc found on the west side of SNR\,292.2$-$0.5
show good correspondence with the thermal X-ray detections with Chandra and XMM-Newton, suggesting a possible SNR-MC interaction.
Our comparative study of our column density estimates with the column density derived from X-ray measurements also supports the TeV source distance $d>8.6$\,kpc.

Combining our column density study as well as our ISM morphology study towards HESS\,J1418$-$609, we argue the TeV source may be located at  $d\sim3.5-5.6$\,kpc.
We also claim that the various molecular clumps are too far from the pulsar PSR\,J1418$-$6058 to affect or contribute to the TeV gamma-ray emission.

Our ISM analysis has not constrained the distance of HESS\,J1303$-$631 as the CO(1--0) emission overlaps the TeV emission at all velocity ranges.
Notably, at $v_\text{lsr}=-22\text{ to }-15$\,km/s, however, we did find a CO dip spatially coincident with the SNR candidate found by \citet{HESSJ1303SNR2017}, inferring a distance $d\sim1.5$\,kpc. 

Although CO(1--0) emission north of the SNR\,G284.3$-$1.8 towards HESS\,J1018$-$589 has been found at $d\sim2.8$\,kpc, no CS clumps  have been found at the pulsar PSR\,J1019$-$5857 dispersion measure
 distance $d\sim 8.6$\,kpc, suggesting an atomic-dominated ISM surrounding HESS\,J1018$-$589B.
 An extension of high resolution H\textsc{i} survey would shed more light concerning this TeV source.
 
 As a conclusion, the arc-minute structure of the ISM plays an important role in understanding the morphology of the TeV gamma-ray emission produced by PWNe and progenitor SNR.
 The angular resolution and sensitivity of the Mopra CO(1--0) and $^{13}$CO surveys \citep{Mopra2015} will refine the structure and dynamics of the diffuse molecular gas.
 
 As future work, we aim to use the various properties of the identified clumps; to model the effect of the diffusion of CRs, and high energy electrons, escaping the PWNe and their progenitor SNRs;
 ; to model the interaction of these particles with the identified molecular cloud clumps; and to identify spectral and/or morphological signatures of the TeV gamma-ray emission. 
 These could then be compared to upcoming observations with CTA \citep{CTAproject}, which could further constrain the nature of the TeV sources.

\section*{Acknowledgments}
The Mopra radio telescope  is part of the Australia Telescope National Facility which is funded by the Australian Government for operation as a National Facility managed by CSIRO.
Operations support was provided by the University of New South Wales and the University of Adelaide.
The University of New South Wales Digital Filter Bank used for the observations with the Mopra Telescope (the UNSW–MOPS) was provided with support from the Australian Research Council LE160.100094 (ARC).

\Urlmuskip=0mu plus 1mu\relax
\bibliography{paperbib}{}

\begin{thebibliography}{}
\makeatletter
\relax
\def\mn@urlcharsother{\let\do\@makeother \do\$\do\&\do\#\do\^\do\_\do\%\do\~}
\definecolor{darkblue}{rgb}{0,0,0.597656}
\def\mndoi{\begingroup\mn@urlcharsother \@ifnextchar [ {\mndoi@} {\mndoi@[]}}
\def\mndoi@[#1]#2{\def\@tempa{#1}\ifx\@tempa\@empty \href
  {http://dx.doi.org/#2} {\textcolor{darkblue}{doi:#2}}\else \href
  {http://dx.doi.org/#2} {\textcolor{darkblue}{#1}}\fi \endgroup}
\def\mn@eprint#1#2{\mn@eprint@#1:#2::\@nil}
\def\mn@eprint@arXiv#1{\href {http://arxiv.org/abs/#1} {{\tt arXiv:#1}}}
\def\mn@eprint@dblp#1{\href {http://dblp.uni-trier.de/rec/bibtex/#1.xml}
  {dblp:#1}}
\def\mn@eprint@#1:#2:#3:#4\@nil{\def\@tempa {#1}\def\@tempb {#2}\def\@tempc
  {#3}\ifx \@tempc \@empty \let \@tempc \@tempb \let \@tempb \@tempa \fi \ifx
  \@tempb \@empty \def\@tempb {arXiv}\fi \@ifundefined
  {mn@eprint@\@tempb}{\@tempb:\@tempc}{\expandafter \expandafter \csname
  mn@eprint@\@tempb\endcsname \expandafter{\@tempc}}}

\bibitem[\protect\citeauthoryear{{Abdalla} et~al.,}{{Abdalla}
  et~al.}{2018a}]{HGPS2018}
{Abdalla} H.,  et~al., 2018a, \mndoi [\aap] {10.1051/0004-6361/201732098},
  \href {http://adsabs.harvard.edu/abs/2018A%26A...612A...1H} {612, A1}

\bibitem[\protect\citeauthoryear{{Abdalla} et~al.,}{{Abdalla}
  et~al.}{2018b}]{HESSPWN2018}
{Abdalla} H.,  et~al., 2018b, \mndoi [\aap] {10.1051/0004-6361/201629377},
  \href {http://adsabs.harvard.edu/abs/2018A%26A...612A...2H} {612, A2}

\bibitem[\protect\citeauthoryear{{Abeysekara} et~al.,}{{Abeysekara}
  et~al.}{2017}]{HAWC2017}
{Abeysekara} A.~U.,  et~al., 2017, \mndoi [\apj] {10.3847/1538-4357/aa7556},
  \href {http://adsabs.harvard.edu/abs/2017ApJ...843...40A} {843, 40}

\bibitem[\protect\citeauthoryear{{Abramowski} et~al.,}{{Abramowski}
  et~al.}{2011}]{J1026HESS}
{Abramowski} A.,  et~al., 2011, \mndoi [\aap] {10.1051/0004-6361/201015290},
  \href {http://adsabs.harvard.edu/abs/2011A%26A...525A..46H} {525, A46}

\bibitem[\protect\citeauthoryear{{Abramowski} et~al.,}{{Abramowski}
  et~al.}{2012a}]{J1018HESS2}
{Abramowski} A.,  et~al., 2012a, \mndoi [\aap] {10.1051/0004-6361/201218843},
  \href {http://adsabs.harvard.edu/abs/2012A%26A...541A...5H} {541, A5}

\bibitem[\protect\citeauthoryear{{Abramowski} et~al.,}{{Abramowski}
  et~al.}{2012b}]{J1303PWN}
{Abramowski} A.,  et~al., 2012b, \mndoi [\aap] {10.1051/0004-6361/201219814},
  \href {http://adsabs.harvard.edu/abs/2012A%26A...548A..46H} {548, A46}

\bibitem[\protect\citeauthoryear{{Abramowski} et~al.,}{{Abramowski}
  et~al.}{2015}]{J1018HESS3}
{Abramowski} A.,  et~al., 2015, \mndoi [\aap] {10.1051/0004-6361/201525699},
  \href {http://adsabs.harvard.edu/abs/2015A%26A...577A.131H} {577, A131}

\bibitem[\protect\citeauthoryear{{Acero} et~al.,}{{Acero}
  et~al.}{2013}]{FermiPWN}
{Acero} F.,  et~al., 2013, \mndoi [\apj] {10.1088/0004-637X/773/1/77}, \href
  {http://adsabs.harvard.edu/abs/2013ApJ...773...77A} {773, 77}

\bibitem[\protect\citeauthoryear{{Acharya} et~al.,}{{Acharya}
  et~al.}{2013}]{CTAproject}
{Acharya} B.~S.,  et~al., 2013, \mndoi [Astroparticle Physics]
  {10.1016/j.astropartphys.2013.01.007}, \href
  {http://adsabs.harvard.edu/abs/2013APh....43....3A} {43, 3}

\bibitem[\protect\citeauthoryear{{Acharya} et~al.,}{{Acharya}
  et~al.}{2017}]{CTAscience}
{Acharya} B.~S.,  et~al., 2017, preprint, \href
  {http://adsabs.harvard.edu/abs/2017arXiv170907997C} {} (\mn@eprint {arXiv}
  {1709.07997})

\bibitem[\protect\citeauthoryear{{Aharonian}}{{Aharonian}}{1991}]{Aha1991}
{Aharonian} F.~A.,  1991, \mndoi [\apss] {10.1007/BF00648185}, \href
  {http://adsabs.harvard.edu/abs/1991Ap%26SS.180..305A} {180, 305}

\bibitem[\protect\citeauthoryear{{Aharonian} \& {Atoyan}}{{Aharonian} \&
  {Atoyan}}{1996}]{Aha1996}
{Aharonian} F.~A.,  {Atoyan} A.~M.,  1996, \aap, \href
  {http://adsabs.harvard.edu/abs/1996A%26A...309..917A} {309, 917}

\bibitem[\protect\citeauthoryear{{Aharonian} et~al.,}{{Aharonian}
  et~al.}{2005}]{HESSJ1303discovery1}
{Aharonian} F.,  et~al., 2005, \mndoi [\aap] {10.1051/0004-6361:20053195},
  \href {http://adsabs.harvard.edu/abs/2005A%26A...439.1013A} {439, 1013}

\bibitem[\protect\citeauthoryear{{Aharonian} et~al.,}{{Aharonian}
  et~al.}{2006}]{HESSJ1418}
{Aharonian} F.,  et~al., 2006, \mndoi [\aap] {10.1051/0004-6361:20065511},
  \href {http://adsabs.harvard.edu/abs/2006A%26A...456..245A} {456, 245}

\bibitem[\protect\citeauthoryear{{Aharonian} et~al.,}{{Aharonian}
  et~al.}{2007}]{HESSJ1809}
{Aharonian} F.,  et~al., 2007, \mndoi [\aap] {10.1051/0004-6361:20077280},
  \href {http://adsabs.harvard.edu/abs/2007A%26A...472..489A} {472, 489}

\bibitem[\protect\citeauthoryear{{Amato}}{{Amato}}{2014}]{Amato}
{Amato} E.,  2014, \mndoi [IJMPS] {10.1142/S2010194514601604}, \href
  {http://adsabs.harvard.edu/abs/2014IJMPS..2860160A} {28, 60160}

\bibitem[\protect\citeauthoryear{{Anada}, {Bamba}, {Ebisawa}  \&
  {Dotani}}{{Anada} et~al.}{2010}]{J1809Suzaku}
{Anada} T.,  {Bamba} A.,  {Ebisawa} K.,   {Dotani} T.,  2010, \mndoi [\pasj]
  {10.1093/pasj/62.1.179}, \href
  {http://adsabs.harvard.edu/abs/2010PASJ...62..179A} {62, 179}

\bibitem[\protect\citeauthoryear{{Araya}}{{Araya}}{2018}]{Araya2018}
{Araya} M.,  2018, \mndoi [\apj] {10.3847/1538-4357/aabd7e}, \href
  {http://adsabs.harvard.edu/abs/2018ApJ...859...69A} {859, 69}

\bibitem[\protect\citeauthoryear{{Bamba}, {Ueno}, {Koyama}  \&
  {Yamauchi}}{{Bamba} et~al.}{2003}]{J1809Bamba}
{Bamba} A.,  {Ueno} M.,  {Koyama} K.,   {Yamauchi} S.,  2003, \mndoi [\apj]
  {10.1086/374354}, \href {http://adsabs.harvard.edu/abs/2003ApJ...589..253B}
  {589, 253}

\bibitem[\protect\citeauthoryear{{Blondin}, {Chevalier}  \&
  {Frierson}}{{Blondin} et~al.}{2001}]{Blon2001}
{Blondin} J.~M.,  {Chevalier} R.~A.,   {Frierson} D.~M.,  2001, \mndoi [\apj]
  {10.1086/324042}, \href {http://adsabs.harvard.edu/abs/2001ApJ...563..806B}
  {563, 806}

\bibitem[\protect\citeauthoryear{{Bolatto}, {Wolfire}  \& {Leroy}}{{Bolatto}
  et~al.}{2013}]{BolattoXco}
{Bolatto} A.~D.,  {Wolfire} M.,   {Leroy} A.~K.,  2013, \mndoi [\araa]
  {10.1146/annurev-astro-082812-140944}, \href
  {http://adsabs.harvard.edu/abs/2013ARA%26A..51..207B} {51, 207}

\bibitem[\protect\citeauthoryear{{Braiding} et~al.,}{{Braiding}
  et~al.}{2015}]{Mopra2015}
{Braiding} C.,  et~al., 2015, \mndoi [\pasa] {10.1017/pasa.2015.20}, \href
  {http://adsabs.harvard.edu/abs/2015PASA...32...20B} {32, e020}

\bibitem[\protect\citeauthoryear{{Braiding} et~al.,}{{Braiding}
  et~al.}{2018}]{Braiding2018}
{Braiding} C.,  et~al., 2018, \mndoi [\pasa] {10.1017/pasa.2018.18}, \href
  {http://adsabs.harvard.edu/abs/2018PASA...35...29B} {35, e029}

\bibitem[\protect\citeauthoryear{{Brand} \& {Blitz}}{{Brand} \&
  {Blitz}}{1993}]{Blitz}
{Brand} J.,  {Blitz} L.,  1993, \aap, \href
  {http://adsabs.harvard.edu/abs/1993A%26A...275...67B} {275, 67}

\bibitem[\protect\citeauthoryear{{Brogan}, {Devine}, {Lazio}, {Kassim}, {Tam},
  {Brisken}, {Dyer}  \& {Roberts}}{{Brogan} et~al.}{2004}]{J1809SNRBrogan2004}
{Brogan} C.~L.,  {Devine} K.~E.,  {Lazio} T.~J.,  {Kassim} N.~E.,  {Tam} C.~R.,
   {Brisken} W.~F.,  {Dyer} K.~K.,   {Roberts} M.~S.~E.,  2004, \mndoi [\aj]
  {10.1086/379856}, \href {http://adsabs.harvard.edu/abs/2004AJ....127..355B}
  {127, 355}

\bibitem[\protect\citeauthoryear{{Bucciantini}, {Arons}  \&
  {Amato}}{{Bucciantini} et~al.}{2011}]{BucciantiniPWN2011}
{Bucciantini} N.,  {Arons} J.,   {Amato} E.,  2011, \mndoi [\mnras]
  {10.1111/j.1365-2966.2010.17449.x}, \href
  {http://adsabs.harvard.edu/abs/2011MNRAS.410..381B} {410, 381}

\bibitem[\protect\citeauthoryear{{Camilo}, {Kaspi}, {Lyne}, {Manchester},
  {Bell}, {D'Amico}, {McKay}  \& {Crawford}}{{Camilo}
  et~al.}{2000}]{CamiloJ1119}
{Camilo} F.,  {Kaspi} V.~M.,  {Lyne} A.~G.,  {Manchester} R.~N.,  {Bell} J.~F.,
   {D'Amico} N.,  {McKay} N.~P.~F.,   {Crawford} F.,  2000, \mndoi [\apj]
  {10.1086/309435}, \href {http://adsabs.harvard.edu/abs/2000ApJ...541..367C}
  {541, 367}

\bibitem[\protect\citeauthoryear{{Camilo} et~al.,}{{Camilo}
  et~al.}{2001}]{PSRJ1016Camilo}
{Camilo} F.,  et~al., 2001, \mndoi [\apjl] {10.1086/323171}, \href
  {http://adsabs.harvard.edu/abs/2001ApJ...557L..51C} {557, L51}

\bibitem[\protect\citeauthoryear{{Castelletti}, {Giacani}  \&
  {Petriella}}{{Castelletti} et~al.}{2016}]{J1809radio2016}
{Castelletti} G.,  {Giacani} E.,   {Petriella} A.,  2016, \mndoi [\aap]
  {10.1051/0004-6361/201527578}, \href
  {http://adsabs.harvard.edu/abs/2016A%26A...587A..71C} {587, A71}

\bibitem[\protect\citeauthoryear{{Caswell}, {McClure-Griffiths}  \&
  {Cheung}}{{Caswell} et~al.}{2004}]{Caswell2004}
{Caswell} J.~L.,  {McClure-Griffiths} N.~M.,   {Cheung} M.~C.~M.,  2004, \mndoi
  [\mnras] {10.1111/j.1365-2966.2004.08030.x}, \href
  {http://adsabs.harvard.edu/abs/2004MNRAS.352.1405C} {352, 1405}

\bibitem[\protect\citeauthoryear{{Cioffi}, {McKee}  \& {Bertschinger}}{{Cioffi}
  et~al.}{1988}]{CioffiSNR}
{Cioffi} D.~F.,  {McKee} C.~F.,   {Bertschinger} E.,  1988, \mndoi [\apj]
  {10.1086/166834}, \href {http://adsabs.harvard.edu/abs/1988ApJ...334..252C}
  {334, 252}

\bibitem[\protect\citeauthoryear{{Cordes}, {Lazio}, {Chatterjee}, {Arzoumanian}
   \& {Chernoff}}{{Cordes} et~al.}{2002}]{Cordes}
{Cordes} J.,  {Lazio} T.,  {Chatterjee} S.,  {Arzoumanian} Z.,   {Chernoff} D.,
   2002, in 34th COSPAR Scientific Assembly.

\bibitem[\protect\citeauthoryear{{Crawford}, {Gaensler}, {Kaspi}, {Manchester},
  {Camilo}, {Lyne}  \& {Pivovaroff}}{{Crawford} et~al.}{2001}]{Crawford2001}
{Crawford} F.,  {Gaensler} B.~M.,  {Kaspi} V.~M.,  {Manchester} R.~N.,
  {Camilo} F.,  {Lyne} A.~G.,   {Pivovaroff} M.~J.,  2001, \mndoi [\apj]
  {10.1086/321328}, \href {http://adsabs.harvard.edu/abs/2001ApJ...554..152C}
  {554, 152}

\bibitem[\protect\citeauthoryear{{Crutcher}, {Wandelt}, {Heiles}, {Falgarone}
  \& {Troland}}{{Crutcher} et~al.}{2010}]{Crutcher2010}
{Crutcher} R.~M.,  {Wandelt} B.,  {Heiles} C.,  {Falgarone} E.,   {Troland}
  T.~H.,  2010, \mndoi [\apj] {10.1088/0004-637X/725/1/466}, \href
  {http://adsabs.harvard.edu/abs/2010ApJ...725..466C} {725, 466}

\bibitem[\protect\citeauthoryear{{Dame}}{{Dame}}{2007}]{J1026Dame}
{Dame} T.~M.,  2007, \mndoi [\apjl] {10.1086/521363}, \href
  {http://adsabs.harvard.edu/abs/2007ApJ...665L.163D} {665, L163}

\bibitem[\protect\citeauthoryear{{Dickey} \& {Lockman}}{{Dickey} \&
  {Lockman}}{1990}]{Dickey1990}
{Dickey} J.~M.,  {Lockman} F.~J.,  1990, \mndoi [\araa]
  {10.1146/annurev.aa.28.090190.001243}, \href
  {http://adsabs.harvard.edu/abs/1990ARA%26A..28..215D} {28, 215}

\bibitem[\protect\citeauthoryear{Djannati-Ata\"i}{Djannati-Ata\"i}{2009}]{Arache2009}
Djannati-Ata\"i A.,  2009, HESS discovery of VHE gamma-ray emission from a
  remarkable young composite SNR

\bibitem[\protect\citeauthoryear{{Fink}}{{Fink}}{1981}]{Fink1981}
{Fink} R.~W.,  1981.
 Vol. 3, Cleveland : CRC press, Boca Raton, Fl

\bibitem[\protect\citeauthoryear{{Fukui} et~al.,}{{Fukui}
  et~al.}{2009}]{FukuiJ1023}
{Fukui} Y.,  et~al., 2009, \mndoi [\pasj] {10.1093/pasj/61.4.L23}, \href
  {http://adsabs.harvard.edu/abs/2009PASJ...61L..23F} {61, L23}

\bibitem[\protect\citeauthoryear{{Fukui}, {Torii}, {Onishi}, {Yamamoto},
  {Okamoto}, {Hayakawa}, {Tachihara}  \& {Sano}}{{Fukui}
  et~al.}{2015}]{FukuiHIgas2015}
{Fukui} Y.,  {Torii} K.,  {Onishi} T.,  {Yamamoto} H.,  {Okamoto} R.,
  {Hayakawa} T.,  {Tachihara} K.,   {Sano} H.,  2015, \mndoi [\apj]
  {10.1088/0004-637X/798/1/6}, \href
  {http://adsabs.harvard.edu/abs/2015ApJ...798....6F} {798, 6}

\bibitem[\protect\citeauthoryear{{Furukawa}, {Dawson}, {Ohama}, {Kawamura},
  {Mizuno}, {Onishi}  \& {Fukui}}{{Furukawa} et~al.}{2009}]{J1026CCC}
{Furukawa} N.,  {Dawson} J.~R.,  {Ohama} A.,  {Kawamura} A.,  {Mizuno} N.,
  {Onishi} T.,   {Fukui} Y.,  2009, \mndoi [\apjl]
  {10.1088/0004-637X/696/2/L115}, \href
  {http://adsabs.harvard.edu/abs/2009ApJ...696L.115F} {696, L115}

\bibitem[\protect\citeauthoryear{{Furukawa} et~al.,}{{Furukawa}
  et~al.}{2014}]{FurukawaJ1023}
{Furukawa} N.,  et~al., 2014, \mndoi [\apj] {10.1088/0004-637X/781/2/70}, \href
  {http://adsabs.harvard.edu/abs/2014ApJ...781...70F} {781, 70}

\bibitem[\protect\citeauthoryear{{Gabici}, {Aharonian}  \& {Blasi}}{{Gabici}
  et~al.}{2007}]{Gabicidiff2006}
{Gabici} S.,  {Aharonian} F.~A.,   {Blasi} P.,  2007, \mndoi [\apss]
  {10.1007/s10509-007-9427-6}, \href
  {http://adsabs.harvard.edu/abs/2007Ap%26SS.309..365G} {309, 365}

\bibitem[\protect\citeauthoryear{{Gabici}, {Casanova}, {Aharonian}  \&
  {Rowell}}{{Gabici} et~al.}{2010}]{Gabici2010W28}
{Gabici} S.,  {Casanova} S.,  {Aharonian} F.~A.,   {Rowell} G.,  2010, in
  {Boissier} S.,  {Heydari-Malayeri} M.,  {Samadi} R.,   {Valls-Gabaud} D.,
  eds, SF2A-2010: Proceedings of the Annual meeting of the French Society of
  Astronomy and Astrophysics. p.~313 (\mn@eprint {arXiv} {1009.5291})

\bibitem[\protect\citeauthoryear{{Gallant} \& {Arons}}{{Gallant} \&
  {Arons}}{1994}]{gallant}
{Gallant} Y.~A.,  {Arons} J.,  1994, \mndoi [\apj] {10.1086/174810}, \href
  {http://adsabs.harvard.edu/abs/1994ApJ...435..230G} {435, 230}

\bibitem[\protect\citeauthoryear{{Ginzburg} \& {Syrovatskii}}{{Ginzburg} \&
  {Syrovatskii}}{1964}]{ginzburg}
{Ginzburg} V.~L.,  {Syrovatskii} S.~I.,  1964, {The Origin of Cosmic Rays}

\bibitem[\protect\citeauthoryear{{Giuliani} et~al.,}{{Giuliani}
  et~al.}{2010}]{GiulianiW282010}
{Giuliani} A.,  et~al., 2010, \mndoi [\aap] {10.1051/0004-6361/201014256},
  \href {http://adsabs.harvard.edu/abs/2010A%26A...516L..11G} {516, L11}

\bibitem[\protect\citeauthoryear{{Gonzalez} \& {Safi-Harb}}{{Gonzalez} \&
  {Safi-Harb}}{2005}]{J1119SafiGonzalez2005}
{Gonzalez} M.,  {Safi-Harb} S.,  2005, \mndoi [\apj] {10.1086/426576}, \href
  {http://adsabs.harvard.edu/abs/2005ApJ...619..856G} {619, 856}

\bibitem[\protect\citeauthoryear{{Gotthelf}, {Halpern}, {Buxton}  \&
  {Bailyn}}{{Gotthelf} et~al.}{2004}]{AXPJ1809}
{Gotthelf} E.~V.,  {Halpern} J.~P.,  {Buxton} M.,   {Bailyn} C.,  2004, \mndoi
  [\apj] {10.1086/382232}, \href
  {http://adsabs.harvard.edu/abs/2004ApJ...605..368G} {605, 368}

\bibitem[\protect\citeauthoryear{{Gusdorf}, {Cabrit}, {Flower}  \& {Pineau Des
  For{\^e}ts}}{{Gusdorf} et~al.}{2008}]{Gusdorf}
{Gusdorf} A.,  {Cabrit} S.,  {Flower} D.~R.,   {Pineau Des For{\^e}ts} G.,
  2008, \mndoi [\aap] {10.1051/0004-6361:20078900}, \href
  {http://adsabs.harvard.edu/abs/2008A%26A...482..809G} {482, 809}

\bibitem[\protect\citeauthoryear{{Hawkes} et~al.,}{{Hawkes}
  et~al.}{2014}]{Hawkes2014}
{Hawkes} J.,  et~al., 2014, \mndoi [IJMPS] {10.1142/S2010194514601987}, \href
  {http://adsabs.harvard.edu/abs/2014IJMPS..2860198H} {28, 1460198}

\bibitem[\protect\citeauthoryear{{Irvine}, {Goldsmith}  \&
  {Hjalmarson}}{{Irvine} et~al.}{1987}]{Irvine1987}
{Irvine} W.~M.,  {Goldsmith} P.~F.,   {Hjalmarson} A.,  1987, in {Hollenbach}
  D.~J.,  {Thronson} Jr. H.~A.,  eds,  Astrophysics and Space Science Library
  Vol. 134, Interstellar Processes. pp 561--609

\bibitem[\protect\citeauthoryear{{Kishishita}, {Bamba}, {Uchiyama}, {Tanaka}
  \& {Takahashi}}{{Kishishita} et~al.}{2012}]{J1418Suzaku}
{Kishishita} T.,  {Bamba} A.,  {Uchiyama} Y.,  {Tanaka} Y.,   {Takahashi} T.,
  2012, \mndoi [\apj] {10.1088/0004-637X/750/2/162}, \href
  {http://adsabs.harvard.edu/abs/2012ApJ...750..162K} {750, 162}

\bibitem[\protect\citeauthoryear{{Kumar}, {Safi-Harb}  \& {Gonzalez}}{{Kumar}
  et~al.}{2012}]{KumarJ1119}
{Kumar} H.~S.,  {Safi-Harb} S.,   {Gonzalez} M.~E.,  2012, \mndoi [\apj]
  {10.1088/0004-637X/754/2/96}, \href
  {http://adsabs.harvard.edu/abs/2012ApJ...754...96K} {754, 96}

\bibitem[\protect\citeauthoryear{{Li} \& {Chen}}{{Li} \&
  {Chen}}{2010}]{LiChenW28Diffusion}
{Li} H.,  {Chen} Y.,  2010, \mndoi [\mnras] {10.1111/j.1745-3933.2010.00944.x},
  \href {http://adsabs.harvard.edu/abs/2010MNRAS.409L..35L} {409, L35}

\bibitem[\protect\citeauthoryear{{Malkov}, {Diamond}, {Sagdeev}, {Aharonian}
  \& {Moskalenko}}{{Malkov} et~al.}{2013}]{Malkov2013}
{Malkov} M.~A.,  {Diamond} P.~H.,  {Sagdeev} R.~Z.,  {Aharonian} F.~A.,
  {Moskalenko} I.~V.,  2013, \mndoi [\apj] {10.1088/0004-637X/768/1/73}, \href
  {http://adsabs.harvard.edu/abs/2013ApJ...768...73M} {768, 73}

\bibitem[\protect\citeauthoryear{{McClure-Griffiths}, {Dickey}, {Gaensler},
  {Green}, {Haverkorn}  \& {Strasser}}{{McClure-Griffiths}
  et~al.}{2005}]{SGPS2005}
{McClure-Griffiths} N.~M.,  {Dickey} J.~M.,  {Gaensler} B.~M.,  {Green} A.~J.,
  {Haverkorn} M.,   {Strasser} S.,  2005, \mndoi [\apjs] {10.1086/430114},
  \href {http://adsabs.harvard.edu/abs/2005ApJS..158..178M} {158, 178}

\bibitem[\protect\citeauthoryear{{McClure-Griffiths}
  et~al.,}{{McClure-Griffiths} et~al.}{2009}]{GASS}
{McClure-Griffiths} N.~M.,  et~al., 2009, \mndoi [\apjs]
  {10.1088/0067-0049/181/2/398}, \href
  {http://adsabs.harvard.edu/abs/2009ApJS..181..398M} {181, 398}

\bibitem[\protect\citeauthoryear{{Mizuno} \& {Fukui}}{{Mizuno} \&
  {Fukui}}{2004}]{Fukui}
{Mizuno} A.,  {Fukui} Y.,  2004, in {Clemens} D.,  {Shah} R.,   {Brainerd} T.,
  eds,  Astronomical Society of the Pacific Conference Series Vol. 317, Milky
  Way Surveys: The Structure and Evolution of our Galaxy. p.~59

\bibitem[\protect\citeauthoryear{{Moriguchi}, {Yamaguchi}, {Onishi}, {Mizuno}
  \& {Fukui}}{{Moriguchi} et~al.}{2001}]{VelaCO}
{Moriguchi} Y.,  {Yamaguchi} N.,  {Onishi} T.,  {Mizuno} A.,   {Fukui} Y.,
  2001, \mndoi [\pasj] {10.1093/pasj/53.6.1025}, \href
  {http://adsabs.harvard.edu/abs/2001PASJ...53.1025M} {53, 1025}

\bibitem[\protect\citeauthoryear{{Nava}, {Gabici}, {Marcowith}, {Morlino}  \&
  {Ptuskin}}{{Nava} et~al.}{2016}]{Navadiffusion2016}
{Nava} L.,  {Gabici} S.,  {Marcowith} A.,  {Morlino} G.,   {Ptuskin} V.~S.,
  2016, \mndoi [\mnras] {10.1093/mnras/stw1592}, \href
  {http://adsabs.harvard.edu/abs/2016MNRAS.461.3552N} {461, 3552}

\bibitem[\protect\citeauthoryear{{Ng}, {Roberts}  \& {Romani}}{{Ng}
  et~al.}{2005}]{NgRobertsJ1418Chandra}
{Ng} C.-Y.,  {Roberts} M.~S.~E.,   {Romani} R.~W.,  2005, \mndoi [\apj]
  {10.1086/430632}, \href {http://adsabs.harvard.edu/abs/2005ApJ...627..904N}
  {627, 904}

\bibitem[\protect\citeauthoryear{{Ng}, {Kaspi}, {Ho}, {Weltevrede}, {Bogdanov},
  {Shannon}  \& {Gonzalez}}{{Ng} et~al.}{2012}]{J1119XMM}
{Ng} C.-Y.,  {Kaspi} V.~M.,  {Ho} W.~C.~G.,  {Weltevrede} P.,  {Bogdanov} S.,
  {Shannon} R.,   {Gonzalez} M.~E.,  2012, \mndoi [\apj]
  {10.1088/0004-637X/761/1/65}, \href
  {http://adsabs.harvard.edu/abs/2012ApJ...761...65N} {761, 65}

\bibitem[\protect\citeauthoryear{{Nicholas}, {Rowell}, {Burton}, {Walsh},
  {Fukui}, {Kawamura}  \& {Maxted}}{{Nicholas} et~al.}{2012}]{Nicholas7mm}
{Nicholas} B.~P.,  {Rowell} G.,  {Burton} M.~G.,  {Walsh} A.~J.,  {Fukui} Y.,
  {Kawamura} A.,   {Maxted} N.~I.,  2012, \mndoi [\mnras]
  {10.1111/j.1365-2966.2011.19688.x}, \href
  {http://adsabs.harvard.edu/abs/2012MNRAS.419..251N} {419, 251}

\bibitem[\protect\citeauthoryear{{Parsons}, {Thompson}  \&
  {Chrysostomou}}{{Parsons} et~al.}{2009}]{Parsons2009}
{Parsons} H.,  {Thompson} M.~A.,   {Chrysostomou} A.,  2009, \mndoi [\mnras]
  {10.1111/j.1365-2966.2009.15375.x}, \href
  {http://adsabs.harvard.edu/abs/2009MNRAS.399.1506P} {399, 1506}

\bibitem[\protect\citeauthoryear{{Pivovaroff}, {Kaspi}, {Camilo}, {Gaensler}
  \& {Crawford}}{{Pivovaroff} et~al.}{2001}]{J1119Pivovaroff2001}
{Pivovaroff} M.~J.,  {Kaspi} V.~M.,  {Camilo} F.,  {Gaensler} B.~M.,
  {Crawford} F.,  2001, \mndoi [\apj] {10.1086/321340}, \href
  {http://adsabs.harvard.edu/abs/2001ApJ...554..161P} {554, 161}

\bibitem[\protect\citeauthoryear{{Ray} et~al.,}{{Ray} et~al.}{2011}]{Ray1}
{Ray} P.~S.,  et~al., 2011, \mndoi [\apjs] {10.1088/0067-0049/194/2/17}, \href
  {http://adsabs.harvard.edu/abs/2011ApJS..194...17R} {194, 17}

\bibitem[\protect\citeauthoryear{{Reynolds}}{{Reynolds}}{2008}]{ReynoldsSNR}
{Reynolds} S.~P.,  2008, \mndoi [\araa]
  {10.1146/annurev.astro.46.060407.145237}, \href
  {http://adsabs.harvard.edu/abs/2008ARA%26A..46...89R} {46, 89}

\bibitem[\protect\citeauthoryear{{Roberts} \& {Romani}}{{Roberts} \&
  {Romani}}{1998}]{RobertsJ1418}
{Roberts} M.~S.~E.,  {Romani} R.~W.,  1998, \mndoi [\apj] {10.1086/305426},
  \href {http://adsabs.harvard.edu/abs/1998ApJ...496..827R} {496, 827}

\bibitem[\protect\citeauthoryear{{Roberts}, {Brogan}, {Gaensler}, {Hessels},
  {Ng}  \& {Romani}}{{Roberts} et~al.}{2005}]{RobertsRPWNJ1418}
{Roberts} M.~S.~E.,  {Brogan} C.~L.,  {Gaensler} B.~M.,  {Hessels} J.~W.~T.,
  {Ng} C.-Y.,   {Romani} R.~W.,  2005, \mndoi [\apss]
  {10.1007/s10509-005-7579-9}, \href
  {http://adsabs.harvard.edu/abs/2005Ap%26SS.297...93R} {297, 93}

\bibitem[\protect\citeauthoryear{{Ruiz} \& {May}}{{Ruiz} \&
  {May}}{1986}]{RuizMay1986}
{Ruiz} M.~T.,  {May} J.,  1986, \mndoi [\apj] {10.1086/164634}, \href
  {http://adsabs.harvard.edu/abs/1986ApJ...309..667R} {309, 667}

\bibitem[\protect\citeauthoryear{{Safi-Harb} \& {Kumar}}{{Safi-Harb} \&
  {Kumar}}{2008}]{J1119Safi2008}
{Safi-Harb} S.,  {Kumar} H.~S.,  2008, \mndoi [\apj] {10.1086/590359}, \href
  {http://adsabs.harvard.edu/abs/2008ApJ...684..532S} {684, 532}

\bibitem[\protect\citeauthoryear{{Sano} et~al.,}{{Sano}
  et~al.}{2013}]{SanoBamplification2013}
{Sano} H.,  et~al., 2013, \mndoi [\apj] {10.1088/0004-637X/778/1/59}, \href
  {http://adsabs.harvard.edu/abs/2013ApJ...778...59S} {778, 59}

\bibitem[\protect\citeauthoryear{{Schilke}, {Walmsley}, {Pineau des Forets}  \&
  {Flower}}{{Schilke} et~al.}{1997}]{Schilke}
{Schilke} P.,  {Walmsley} C.~M.,  {Pineau des Forets} G.,   {Flower} D.~R.,
  1997, \aap, \href {http://adsabs.harvard.edu/abs/1997A%26A...321..293S} {321,
  293}

\bibitem[\protect\citeauthoryear{{Slane} et~al.,}{{Slane}
  et~al.}{2018}]{Velaslane}
{Slane} P.,  et~al., 2018, \mndoi [\apj] {10.3847/1538-4357/aada12}, \href
  {http://adsabs.harvard.edu/abs/2018ApJ...865...86S} {865, 86}

\bibitem[\protect\citeauthoryear{{Sushch}, {Oya}, {Schwanke}, {Johnston}  \&
  {Dalton}}{{Sushch} et~al.}{2017}]{HESSJ1303SNR2017}
{Sushch} I.,  {Oya} I.,  {Schwanke} U.,  {Johnston} S.,   {Dalton} M.~L.,
  2017, \mndoi [\aap] {10.1051/0004-6361/201527871}, \href
  {http://adsabs.harvard.edu/abs/2017A%26A...605A.115S} {605, A115}

\bibitem[\protect\citeauthoryear{{Taylor} \& {Cordes}}{{Taylor} \&
  {Cordes}}{1993}]{Taylorcordes1993}
{Taylor} J.~H.,  {Cordes} J.~M.,  1993, \mndoi [\apj] {10.1086/172870}, \href
  {http://adsabs.harvard.edu/abs/1993ApJ...411..674T} {411, 674}

\bibitem[\protect\citeauthoryear{{Torii}, {Tsunemi}, {Dotani}, {Mitsuda},
  {Kawai}, {Kinugasa}, {Saito}  \& {Shibata}}{{Torii} et~al.}{1999}]{Torii1999}
{Torii} K.,  {Tsunemi} H.,  {Dotani} T.,  {Mitsuda} K.,  {Kawai} N.,
  {Kinugasa} K.,  {Saito} Y.,   {Shibata} S.,  1999, \mndoi [\apjl]
  {10.1086/312251}, \href {http://adsabs.harvard.edu/abs/1999ApJ...523L..69T}
  {523, L69}

\bibitem[\protect\citeauthoryear{{Urquhart} et~al.,}{{Urquhart}
  et~al.}{2010}]{Urquhart2010}
{Urquhart} J.~S.,  et~al., 2010, \mndoi [\pasa] {10.1071/AS10002}, \href
  {http://adsabs.harvard.edu/abs/2010PASA...27..321U} {27, 321}

\bibitem[\protect\citeauthoryear{{Vall{\'e}e}}{{Vall{\'e}e}}{2013}]{Vallee2013}
{Vall{\'e}e} J.~P.,  2013, \mndoi [International Journal of Astronomy and
  Astrophysics] {10.4236/ijaa.2013.31003}, \href
  {http://adsabs.harvard.edu/abs/2013IJAA....3...20V} {3, 20}

\bibitem[\protect\citeauthoryear{{Voisin}, {Rowell}, {Burton}, {Walsh}, {Fukui}
   \& {Aharonian}}{{Voisin} et~al.}{2016}]{voisin2016a}
{Voisin} F.,  {Rowell} G.,  {Burton} M.~G.,  {Walsh} A.,  {Fukui} Y.,
  {Aharonian} F.,  2016, \mndoi [\mnras] {10.1093/mnras/stw473}, \href
  {http://adsabs.harvard.edu/abs/2016MNRAS.458.2813V} {458, 2813}

\bibitem[\protect\citeauthoryear{{Voronkov}, {Caswell}, {Ellingsen}, {Green}
  \& {Breen}}{{Voronkov} et~al.}{2014}]{Voronkov}
{Voronkov} M.~A.,  {Caswell} J.~L.,  {Ellingsen} S.~P.,  {Green} J.~A.,
  {Breen} S.~L.,  2014, \mndoi [\mnras] {10.1093/mnras/stu116}, \href
  {http://adsabs.harvard.edu/abs/2014MNRAS.439.2584V} {439, 2584}

\bibitem[\protect\citeauthoryear{{Wang}}{{Wang}}{2011}]{PSRJ1826distance}
{Wang} W.,  2011, \mndoi [Research in Astronomy and Astrophysics]
  {10.1088/1674-4527/11/7/007}, \href
  {http://adsabs.harvard.edu/abs/2011RAA....11..824W} {11, 824}

\bibitem[\protect\citeauthoryear{{Weaver}, {McCray}, {Castor}, {Shapiro}  \&
  {Moore}}{{Weaver} et~al.}{1977}]{Weaver1977}
{Weaver} R.,  {McCray} R.,  {Castor} J.,  {Shapiro} P.,   {Moore} R.,  1977,
  \mndoi [\apj] {10.1086/155692}, \href
  {http://adsabs.harvard.edu/abs/1977ApJ...218..377W} {218, 377}

\bibitem[\protect\citeauthoryear{{Weltevrede}, {Johnston}  \&
  {Espinoza}}{{Weltevrede} et~al.}{2011}]{WelJ1119}
{Weltevrede} P.,  {Johnston} S.,   {Espinoza} C.~M.,  2011, \mndoi [\mnras]
  {10.1111/j.1365-2966.2010.17821.x}, \href
  {http://adsabs.harvard.edu/abs/2011MNRAS.411.1917W} {411, 1917}

\bibitem[\protect\citeauthoryear{{Yang}, {de O{\~n}a Wilhelmi}  \&
  {Aharonian}}{{Yang} et~al.}{2017}]{YangWesterlund2_2017}
{Yang} R.-z.,  {de O{\~n}a Wilhelmi} E.,   {Aharonian} F.,  2017, preprint,
  \href {http://adsabs.harvard.edu/abs/2017arXiv171002803Y} {} (\mn@eprint
  {arXiv} {1710.02803})

\bibitem[\protect\citeauthoryear{{Zinchenko}, {Forsstroem}, {Lapinov}  \&
  {Mattila}}{{Zinchenko} et~al.}{1994}]{Zitchenko1994}
{Zinchenko} I.,  {Forsstroem} V.,  {Lapinov} A.,   {Mattila} K.,  1994, \aap,
  \href {http://adsabs.harvard.edu/abs/1994A%26A...288..601Z} {288, 601}

\makeatother
\end{thebibliography}
\bibliographystyle{pasa-mnras}

\appendix
\renewcommand{\theequation}{\Alph{section}.\arabic{equation}}
\renewcommand{\thefigure}{\Alph{section}.\arabic{figure}}
\renewcommand{\thetable}{\Alph{section}.\arabic{table}}
\section[]{Cleaning and integrated intensity mapping methods}
\label{sec:integratedmap}
\setcounter{figure}{0}
A flow chart which outlines the various processes used to clean the various data-cubes and produce improved integrated intensity maps is shown in Figure\,\ref{flowchartfigure}.
The procedure to produce clean data-cubes is explained in Section\,\ref{sec:ISManalysis}.
\begin{figure}
 \includegraphics[width=0.5\textwidth]{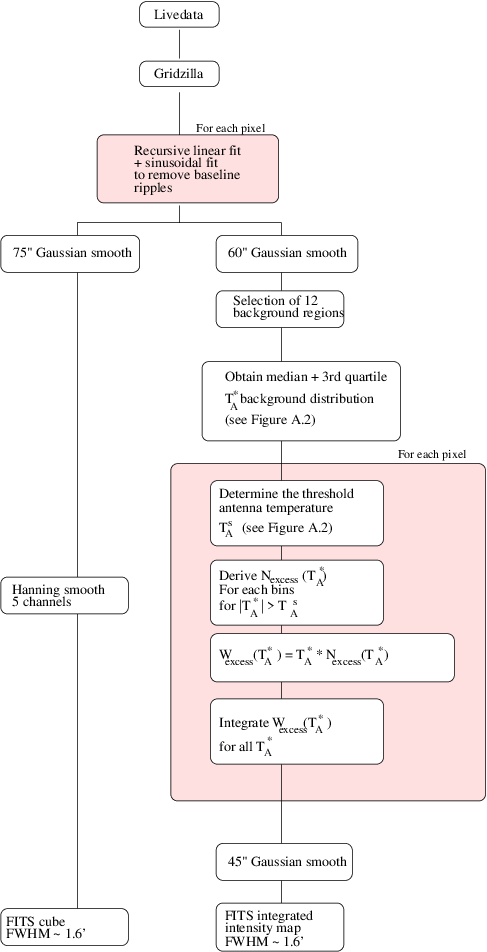}
 \caption{Flow chart describing the procedure used to reduce our Mopra 7mm observation data.
 The left path indicates the procedure used to create cleaned data-cubes while the right paths shows the procedure to create integrated intensity maps.}
\label{flowchartfigure}
\end{figure}

\subsection{Integrated intensity maps}

\begin{figure}
 \includegraphics[height=0.5\textwidth, angle=270]{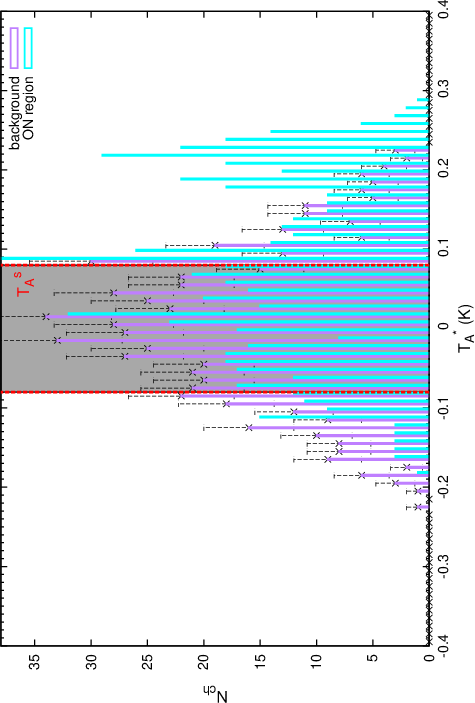}
 \caption{Binned distribution of the number of channels $N_\text{ch}\left(T_\text{A}^{*}\right)$ as a function of the antenna temperature $T_\text{A}^{*}$.
  The purple histogram and error-bars represent the median and third-quartile background distribution respectively while the cyan histogram represents the distribution of the `ON' region.
 The red dashed vertical lines highlight the threshold antenna temperature $T_\text{A}^{s}$.
 Finally, the channels within the grey shaded area are rejected in our analysis.
  }
 \label{binneddistributionfigure}
\end{figure}

From the unsmoothed cleaned data-cubes, we first apply a Gaussian smooth with FWHM$\sim1^{\prime}$ using the {\tt Miriad} task \textit{Convol}.
Then, from our cleaned data-cubes, we choose twelve background regions, each consisting of nine pixels, across the map, with no CS(1--0) emission.
From the spectra obtained from these background regions,  a binned distribution of the number of channels $N_\text{ch}\left(T_\text{A}^{*}\right)$ within the antenna temperature $T_\text{A}^{*}$ and $T_\text{A}^{*}+\Delta T_\text{A}^{*}$ is then 
plotted ($\Delta T_\text{A}^{*}=0.01$\,K in this work).
We then record the median $N_\text{ch}$ value for each bin to obtain our background distribution (see purple histogram in Figure\,\ref{binneddistributionfigure}).
Note that we also record the third quartile of the background distribution (see error bars in Figure\,\ref{binneddistributionfigure}) to look at the fluctuation of the background distribution.

Now that we have estimated the background level distribution at the studied velocity range, the integrated intensity  of each pixel is derived using the following method:
\begin{enumerate}
 \item Obtain the binned distribution of the region, consisting of 9 pixels, centred on the current pixel (see blue histogram in Figure\,\ref{binneddistributionfigure}).
 \item Determine the threshold antenna temperature $T_\text{A}^{s}$ so that channels with antenna temperature $|T_\text{A}^{*}|<T_\text{A}^{s}$ are automatically discarded (see grey shaded area in Figure\,\ref{binneddistributionfigure}).
  We here define $T_\text{A}^{s}$ as the lowest antenna temperature of the 'ON' whose $N_{\text{ch}}\left(T_\text{A}^{*}\right)$  and $N_{\text{ch}}\left(T_\text{A}^{*}+\Delta T_\text{A}^{*} \right)$ exceed the third quartile of the background distribution 
  (to account for fluctuation of the background distribution). It should be noted that the antenna temperature $T_\text{A}^{*}<-T_\text{A}^{s}$ are not discarded in order to mitigate 
  potential excess from a region with increased $T_\text{rms}$.
  \item For each bin, we subtract  the background distribution from $N_\text{ch}^{\text{ON}}$ to obtain the excess number of channels $N_\text{excess}\left(T_\text{A}^{*}\right)$.
  If the value is negative, then we define $N_\text{excess}\left(T_\text{A}^{*}\right)=0$.
 \item We multiply $N_\text{excess}\left(T_\text{A}^{*}\right)$ with $T_\text{A}^{*}$ and the velocity channels spacing ($\Delta v_\text{lsr}\sim0.2$\,km/s for 7mm Mopra observation) to 
 obtain the excess integrated intensity $W_\text{excess}\left(T_\text{A}^{*}\right)$ at each bin.
 \item We finally derive the integrated intensity at a given pixel, by first summing $W_\text{excess}\left(T_\text{A}^{*}\right)$ and then dividing by the number of pixels.
\end{enumerate}
To account for possible spatial fluctuations, we then smooth our integrated intensity map with a Gaussian with FWHM$\sim45^{\prime\prime}$, resulting in a total smoothing of $1.25^{\prime}$ 
(as per the data cubes).

Now that the integrated intensity maps are made, we need to establish the background level for each maps.
Indeed, our new integrated intensity maps are not devoid of noise despite our aforementioned cuts.
To do so, we simulate data consisting of noise with the same $T_\text{rms}$ as per our data-cubes.
Using the same aforementioned steps, we produce integrated intensity maps within the same velocity ranges.
We then define our noise level for our integrated intensity maps as the 99\% distribution value of the `noise maps' distribution.
\subsection[]{comparison of methods}
\begin{figure*}
 \includegraphics[width=\textwidth]{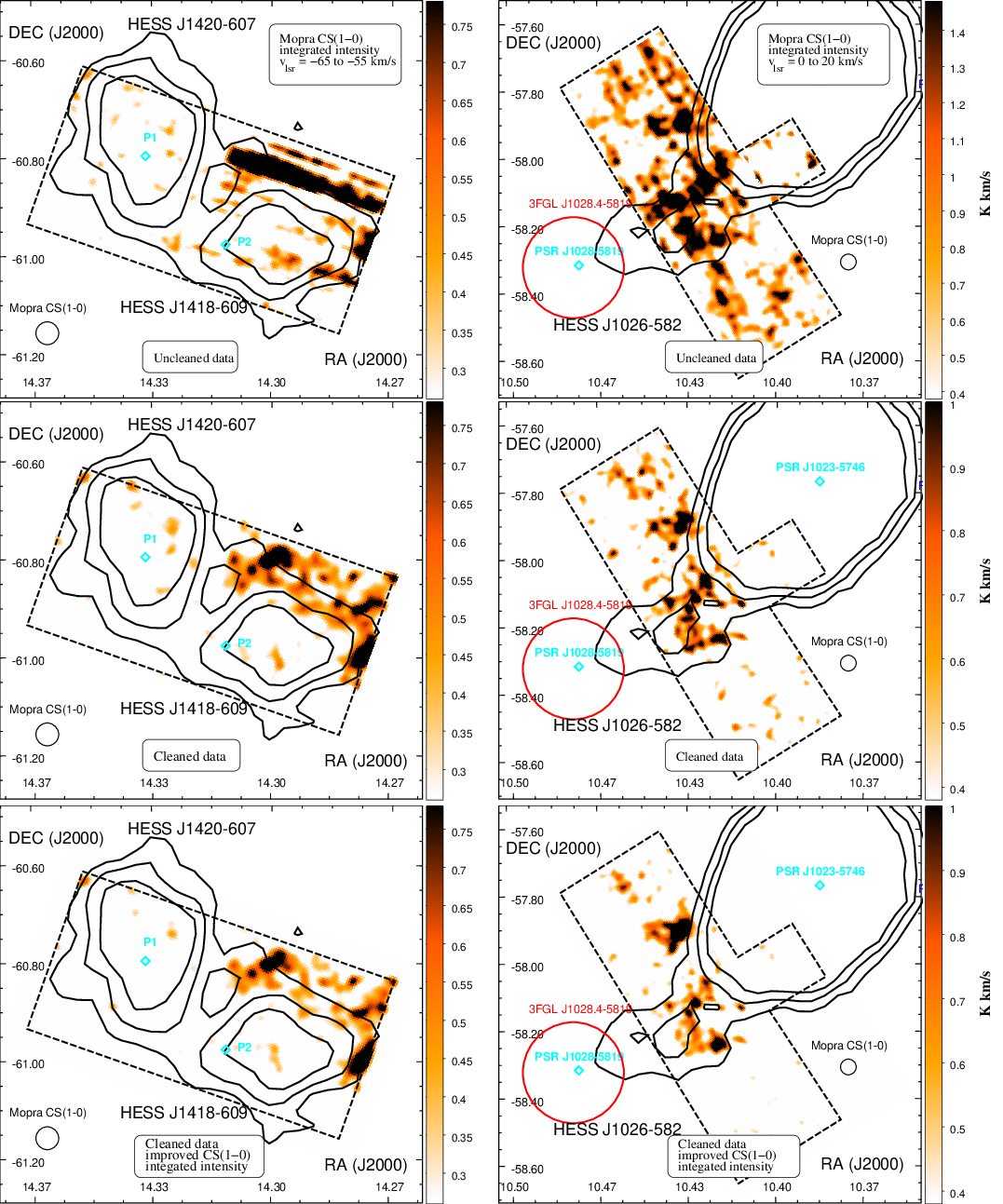}
 \caption{Mopra CS(1--0) integrated intensity maps towards HESS\,J1418$-$609 and HESS\,J1420$-$607 (left panels) and HESS\,J1026$-$582 (right panels).
 Top panels illustrate the standard integrated intensity using uncleaned data-cubes. The middle panels highlight the CS(1--0) integrated intensity after removing baseline ripples (see Figure\,\ref{flowchartfigure}).
 Finally, the bottom panels shows the  final product from our method described in Appendix\,\ref{sec:integratedmap}.}
 \label{rawvscleanedmap}
\end{figure*}

Figure\,\ref{rawvscleanedmap} compares the  CS(1--0) integrated intensity maps using uncleaned data-cubes with the standard intensity maps after removing baseline ripples, and with the method used in this work, 
to obtain integrated intensity maps. 
We remark the striking difference between the uncleaned and cleaned data.
We finally notice the removal of additional noise using our method.
\section[]{Galactic model}
\setcounter{figure}{0}
We used the Galactic model from \citet{Vallee2013} (see Figure\,\ref{galactic_model}) to help identify the Galactic arm in which each molecular cloud is located.
\label{sec:galacticmodel}
\begin{figure}
 \includegraphics[width=0.5\textwidth]{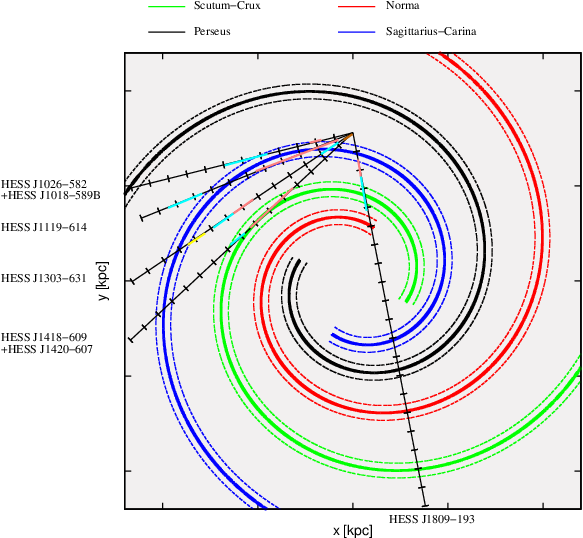}
 \caption{Diagram indicating the position of the spiral arm in our Galaxy based on \citet{Vallee2013} model.
 The various lines with ticks represent the direction of the various TeV sources studied in this paper. 
 For each TeV source, the brown, cyan, yellow, and brown regions indicates the distances of velocity ranges assumed for each source (see Figures\,\ref{J1809spectra}, \ref{J1026COCS}, \ref{J1119COCS}, \ref{J1420spec}, \ref{J1303spec}, \ref{J1018CO})}
 \label{galactic_model}
\end{figure}

 \section[]{Fit parameters}
 \label{sec:Fit parameter}
 Tables\,\ref{J1809table} to \ref{tableJ1018} show the (RA, Dec) position, the semi-major axes and the Gaussian fitting parameters of the CS(1--0)
, C$^{34}$S(1--0), HC$_3$N(5--4,F=4--3), SiO(1--0, v=0) and CO(1--0) detections found towards the studied TeV sources. 
 \setcounter{figure}{0}
 \begin{table*}
  \centering
  \begin{minipage}{\textwidth}
  \caption{Position, size and fitting parameters of the emission traced by CS(1--0), its isotopologue C$^{34}$S(1--0), CO(1--0) and HC$_{3}$N(5--4,F=4--3) towards \mbox{HESS\,J1809--193}.
  T$_{\text{A}}^{*}$ denotes the peak temperature at $v_{\text{lsr}}=v_{\text{cent}}$, $\Delta v$ indicates the full width half maximum (FWHM) of a Gaussian fit. $W$ represents the main beam integrated intensity
  using a main beam efficiency factor $\eta_{\text{mb}}$ (see \citealt{Urquhart2010} and text).
  Finally, the $\big\{$ indicates that the two components with little velocity separation are physically connected.}
   \begin{tabular}{p{3.2cm}p{2cm}clrrrrc}
    \toprule\toprule\\
   Molecular tracer& Region & Position & Area$^{a}$& $T_{\text{A}}^{*}$ & $v_{\text{cent}}$ & $\Delta v$  & $W=\int{T_{\text{mb}}\text{d}v}$ \\
   & & (RA, Dec)& (arcsec$^{2}$)& (K) & (km/s) & (km/s) & (K km/s) \\
    & & (J2000.0)\\
   \midrule \\
    \multirow{3}{*}{Nanten CO(1--0)} & 1 & (272.61$^{\circ}$, -19.38$^{\circ}$)& 819$\times$373& 6.5 & 30.7 & 8.7 & 67.6\\
     & 2 &(272.36$^{\circ}$, -19.71$^{\circ}$)& 663$\times$437 & 6.9 & 31.0 & 13.6 & 111.8 \\
     & 3 &  (272.37$^{\circ}$, -19.34$^{\circ}$)& 600$\times$360 & 7.1 & 32.0 & 13.5 & 114.0 \\
     \\
     \\
    \multirow{14}{*}{Mopra CS(1--0)} & 1 & (272.43$^{\circ}$, -19.74$^{\circ}$)& 819$\times$373 &0.2 & 30.2 & 4.6 & 2.1 \\
    & 1-1 & (272.61$^{\circ}$, -19.38$^{\circ}$)&146$\times$45& 0.6 & 29.4 & 3.1 & 4.6 \\
    & 1-2 & (272.62$^{\circ}$, -19.46$^{\circ}$)&30$\times$30 & 0.3 & 30.0 & 2.3 & 1.9 \\
     & 1-3 & (272.57$^{\circ}$, -19.46$^{\circ}$)&30$\times$30 &0.3 & 29.5 & 2.6 & 2.2 \\
     & 1-4 &(272.54$^{\circ}$, -19.47$^{\circ}$)&30$\times$30 & 0.5 & 29.1 & 3.6 & 4.3 \\
    & 2 & (272.36$^{\circ}$, -19.71$^{\circ}$)& 663$\times$437 &0.2 & 29.2 & 3.9 & 2.0 \\
    & \multirow{2}{*}{2-1} & \multirow{2}{*}{(272.57$^{\circ}$, -19.46$^{\circ}$)}&\multirow{2}{*}{30$\times$30} &\multirow{2}{*}{$ \bigg\{$} 0.5 & 31.5 & 2.7 & 3.3 \\ 
     & & & & 0.3 & 31.7 & 1.1 & 0.8 \\
     & \multirow{2}{*}{2-2}& \multirow{2}{*}{(272.53$^{\circ}$, -19.72$^{\circ}$)}&\multirow{2}{*}{30$\times$30} &\multirow{2}{*}{$ \bigg\{$} 0.5 & 29.4 & 3.9 & 4.4 \\
     & & & & 0.1 & 31.7 & 1.2 & 0.4 \\
     & \multirow{2}{*}{2-3} & \multirow{2}{*}{(272.41$^{\circ}$, -19.77$^{\circ}$)}&\multirow{2}{*}{30$\times$30} &\multirow{2}{*}{$ \bigg\{$} 0.5 & 30.8 & 2.6 & 3.0 \\
     & & & & 0.3 & 28.6 & 2.2 & 1.3 \\
    & \multirow{2}{*}{3} & \multirow{2}{*}{(272.37$^{\circ}$, -19.34$^{\circ}$)}&\multirow{2}{*}{600$\times$360} &\multirow{2}{*}{$ \bigg\{$} 0.1 & 34.2 & 3.6 &1.3 \\
    & & & & 0.1 & 31.3 & 1.5 & 0.4 \\
    \\
    \\
    \multirow{8}{*}{Mopra C$^{34}$S(1--0)$^{a}$} & 1-1 & (272.61$^{\circ}$, -19.38$^{\circ}$)&146$\times$45& 0.1  & 30.0  & 3.7  & 0.9  \\
     & 1-2 & (272.62$^{\circ}$, -19.46$^{\circ}$)&30$\times$30 & 0.2  & 30.0  & 1.2 & 0.5 \\
     & 1-3 & (272.57$^{\circ}$, -19.46$^{\circ}$)&30$\times$30 & 0.2  & 29.4  & 0.8  & 0.4  \\
     & 1-4 &(272.54$^{\circ}$, -19.47$^{\circ}$)&30$\times$30 & 0.2  & 29.3  & 0.8 & 0.4  \\
     & \multirow{2}{*}{2-1} & \multirow{2}{*}{(272.57$^{\circ}$, -19.46$^{\circ}$)}&\multirow{2}{*}{30$\times$30} &\multirow{2}{*}{$ \bigg\{$} 0.2  & 31.6 & 1.0  & 0.4  \\ 
     & & & & 0.1  & 30.1  & 1.0  & 0.3  \\
     & 2-2 & (272.53$^{\circ}$, -19.72$^{\circ}$)& 30$\times$30 & 0.1  & 30.5  & 1.6  & 0.6  \\
     & 2-3 & (272.41$^{\circ}$, -19.77$^{\circ}$)& 30$\times$30 & 0.2  & 30.8  & 1.0  & 0.4  \\
     \\
     \\
     \multirow{4}{*}{Mopra \scriptsize{HC$_3$N(5--4,F=4--3)}}& HC1 & (272.69$^{\circ}$, -19.27$^{\circ}$)&187$\times$60 & 0.2 & 31.7 & 1.8 & 1.0 \\
     & HC2 &(272.61$^{\circ}$, -19.38$^{\circ}$)&205$\times$67 & 0.2 & 29.3 & 2.1 & 1.1 \\
     & HC3 &(272.53$^{\circ}$, -19.46$^{\circ}$)&88$\times$88 &  0.2 & 29.2& 2.1 & 1.2 \\
     & HC4 & (272.49$^{\circ}$, -19.74$^{\circ}$)&257$\times$214 & 0.1 & 29.1 & 5.1 & 1.3\\
     \\
     \\
     \multirow{3}{*}{Mopra \scriptsize{SiO(1--0,v=0)}}& S1 & (272.62$^{\circ}$, -19.37$^{\circ}$)& 30$\times$30 & 0.06 & 30.2 & 4.0 & 0.6 \\
     & S2 & (272.54$^{\circ}$, -19.46$^{\circ}$)&40$\times$40 & 0.05 & 26.9 & 3.6 & 0.4 \\
     & S3 & (272.44$^{\circ}$, -19.70$^{\circ}$)&66$\times$66 & 0.06 & 29.4 & 4.6 & 0.7 \\
     \bottomrule
     \multicolumn{8}{p{18cm}}{\footnotesize{$^{a}$:The values represent the semi-major and semi-minor axes respectively.}}\\
     \end{tabular}
   \label{J1809table}
\end{minipage}
\end{table*}

\begin{table*}
    \centering
    \begin{minipage}{\textwidth}
    \caption{Position, size and fitting parameters of the emission traced by CS(1--0) and CO(1--0) towards \mbox{HESS\,J1026--582}.
  T$_{\text{A}}^{*}$ denotes the peak temperature at $v_{\text{lsr}}=v_{\text{cent}}$, $\Delta v$ indicates the full width half maximum (FWHM) of a Gaussian fit. $W$ represents the main beam integrated intensity using a 
  main beam efficiency factor $\eta_{\text{mb}}$ (see \citealt{Urquhart2010} for 7mm tracers)).
  Finally, the $\big\{$ indicates that the two components with little velocity separation are physically connected.}
    \begin{tabular}{p{3cm}p{2cm}ccrrrr}
    \toprule\toprule\\
   Molecular tracer& Region &  Position & Area$^{a}$ & $T_{\text{A}}^{*}$ & $v_{\text{cent}}$ & $\Delta v$  & $W=\int{T_{\text{mb}}\text{d}v}$ \\
   & &  (RA, Dec)& (arcsec$^2$) &  (K) & (km/s) & (km/s) & K km/s \\
   & & (J2000.0) \\
   \midrule \\
    \\ 
     \multirow{2}{*}{Nanten CO(1--0)} & A & (157.47$^{\circ}$, -58.58$^{\circ}$)&1096$\times$1096 & 1.1 & -15.2 & 4.3 & 5.7\\
     & B & (156.35$^{\circ}$, -58.20$^{\circ}$)&773$\times$773 &2.3 & 3.2 & 4.3 & 11.9 \\
     \\
     \multirow{9}{*}{Mopra CS(1--0)} & 1 &(156.36$^{\circ}$, -58.15$^{\circ}$)&87$\times$87 &0.3 & -19.2 & 1.4 & 1.0 \\
     & 2 & (156.15$^{\circ}$, -58.10$^{\circ}$)&87$\times$87 &0.2 & -16.57& 1.2 & 0.7  \\
     & 3 & (156.85$^{\circ}$, -57.77$^{\circ}$)&85$\times$260 &0.2 & 11.4 & 3.1 & 1.3  \\
     & 4 & (156.66$^{\circ}$, -57.79$^{\circ}$)&173$\times$272 &0.3 & 11.3 & 3.6 & 2.3  \\
     & 5 & (156.57$^{\circ}$, -57.89$^{\circ}$)&153$\times$61 &0.2  & 1.2 & 2.2 & 1.3  \\ 
     & \multirow{2}{*}{6} & \multirow{2}{*}{(156.46$^{\circ}$, -58.20$^{\circ}$)}&\multirow{2}{*}{57$\times$234}& \multirow{2}{*}{$ \bigg\{$} 0.2  & 3.7 & 1.8 & 0.8  \\
     & & & & 0.2 & 2.0 & 1.0 & 0.5 \\
     & 7 & (156.33$^{\circ}$, -58.24$^{\circ}$)&98$\times$98& 0.5  & 3.6 & 1.6 & 1.9  \\
     \bottomrule
     \multicolumn{8}{p{18cm}}{\footnotesize{$^{a}$:The values represent the semi-major and semi-minor axes respectively.}}\\
   \end{tabular}
   \label{tableJ1026}
  \end{minipage}
 \end{table*}

 \begin{table*}
    \centering
    \begin{minipage}{\textwidth}
    \caption{Position, size and fitting parameters of the emission traced by CO(1--0) towards \mbox{HESS\,J1119--582}.
  T$_{\text{A}}^{*}$ denotes the peak temperature at $v_{\text{lsr}}=v_{\text{cent}}$, $\Delta v$ indicates the full width half maximum (FWHM) of a Gaussian fit. $W$ represents the main beam integrated intensity using 
  the main beam efficiency factor $\eta_{\text{mb}}$ (see \citealt{Urquhart2010} for 7mm tracers).
  Finally, the $\big\{$ indicates that two the two components with little velocity separation are physically connected.}
    \begin{tabular}{p{3cm}p{2cm}ccrrrr}
    \toprule\toprule\\
   Molecular tracer& Region & Position & Area$^{a}$ & $T_{\text{A}}^{*}$ & $v_{\text{cent}}$ & $\Delta v$  & $W=\int{T_{\text{mb}}\text{d}v}$ \\
   & &   (RA, Dec)& (arcsec$^2$) & (K) & (km/s) & (km/s) & (K km/s) \\
   & & (J2000.0) \\
   \midrule \\
    \\
    \multirow{4}{*}{Nanten CO(1--0)} & A & (170.20$^{\circ}$, -61.35$^{\circ}$)&522$\times$522 & 1.4 & -12.3 & 5.9 & 9.6\\
      & B & (169.40$^{\circ}$, -61.48$^{\circ}$)&495$\times$159 & 0.6 & -19.1 & 5.1 & 3.8 \\
     & \multirow{2}{*}{C} &\multirow{2}{*}{170.08$^{\circ}$, -61.19$^{\circ}$)}&\multirow{2}{*}{219$\times$219} &\multirow{2}{*}{$ \bigg\{$} 1.1 & 20.4 & 8.2 & 4.1 \\
     & & & &$\ \ $ 1.0 & 35.1 & 2.2 & 1.5 \\
     & D & (169.38$^{\circ}$, -61.44$^{\circ}$)&705$\times$480 & 1.7 & 30.2 & 4.7 & 4.4 \\
     \bottomrule
     \multicolumn{8}{p{18cm}}{\footnotesize{$^{a}$:The values represent the semi-major and semi-minor axes respectively.}}\\
   \end{tabular}
   \label{tableJ1119}
  \end{minipage}
 \end{table*}

 \begin{table*}
    \centering
    \begin{minipage}{\textwidth}
    \caption{Position, size and fitting parameters of the emission traced by CS(1--0) and CO(1--0) towards \mbox{HESS\,J1418--609} and \mbox{HESS\,J1420--607}.
  T$_{\text{A}}^{*}$ denotes the peak temperature at $v_{\text{lsr}}=v_{\text{cent}}$, $\Delta v$ indicates the full width half maximum (FWHM) of a Gaussian fit. $W$ represents the main beam integrated intensity using the main beam efficiency factor $\eta_{\text{mb}}$ (see \citealt{Urquhart2010} for tracers).
  Finally, the $\big\{$ indicates that  the two components with little velocity separation are physically connected.}
    \begin{tabular}{p{3cm}p{2cm}ccrrrr}
    \toprule\toprule\\
   Molecular tracer tracer& Regions & Position & Area$^{a}$ & $T_{\text{A}}^{*}$ & $v_{\text{cent}}$ & $\Delta v$  & $W=\int{T_{\text{mb}}\text{d}v}$ \\
   & & (RA, Dec)& (arcsec$^{2}$) & (K) & (km/s) & (km/s) & (K km/s) \\
   & & (J2000.0)\\
   \midrule \\
    \\
    \multirow{7}{*}{Nanten CO(1--0)} & A & (214.0$^{\circ}$, -61.1$^{\circ}$)&364$\times$364 & 5.9 & -45.0 & 10.2 & 71.3 \\
    & B & (214.47$^{\circ}$, -61.11$^{\circ}$)&352$\times$352 & 1.9 & -3.8 & 6.9 & 15.3 \\
    & \multirow{2}{*}{6} & \multirow{2}{*}{(215.33$^{\circ}$, -60.84$^{\circ}$)} &\multirow{2}{*}{110$\times$288} & \multirow{2}{*}{$ \bigg\{$} 4.8  & -50.1 & 8.8 & 50.4  \\
    & & & & 3.6 & -43.3 & 6.3 & 27.3 \\
     & 7 & (215.08$^{\circ}$, -60.85$^{\circ}$)&94$\times$171 & 5.3 & -49.1 & 7.2 & 45.9  \\
     & 8 & (215.11$^{\circ}$, -60.93$^{\circ}$)&58$\times$173 & 4.7 & -48.2 & 9.0  & 50.5  \\
     & 9 & (215.03$^{\circ}$, -60.79$^{\circ}$)&87$\times$87 & 4.8 & -49.0 & 8.7 & 49.7  \\
     & 10 & (215.05$^{\circ}$, -60.69$^{\circ}$) & 179$\times$143 & 5.0 & -47.2 & 7.0 & 41.9 \\
    \\
    \multirow{13}{*}{Mopra CS(1--0)} & 1 & (214.52$^{\circ}$, -60.80$^{\circ}$)&71$\times$217 &  0.3 & -59.1 & 1.4 & 1.1 \\
     & 2 & (214.37$^{\circ}$, -60.85$^{\circ}$)&85$\times$207 & 0.1 & -56.7 & 0.9 & 0.3  \\
     & 3 & (214.21$^{\circ}$, -60.84$^{\circ}$)&61$\times$140 & 0.3 & -56.7 & 1.1 & 0.7  \\
     & 4 & (214.08$^{\circ}$, -60.88$^{\circ}$)&90$\times$250 &0.2 & -57.5 & 2.6 & 1.1  \\
     & \multirow{2}{*}{5} & \multirow{2}{*}{(214.11$^{\circ}$, -60.99$^{\circ}$)}& \multirow{2}{*}{82$\times$201} & \multirow{2}{*}{$ \bigg\{$} 0.2  & -62.2 & 2.2 & 1.2  \\ 
     & & & &  0.2 & -60.3 & 1.4 & 0.6 \\
     & 6 & (215.33$^{\circ}$, -60.84$^{\circ}$)&110$\times$288 & 0.3  & -44.8 & 1.3 & 1.0  \\
     & 7 & (215.08$^{\circ}$, -60.85$^{\circ}$)&94$\times$171 &0.2  & -48.8 & 1.6 & 0.8  \\
     & 8 & (215.11$^{\circ}$, -60.93$^{\circ}$)&58$\times$173 &0.2 & -48.6 & 1.6  & 0.9  \\
     & 9 & (215.03$^{\circ}$, -60.79$^{\circ}$)&87$\times$87 & 0.2  & -48.6 & 1.2 & 0.6  \\
     & 10 & (215.05$^{\circ}$, -60.69$^{\circ}$) & 179$\times$143 & 0.3 & -46.7 & 2.6 & 1.8 \\
     & \multirow{2}{*}{11} & \multirow{2}{*}{(214.75$^{\circ}$, -61.00$^{\circ}$)}&\multirow{2}{*}{87$\times$87} & \multirow{2}{*}{$ \bigg\{$} 0.2  & -51.3 & 1.7 & 0.6  \\
     & & & & 0.1 & -42.4 & 0.7 & 0.3 \\
     & 12 & (214.46$^{\circ}$, -61.01$^{\circ}$)&87$\times$87 & 0.3 & -47.2 & 2.8  & 1.8  \\
     & 13 & (214.31$^{\circ}$, -61.09$^{\circ}$)&87$\times$87 & 0.2 & -50.5 & 1.0  & 0.6  \\
     & \multirow{2}{*}{14} & \multirow{2}{*}{(214.88$^{\circ}$, -60.85$^{\circ}$)}&\multirow{2}{*}{109$\times$178} & \multirow{2}{*}{$ \bigg\{$} 0.3  & -3.4 & 4.3 & 3.1  \\
     & & & & 0.1 & -4.9 & 0.7 & 0.3 \\
     & 15 & (214.99$^{\circ}$, -60.80$^{\circ}$)&87$\times$87 & 0.2 & 0.9 & 1.6  & 0.7  \\
     \bottomrule
     \multicolumn{8}{p{18cm}}{\footnotesize{$^{a}$:The values represent the semi-major and semi-minor axes respectively.}}\\
   \end{tabular}
\label{tableJ1418}
  \end{minipage}
 \end{table*}
 
 \begin{table*}
  \begin{minipage}{\textwidth}
  \caption{Position, size and fitting parameters of the emission traced by  CS(1--0) and CO(1--0) transitions towards \mbox{HESS\,J1303--631}.
  T$_{\text{A}}^{*}$ denotes the peak temperature at $v_{\text{lsr}}=v_{\text{cent}}$, $\Delta v$ indicates the full width half maximum (FWHM) of a Gaussian fit. $W$ represents the main beam integrated intensity using 
  the main beam efficiency factor $\eta_{\text{mb}}$ (see \citealt{Urquhart2010} for 7mm tracers). Finally, the $\big\{$ indicates that two the two components with little velocity separation are physically connected.}
   \centering
   \begin{tabular}{p{3cm}p{2cm}clrrrr}
    \toprule\toprule\\
   Molecular tracer& Region & Position & Area$^{a}$ & $T_{\text{A}}^{*}$ & $v_{\text{cent}}$ & $\Delta v$  & $W=\int{T_{\text{mb}}\text{d}v}$ \\
   & & (RA, Dec)& arcsec$^{2}$ &  (K) & (km/s) & (km/s) & K km/s \\
   & & (J2000.0)\\
   \midrule \\
    \multirow{4}{*}{Nanten CO(1--0)} & A & (195.63$^{\circ}$, -63.46$^{\circ}$)&512$\times$300 & 2.8 & -30.0 & 10.3 & 34.7 \\
    & B & (196.06$^{\circ}$, -63.21$^{\circ}$)&438$\times$438& 3.8 & -19.7 & 7.9 & 35.6 \\
    & \multirow{2}{*}{C} & \multirow{2}{*}{(195.89$^{\circ}$, -63.17$^{\circ}$)}&\multirow{2}{*}{130$\times$131}&\multirow{2}{*}{$ \bigg\{$} 5.6 & 33.4 & 5.4 &36.6 \\
    & & & &$\ \ $ 1.0 & 24.7 & 3.1 & 3.8 \\
    & \multirow{2}{*}{D} & \multirow{2}{*}{(195.22$^{\circ}$, -63.15$^{\circ}$)}&\multirow{2}{*}{109$\times$208}&\multirow{2}{*}{$ \bigg\{$} 2.2 & 32.7 & 7.4 &19.4 \\
    & & & &$\ \ $ 1.6 & 26.3 & 4.2 & 8.2 \\
    \\
    \multirow{8}{*}{Mopra CS(1--0)} & 1 & (195.52$^{\circ}$, -63.23$^{\circ}$)&87$\times$87 &  0.3 & -30.8 & 2.8 & 1.8 \\
    & 2 & (195.62$^{\circ}$, -63.18$^{\circ}$)&87$\times$87 & 0.1 & -30.1 & 3.1 & 1.0 \\
    & 3 & (195.71$^{\circ}$, -62.99$^{\circ}$)&87$\times$87 & 0.1 & -29.7 & 3.0 & 0.7 \\
    & 4 & (195.78$^{\circ}$, -63.22$^{\circ}$)&87$\times$87 & 0.3 & -23.0 & 2.1 & 1.5 \\
    & 5 & (195.60$^{\circ}$, -63.00$^{\circ}$)&87$\times$87 & 0.2 & -18.1 & 1.8 & 0.9 \\
    & 6 & (195.46$^{\circ}$, -63.08$^{\circ}$)&87$\times$87 & 0.3 & -30.6 & 2.1 & 1.3\\
    & 7 & (195.49$^{\circ}$, -62.96$^{\circ}$)&87$\times$87 & 0.2 & -23.7 & 2.0 & 1.1 \\
    & 8 & (195.46$^{\circ}$, -63.07$^{\circ}$)&92$\times$92 & 0.1 & 30.7 & 2.7 & 0.9 \\
    & 9 & (195.48$^{\circ}$, -62.95$^{\circ}$)&47$\times$47 & 0.2 & 24.1 & 3.1 & 1.5\\
    & 10 & (195.20$^{\circ}$, -63.14$^{\circ}$)&36$\times$36 & 0.2 & 25.9 & 1.6 & 0.9 \\
    \bottomrule
    \multicolumn{8}{p{18cm}}{\footnotesize{$^{a}$:The values represent the semi-major and semi-minor axes respectively.}}\\
    \\
    \end{tabular}
    \label{J1303table}
    \end{minipage}
    \end{table*}

    \begin{table*}
    \centering
    \begin{minipage}{\textwidth}
    \caption{Position, size and fitting parameters of the emission traced by CO(1--0) towards \mbox{HESS\,J1018--589B}.
  T$_{\text{A}}^{*}$ denotes the peak temperature at $v_{\text{lsr}}=v_{\text{cent}}$, $\Delta v$ indicates the full width half maximum (FWHM) of a Gaussian fit. $W$ represents the main beam integrated intensity using a 
  main beam efficiency factor $\eta_{\text{mb}}$ (see \citealt{Urquhart2010} for 7mm tracers).}
    \begin{tabular}{p{3cm}p{2cm}ccrrrr}
    \toprule\toprule\\
   Molecular tracer& Region & Position & Area$^{a}$ & $T_{\text{A}}^{*}$ & $v_{\text{cent}}$ & $\Delta v$  & $W=\int{T_{\text{mb}}\text{d}v}$ \\
   & &   (RA, Dec)& arcsec$^2$ & (K) & (km/s) & (km/s) & K km/s \\
   & & (J2000.0)\\
   \midrule \\
    \\
    \multirow{1}{*}{Nanten CO(1--0)} & A & (154.68$^{\circ}$, $-58.82^{\circ}$)&400$\times$400 & 0.9 & $-15.9$ & 12.0 & 13.8 \\
     \bottomrule
     \multicolumn{8}{p{18cm}}{\footnotesize{$^{a}$:The values represent the semi-major and semi-minor axes respectively.}}\\
   \end{tabular}
\label{tableJ1018}
  \end{minipage}
 \end{table*}
 \section[]{Physical parameters}
 \label{sec:Mass}
 \setcounter{figure}{0}
 \subsection{CS(1-0)}
The averaged optical depth $\tau_{\text{CS(1--0)}}$ can be derived using the integrated intensity ratio $W_{\text{CS(1--0)}}$ and $W_{\text{C}^{34}\text{S(1--0)}}$ via :
\begin{equation}
\label{tauCS}\frac{W_{CS(1-0)}}{W_{C^{34}S(1-0)}} = \frac{1-e^{-\tau_{CS(1-0)}}}{1-e^{\alpha\tau_{CS(1-0)}}}  
\end{equation}
where $\alpha$=[$^{32}$S]/[$^{34}$S]=22 represents the abundance ratio between the two isotopologues based on terrestrial measurements \citep{Fink1981}.
We obtain the column density of the upper state $N_{\text{CS}_1}$:
\begin{equation}
\label{CSeq2} N_{\text{CS}_1} = \frac{8k \pi\nu_{10}^{2}}{A_{10}hc^{3}}\left(\frac{\Delta\Omega_{\text{A}}}{\Delta\Omega_{\text{S}}}\right)\left(\frac{\tau_{\text{CS(1-0)}}}{1-e^{-\tau_{\text{CS(1-0)}}}}\right)\int{T_{\text{mb}}\left(v\right)dv} 
\end{equation}
Assuming the gas to be in local thermal equilibrium at temperature $T_\text{kin}$, we can thus obtain the total CS column density $N_{\text{CS}}$: 
\begin{equation}
 \label{CSeq3} N_{CS} = N_{CS_1}\left(1+\frac{1}{3}e^{2.35/T_{\text{kin}}}+\frac{5}{3}e^{-4.7/T_{\text{kin}}}+\text{...}\right) 
\end{equation}
 \subsection[]{Mass and density}
 \begin{equation}
M_{\text{H}_2}=\mu m_{\text{H}}\pi abN_{\text{H}_2}
\label{H2mass}
\end{equation}
with $\mu=2.8$ being the averaged atomic weight (accounting for 20\% Helium), $m_{\text{H}}$ the hydrogen mass and $a$,$b$ the semi-minor and semi-major axis of the selected elliptic regions.

 \begin{equation}
 \label{H2dens} n_{\text{H}_2}=\frac{M}{4/3\left(\mu m_{\text{H}}\right)\pi ab^2}
\end{equation}
\subsection[]{Physical parameters of individual sources}
\setcounter{table}{0}
\setcounter{figure}{0}
Tables\,\ref{MASSJ1809} to \ref{MASSJ1018} show the column density $N_{\text{H}_2}$ the atomic and molecular mass $M_{\text{H}_\textsc{i}}$ and $M_{\text{H}_2}$ and, molecular density  $n_{\text{H}_2}$ estimates based on the 
CS, CO, and H$_{\textsc{i}}$ emission found towards the regions listed in Section\,\ref{sec:Fit parameter}.
\begin{table*}
    \centering
    \begin{minipage}{\textwidth}
    \caption{Physical parameters obtained from our CS and CO analyses for the different selected regions located towards \mbox{HESS\,J1809--193}. In the case where C$^{34}$S(1--0) is detected, the derived optical depth 
    $\tau_{\text{CS(1--0)}}$ is shown as superscript next to the CS column density $N_{\text{CS}}$. Otherwise, an optical thin scenario is assumed and the derived column densities $N_{\text{H}_2}$ and $N_{\text{CS}}$, mass $M_{\text{H}_2}$(CS) and H$_2$ averaged density $n_{\text{H}_2}$(CS) act as lower limits.}
    \begin{tabular}{cccccccc}
    \toprule
    \multicolumn{6}{c}{HESS\,J1809--193}\\
    \toprule
    \multicolumn{6}{c}{CS(1--0)} \\
   Reg. & Distance & $N_{\text{CS}}\left[10^{12}\right]^{a}$ & $N_{\text{H}_2}\left[10^{20}\right]^{ab}$ & $M_{\text{H}_2}$(CS)$^{abc}$ & $n_{\text{H}_2}$(CS)$^{abc}$\\
   & (kpc)& (cm$^{-2}$)& (cm$^{-2}$) & ($\solmass$) & (cm$^{-3}$) \\
   \midrule
    1 & 3.7 & 20 & 52 & $3.2\times10^{4}$ & $1.7\times10^{2}$\\
    1-1 & 3.7 &210$^{\left(4\right)}$ & 520 & $7.5\times10^{3}$& $1.8\times10^{4}$\\
    1-2 & 3.7 &270$^{\left(7\right)}$ & 670 & $1.3\times10^{3}$ & $3.2\times10^{4}$\\
    1-3 & 3.7 &180$^{\left(4\right)}$ & 460 & $6.0\times10^{3}$ & $2.2\times10^{4}$\\
    1-4 & 3.7 &190$^{\left(2\right)}$ & 460 & $9.0\times10^{4}$ & $2.2\times10^{4}$\\
    2 & 3.7 & 19 & 48 & $3.2\times10^{4}$& $1.5\times10^{2}$ \\
    2-1& 3.7& 380$^{\left(5\right)}$& 950& $1.7\times10^{3}$ & $4.4\times10^{4}$\\
    2-2 & 3.7& 290$^{\left(3\right)}$& 720& $1.3\times10^{3}$ & $3.3\times10^{4}$\\
    2-3 & 3.7& 200$^{\left(2\right)}$& 510& $1.0\times10^{3}$  & $2.3\times10^{4}$\\
    3 & 2.7 & 15 & 30 & $6.2\times10^{3}$ & $2.5\times10^{2}$ \\
     \midrule 
     \vspace{-0.5cm}\\
     \cmidrule{1-5}
     \multicolumn{4}{c}{CO(1--0)} & H\textsc{i} \\
     \cmidrule{1-5}
     Reg. & Distance & $M_{\text{H}_2}$(CO)$^{cd}$ & $n_{\text{H}_2}$(CO)$^{cd}$ & $M_{\text{H}_I}$ \\
     & (kpc)& ($\solmass$) & (cm$^{-3}$) & ($\solmass$) \\
     1 & 3.7 & $8.1\times10^{4}$ & $4.4\times10^{2}$ & $5.5\times10^{3}$ \\
     2 & 3.7 & $2.3\times10^{5}$ & $1.7\times10^{2}$ & $6.0\times10^{3}$ \\
     3 & 2.7 & $5.0\times10^{4}$ & $1.0\times10^{3}$ & $2.6\times10^{3}$\\
     \cmidrule{1-5}
     \multicolumn{7}{p{14cm}}{\footnotesize{$^{a}$:Parameters have been derived using the LTE assumption.}}\\
    \multicolumn{7}{p{14cm}}{\footnotesize{$^{b}$:The H$_{2}$ physical parameters derived using a CS abundance ratio $\chi_{\text{CS}}=4\times10^{-9}$ }}\\
   \multicolumn{7}{p{14cm}}{\footnotesize{$^{c}$:A prolate geometry has been used in order to derive the mass and density.  }}\\
   \multicolumn{7}{p{14cm}}{\footnotesize{$^{d}$:The $X_{\text{CO}}=2.0\times10^{20}$ cm$^{-2}$/(K km/s)  have been used to convert the integrated intensity $W_{\text{CO}}$ into H$_2$ column density $N_{\text{H}_2}$ }}
  
   \end{tabular}
\label{MASSJ1809}
  \end{minipage}

 \end{table*}

 \begin{table*}
    \centering
    \begin{minipage}{\textwidth}
    \caption{Physical parameters obtained from our CS and CO analyses for the different selected regions located towards \mbox{HESS\,J1026--582}. In the case of our CS analysis, we assumed a optically thin scenario
      and the derived column densities $N_{\text{H}_2}$ and $N_{\text{CS}}$, mass $M_{\text{H}_2}$(CS) and H$_2$ averaged density $n_{\text{H}_2}$(CS) act as lower limits.}
    \begin{tabular}{cccccccc}
    \toprule
    \multicolumn{6}{c}{HESS\,J1026--582}\\
    \toprule
    \multicolumn{6}{c}{CS(1--0)} \\
   Reg. & Distance & $N_{\text{CS}}\left[10^{12}\right]^{a}$ & $N_{\text{H}_2}\left[10^{20}\right]^{ab}$ & $M_{\text{H}_2}$(CS)$^{abc}$ & $n_{\text{H}_2}$(CS)$^{abc}$ \\
   & (kpc)& (cm$^{-2}$)& (cm$^{-2}$) & ($\solmass$) & (cm$^{-3}$) \\
   \midrule
    1 & 2.3 & 10 & 26 & $1.7\times10^{2}$ & $6.3\times10^{2}$\\
    2 & 2.3 & 7 & 18 & $1.2\times10^{2}$& $4.6\times10^{2}$\\
    3 & 6.1 & 13 & 33 & $4.5\times10^{3}$ & $3.2\times10^{2}$ \\
    4 & 6.1 & 23 & 58 & $1.7\times10^{4}$ & $2.8\times10^{2}$ \\
    5 & 4.9 & 14 & 35  & $1.3\times10^{3}$ & $5.9\times10^{2}$ \\
    6 & 4.9 & 14 & 35 & $1.8\times10^{3}$ & $7.2\times10^{2}$ \\
    7 & 4.9 & 19 & 49 & $1.9\times10^{3}$ & $5.1\times10^{2}$ \\
     \midrule 
     \vspace{-0.5cm}\\
     \cmidrule{1-5}
     \multicolumn{4}{c}{CO(1--0)} &  H\textsc{i} \\
     \cmidrule{1-5}
     Reg. & Distance & $M_{\text{H}_2}$(CO)$^{cd}$ & $n_{\text{H}_2}$(CO)$^{cd}$ & $M_{\text{H}_I}$ \\
     & (kpc)& ($\solmass$) & (cm$^{-3}$)& ($\solmass$) \\
     A & 2.3 &  $1.1\times10^{4}$ & $2.2\times10^{1}$ & $5.0\times10^{3}$ \\
     B & 4.9 & $5.4\times10^{4}$ & $3.2\times10^{1}$ & $3.2\times10^{4}$ \\
     \cmidrule{1-5}
     \multicolumn{7}{p{14cm}}{\footnotesize{$^{a}$:Parameters have been derived using the LTE assumption.}}\\
    \multicolumn{7}{p{14cm}}{\footnotesize{$^{b}$:The H$_{2}$ physical parameters derived using a CS abundance ratio $\chi_{\text{CS}}=4\times10^{-9}$ }}\\
   \multicolumn{7}{p{14cm}}{\footnotesize{$^{c}$:A prolate geometry has been used in order to derive the mass and density.  }}\\
   \multicolumn{7}{p{14cm}}{\footnotesize{$^{d}$:The $X_{\text{CO}}=2.0\times10^{20}$ cm$^{-2}$/(K km/s)  have been used to convert the integrated intensity $W_{\text{CO}}$ into H$_2$ column density $N_{\text{H}_2}$ }}
  
   \end{tabular}
   \label{MASSJ1026}
  \end{minipage}
 \end{table*}

 \begin{table*}
    \centering
    \begin{minipage}{\textwidth}
    \caption{Physical parameters obtained from our CO analysis for the different selected regions located towards \mbox{HESS\,J1119--582}.}
    \begin{tabular}{cccccccc}
    \toprule
    \multicolumn{4}{c}{HESS\,J1119--582}\\
    \toprule
     \multicolumn{4}{c}{CO(1--0)} & H\textsc{i} \\
     \cmidrule{1-5}
     Reg. & Distance & $M_{\text{H}_2}$(CO)$^{cd}$ & $n_{\text{H}_2}$(CO)$^{cd}$ & $M_{\text{H}_I}$ \\
     & (kpc)& ($\solmass$) & (cm$^{-3}$) & ($\solmass$) \\
     A & 5.0 & $2.2\times10^{4}$ & $5.0\times10^{1}$ & $6.3\times 10^{3}$ \\
     B & 4.2 & $7.1\times10^{3}$ & $5.0\times10^{1}$ & $3.6\times 10^{3}$\\
     C & 8.6 & $1.3\times10^{4}$ & $6.1\times10^{1}$ & $5.8\times 10^{3}$\\
     D & 9.7 & $1.3\times10^{5}$ & $1.4\times10^{1}$ & $7.8\times 10^{4}$\\
     \cmidrule{1-5}
     \multicolumn{7}{p{14cm}}{\footnotesize{$^{a}$:Parameters have been derived using the LTE assumption.}}\\
    \multicolumn{7}{p{14cm}}{\footnotesize{$^{b}$:The H$_{2}$ physical parameters derived using a CS abundance ratio $\chi_{\text{CS}}=4\times10^{-9}$ }}\\
   \multicolumn{7}{p{14cm}}{\footnotesize{$^{c}$:A prolate geometry has been used in order to derive the mass and density.  }}\\
   \multicolumn{7}{p{14cm}}{\footnotesize{$^{d}$:The $X_{\text{CO}}=2.0\times10^{20}$ cm$^{-2}$/(K km/s)  have been used to convert the integrated intensity $W_{\text{CO}}$ into H$_2$ column density $N_{\text{H}_2}$ }}
   \end{tabular}
   \label{MASSJ1119}
  \end{minipage}
 \end{table*}

 \begin{table*}
    \centering
    \begin{minipage}{\textwidth}
    \caption{Physical parameters obtained from our CS and CO analyses for the different selected regions located towards \mbox{HESS\,J1303--631}. In the case of our CS analysis, we assumed a optically thin scenario
      and the derived column densities $N_{\text{H}_2}$ and $N_{\text{CS}}$, mass $M_{\text{H}_2}$(CS) and H$_2$ averaged density $n_{\text{H}_2}$(CS) act as lower limits.}
    \begin{tabular}{cccccccc}
    \toprule
    \multicolumn{6}{c}{HESS\,J1303--631}\\
    \toprule
    \multicolumn{6}{c}{CS(1--0)} \\
   Reg. & Distance & $N_{\text{CS}}\left[10^{12}\right]^{a}$ & $N_{\text{H}_2}\left[10^{20}\right]^{ab}$ & $M_{\text{H}_2}$(CS)$^{abc}$ & $n_{\text{H}_2}$(CS)$^{abc}$ \\
   & (kpc)& (cm$^{-2}$)& (cm$^{-2}$) & ($\solmass$) & (cm$^{-3}$) \\
   \midrule
    1 & 6.6 & 18 & 45 & $2.4\times10^{3}$ & $3.8\times10^{2}$ \\
    2 & 6.6 & 10 & 25 & $1.4\times10^{3}$& $2.2\times10^{2}$ \\
    3 & 6.6 & 7 & 19 & $1.0\times10^{3}$ & $1.6\times10^{2}$ \\
    4 & 1.5 & 15 & 37  & $1.1\times10^{2}$ & $1.4\times10^{3}$ \\
    5 & 1.5 & 9 & 22 & $6.2\times10^{2}$ & $8.6\times10^{2}$\\
    6 & 12.6 & 13 & 33 & $3.1\times10^{3}$ & $1.6\times10^{2}$ \\
    7 & 12.6 & 11 & 28 & $5.1\times10^{3}$ & $1.3\times10^{2}$ \\
     \midrule 
     \vspace{-0.5cm}\\
     \cmidrule{1-5}
     \multicolumn{4}{c}{CO(1--0)} & H\textsc{i}  \\
     \cmidrule{1-5}
     Reg. & Distance & $M_{\text{H}_2}$(CO)$^{cd}$ & $n_{\text{H}_2}$(CO)$^{cd}$ & $M_{\text{H}_I}$  \\
     & (kpc)& ($\solmass$) & (cm$^{-3}$)& ($\solmass$) \\
     A & 6.6 & $7.0\times10^{4}$ & $1.8\times10^{2}$  & $8.4\times10^{3}$ \\
     B & 1.5 & $4.8\times10^{3}$ & $5.4\times10^{2}$ & $7.4\times10^{2}$ \\
     C & 12.6 & $6.8\times10^{4}$ & $1.8\times10^{2}$ & $9.4\times10^{3}$ \\
     D & 12.6 & $4.1\times10^{4}$ & $2.5\times10^{2}$ & $8.5\times10^{3}$ \\
     \cmidrule{1-5}
     \multicolumn{7}{p{14cm}}{\footnotesize{$^{a}$:Parameters have been derived using the LTE assumption.}}\\
    \multicolumn{7}{p{14cm}}{\footnotesize{$^{b}$:The H$_{2}$ physical parameters derived using a CS abundance ratio $\chi_{\text{CS}}=4\times10^{-9}$ }}\\
   \multicolumn{7}{p{14cm}}{\footnotesize{$^{c}$:A prolate geometry has been used in order to derive the mass and density.  }}\\
   \end{tabular}
   \label{MASSJ1303}
  \end{minipage}
 \end{table*}
 
 \begin{table*}
    \centering
    \begin{minipage}{\textwidth}
    \caption{Physical parameters obtained from our CS and CO analyses for the different selected regions located towards \mbox{HESS\,J1420--607} and \mbox{HESS\,J1418--609}. In the case of our CS analysis, we assumed a optically thin scenario
      and the derived column densities $N_{\text{H}_2}$ and $N_{\text{CS}}$  , mass $M_{\text{H}_2}$(CS) and H$_2$ averaged density $n_{\text{H}_2}$(CS) act as lower limits.}
    \begin{tabular}{cccccccc}
    \toprule
    \multicolumn{6}{c}{HESS\,J1420-607+HESS\,J1418-609}\\
    \toprule
    \multicolumn{6}{c}{CS(1--0)} \\
   Reg. & Distance & $N_{\text{CS}}\left[10^{12}\right]^{a}$ & $N_{\text{H}_2}\left[10^{20}\right]^{ab}$ & $M_{\text{H}_2}$(CS)$^{abc}$ & $n_{\text{H}_2}$(CS)$^{abc}$\\
   & (kpc)& (cm$^{-2}$)& (cm$^{-2}$) & ($\solmass$) & (cm$^{-3}$) \\
   \midrule
    1 & 5.6 & 11 & 27 & $2.2\times10^{3}$ & $3.5\times10^{2}$ \\
    2 & 5.6 & 3 & 7 & $6.4\times10^{2}$& $7.5\times10^{1}$ \\
    3 & 5.6 & 7 & 17 & $7.8\times10^{2}$ & $2.6\times10^{2}$ \\
    4 & 5.6 & 11 & 28 & $3.3\times10^{3}$ & $2.8\times10^{2}$ \\
    5 & 5.6 & 17 & 45 & $3.8\times10^{3}$ & $5.0\times10^{2}$ \\
    6 & 3.5 & 10 & 25 & $1.6\times10^{3}$ & $3.3\times10^{2}$ \\
    7 & 3.5 & 8 & 19 & $6.2\times10^{2}$ & $2.8\times10^{2}$ \\
    8 & 3.5 & 10 & 25 & $5.1\times10^{2}$ & $6.2\times10^{2}$ \\
    9 & 3.5 & 6 & 15 & $2.1\times10^{2}$ & $2.4\times10^{2}$ \\
    10 & 3.5 & 18 & 46 & $2.3\times10^{3}$ & $4.5\times10^{2}$ \\
    11 & 3.5 & 6 & 16 & $2.5\times10^{2}$ & $2.6\times10^{2}$ \\
    12 & 3.5 & 18 & 45 & $7.0\times10^{2}$ & $7.4\times10^{2}$ \\
    13 & 3.5 & 6 & 14 & $2.1\times10^{2}$ & $2.3\times10^{2}$\\
    14 & 0.1 & 33 & 83 & $2.7\times10^{0}$ & $3.8\times10^{4}$ \\
    15 & 0.1 & 7 & 18 & $2.3\times10^{-1}$ & $1.1\times10^{4}$ \\
     \midrule 
     \vspace{-0.5cm}\\
     \cmidrule{1-5}
     \multicolumn{4}{c}{CO(1--0)} & H\textsc{i}  \\
     \cmidrule{1-5}
     Reg. & Distance & $M_{\text{H}_2}$(CO)$^{cd}$ & $n_{\text{H}_2}^{cd}$ & $M_{\text{H}_I}$   \\
     & (kpc)& ($\solmass$) & (cm$^{-3}$) & ($\solmass$)\\
     A & 3.5 & $3.8\times10^{4}$ & $5.6\times10^{2}$ & $2.2\times10^{3}$ \\
     6 & 3.5 & $1.0\times10^{4}$ & $2.0\times10^{3}$ & $4.6\times10^{2}$\\
     7 & 3.5 & $3.0\times10^{3}$ & $1.4\times10^{3}$ & $2.4\times10^{2}$ \\
     8 & 3.5 & $2.1\times10^{3}$ & $2.5\times10^{3}$ & $1.5\times10^{2}$ \\
     9 & 3.5 & $1.5\times10^{3}$ & $1.6\times10^{3}$ & $1.1\times10^{2}$ \\
     10 & 3.5 & $4.4\times10^{3}$ & $8.3\times10^{2}$ & $3.6\times10^{2}$ \\  
     B & 0.1 & $6.8\times10^{0}$ & $1.9\times10^{3}$ & $1.0\times10^{0}$ \\
     \cmidrule{1-5}
     \multicolumn{7}{p{14cm}}{\footnotesize{$^{a}$:Parameters have been derived using the LTE assumption.}}\\
    \multicolumn{7}{p{14cm}}{\footnotesize{$^{b}$:The H$_{2}$ physical parameters derived using a CS abundance ratio $\chi_{\text{CS}}=4\times10^{-9}$ }}\\
   \multicolumn{7}{p{14cm}}{\footnotesize{$^{c}$:A prolate geometry has been used in order to derive the mass and density.  }}\\
   \multicolumn{7}{p{14cm}}{\footnotesize{$^{d}$:The $X_{\text{CO}}=2.0\times10^{20}$ cm$^{-2}$/(K km/s)  have been used to convert the integrated intensity $W_{\text{CO}}$ into H$_2$ column density $N_{\text{H}_2}$ }} 
   \end{tabular}
   \label{MASSJ1420}
  \end{minipage}
 \end{table*}

 \begin{table*}
    \centering
    \begin{minipage}{\textwidth}
    \caption{Physical parameters obtained from our  CO analysis for the different selected regions located towards \mbox{HESS\,J1018--589B}.}
    \begin{tabular}{cccccccc}
    \toprule
    \multicolumn{4}{c}{HESS\,J1018--589B}\\
    \toprule
     \multicolumn{4}{c}{CO(1--0)} &  H\textsc{i}\\
     \cmidrule{1-5}
     Reg. & Distance & $M_{\text{H}_2}$(CO)$^{cd}$ & $n_{\text{H}_2}$(CO)$^{cd}$ & $M_{\text{H}_I}$ \\
     & (kpc)& ($\solmass$) & (cm$^{-3}$) & ($\solmass$) \\
     A & 2.0 & $2.9\times10^{3}$ & $1.8\times10^{2}$ & $8.0\times10^{2}$ \\
     \cmidrule{1-5}
     \multicolumn{7}{p{14cm}}{\footnotesize{$^{a}$:Parameters have been derived using the LTE assumption.}}\\
    \multicolumn{7}{p{14cm}}{\footnotesize{$^{b}$:The H$_{2}$ physical parameters derived using a CS abundance ratio $\chi_{\text{CS}}=4\times10^{-9}$ }}\\
   \multicolumn{7}{p{14cm}}{\footnotesize{$^{c}$:A prolate geometry has been used in order to derive the mass and density.  }}\\
   \multicolumn{7}{p{14cm}}{\footnotesize{$^{d}$:The $X_{\text{CO}}=2.0\times10^{20}$ cm$^{-2}$/(K km/s)  have been used to convert the integrated intensity $W_{\text{CO}}$ into H$_2$ column density $N_{\text{H}_2}$ }}
   \end{tabular}
   \label{MASSJ1018}
  \end{minipage}
 \end{table*}
 \clearpage
 \section{Column densities and distance studies}
\label{sec:distance}
\setcounter{figure}{0} 
 \begin{figure*}
\centering
 \begin{minipage}{0.8\textwidth}
  \includegraphics[width=\textwidth]{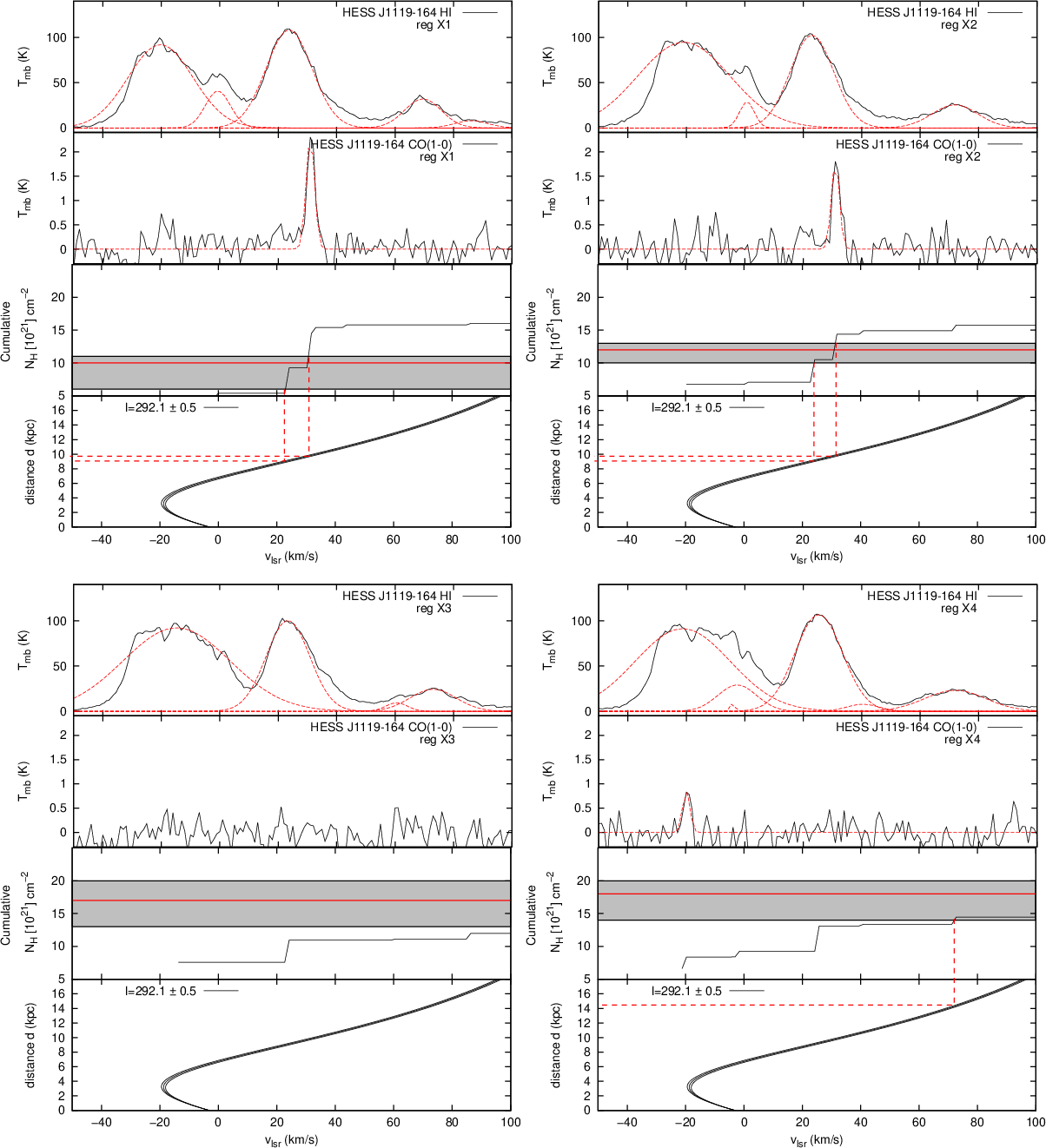}
  \caption{Averaged SGPS \textit{(first panels)} and Nanten CO(1--0) emission \textit{(second panels)} towards the region `A' \textit{(left)} and the western part of the SNR (\textit{right}) (see Figure\,\ref{J1119COCS}).
  The different Gaussian fits are shown as red dashed lines. At each Gaussian peak, the H\textsc{i} and CO(1--0) emission have been integrated and converted into column density, $N_{\text{H}}$ \textit{(third panel)}, 
  via the $X_{\text{CO}}$ and the $X_{\text{HI}}$ factors (see text). Here, all emission are assumed to be in the far distance.
  The grey regions show the X-ray absorbed column density  range obtained by \citet{J1119Pivovaroff2001}.
  The \textit{fourth panels} indicate the Galactic rotation curve towards the position (RA, Dec)=$\left(291.1\pm0.5,-0.3\right)$ with the red dashed lines delimiting the distance where our column densities match with the X-ray column densities
   while the blue dashed line indicates the distance where the column density in the western region roughly equals the column density in the eastern region.}
  \label{J1119distance}
 \end{minipage}
\end{figure*} 

 \begin{figure*}
 \begin{minipage}{\textwidth}
\centering
  \includegraphics[width=0.5\textwidth]{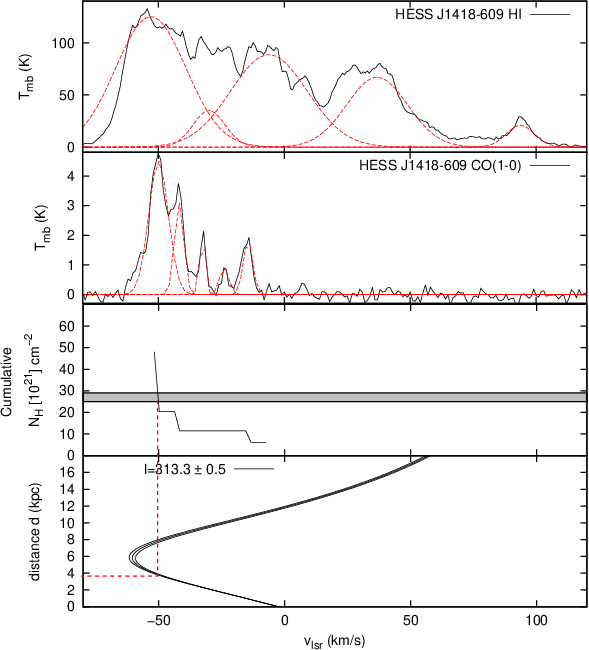}
  \caption{Averaged SGPS H\textsc{i} and Nanten CO(1--0) emission towards  HESS\,J1418$-$609 (see Fig.\,\ref{J1420COCS}) in black solid lines. In both panels, the Gaussian fits are shown as red dashed lines. 
  At each Gaussian peak, the H\textsc{i} and CO(1--0) emission have been integrated and converted into $N_{\text{H}}$ column density via the $X_{\text{CO}}$ and $X_{\text{HI}}$ respectively (see text).
  The bottom panel indicates the evolution of the distance towards the position $\left(313\pm0.5,0.1\right)$ as a function of kinematic velocity.}
  \label{J1418emissiondistance}
  \end{minipage}
\end{figure*}

 \begin{figure*}
 \begin{minipage}{\textwidth}
\centering
  \includegraphics[width=0.5\textwidth]{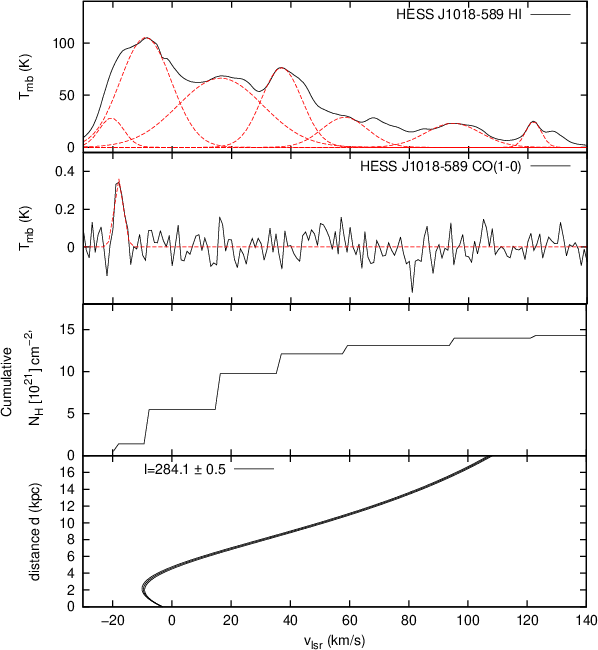}
  \caption{Averaged GASS H\textsc{i} and Nanten CO(1--0) emission towards  HESS\,J1018$-$589 (see Fig.\,\ref{J1018CO}) in black solid lines. In both panels, the Gaussian fits are shown as red dashed lines.
  At each Gaussian peak, the H\textsc{i} and CO(1--0) emission have been integrated and converted into $N_{\text{H}}$ cumulative column density via the $X_{\text{CO}}$ and $X_{\text{HI}}$ respectively (see text).
  The bottom panel indicates the kinematic distance towards the position (RA, Dec)=$\left(284\pm0.5,-1.7\right)$ as a function of kinematic velocity $v_{\text{lsr}}$.}
  \label{J1018emissiondistance}
  \end{minipage}
\end{figure*}
\clearpage
\section[]{HESS\,J1809$-$193 additional figures}
\setcounter{figure}{0}
\label{J1809appendix}
\begin{figure*}
\centering
\begin{minipage}{\textwidth}
 \hbox{
 \includegraphics[width=0.5\textwidth]{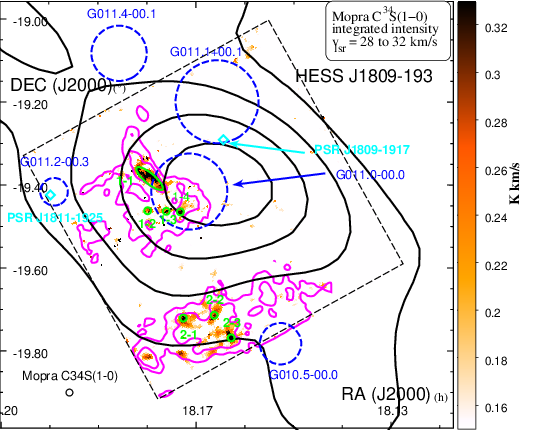}
 \includegraphics[width=0.5\textwidth]{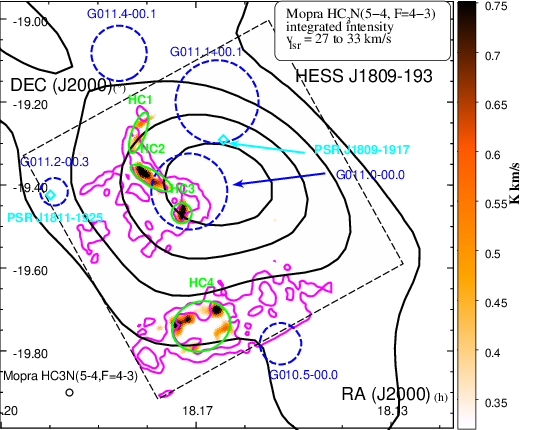}
}
\caption{Mopra C$^{34}$S(1--0) (Left panel) and HC$_3$N(5--4,F=4--3) (right panel) integrated intensity between $v_\text{lsr}=28\text{ to }32$\,km/s and$v_\text{lsr}=27\text{ to }32$\,km/s respectively towards
   HESS\,J1809$-$193. The various C$^{34}$S detections labelled `1--1 to  1--4' and `2--1 to 2--3' (left panel) and the HC$_3$N detections labelled `HC1 to HC4' are shown in green ellipses. 
In both panels, the CS(1--0) integrated intensity between $v_\text{lsr}=25\text{ to }38$\,km/s are shown in purple. The SNRs are shown as dashed blue circles while the position of the pulsars PSR\,J1809$-$1917 and PSR\,J1811$-$1925.} 
\label{extraJ1809figure}
\end{minipage}
\end{figure*}

\end{document}